\documentclass[twocolumn]{aastex631}
\usepackage{comment}
\usepackage{natbib} 

\newcommand{\kms}{km s$^{-1}$}
\newcommand{\Mjsr}{MJy sr$^{-1}$}

\newcommand{\hii}{H~{\scriptsize II}}

\newcommand{\msun}{M$_{\odot}$}

\newcommand{\eg}{e.g.}

\newcommand{\degree}{$^{\circ}$}

\defcitealias{Butterfield2023}{FIREPLACE I}
\defcitealias{Butterfield2024}{FIREPLACE II}

\begin{document}

\title[FIREPLACE III (DR2)]{\uppercase{SOFIA/HAWC+ Far-Infrared Polarimetric Large-Area CMZ Exploration (FIREPLACE) Survey III: Full Survey Data Set}}

\correspondingauthor{Dylan Par\'e}
\email{dylanpare@gmail.com}

\author[0000-0002-5811-0136]{Dylan Par\'e}
\affiliation{Department of Physics, Villanova University, 800 E. Lancaster Ave., Villanova, PA 19085, USA}

\author[0000-0002-4013-6469]{Natalie O. Butterfield}
\affiliation{National Radio Astronomy Observatory, 520 Edgemont Road, Charlottesville, VA 22903, USA}

\author[0000-0003-0016-0533]{David T. Chuss}
\affiliation{Department of Physics, Villanova University, 800 E. Lancaster Ave., Villanova, PA 19085, USA}

\author[0000-0001-8819-9648]{Jordan A. Guerra}
\affiliation{Department of Physics, Villanova University, 800 E. Lancaster Ave., Villanova, PA 19085, USA}
\affil{Cooperative Institute for Research in Environmental Sciences (CIRES), University of Colorado, Boulder, CO 80309 USA}

\author[0000-0001-7466-0317]{Jeffrey Inara Iuliano}
\affiliation{Department of Physics, Villanova University, 800 E. Lancaster Ave., Villanova, PA 19085, USA}

\author[0009-0006-4830-163X]{Kaitlyn Karpovich}
\affiliation{Department of Physics, Villanova University, 800 E. Lancaster Ave., Villanova, PA 19085, USA}

\author[0000-0002-6753-2066]{Mark R. Morris}
\affiliation{Department of Physics and Astronomy, University of California, Los Angeles, Box 951547, Los Angeles, CA 90095-1547 USA}

\author[0000-0002-7567-4451]{Edward J. Wollack}
\affiliation{NASA Goddard Space Flight Center, Mail Code: 665, Greenbelt, MD 20771}

\begin{abstract}

We present the second data release (DR2) of the Far-InfraREd Polarimetric Large-Area CMZ Exploration (FIREPLACE) survey. This survey utilized the Stratospheric Observatory for Infrared Astronomy (SOFIA) High-resolution Airborne Wideband Camera plus (HAWC+) instrument at 214 \micron\ (E-band) at a resolution of 19.6\arcsec\ to observe thermal polarized dust emission throughout the Central Molecular Zone (CMZ). DR2 consists of observations obtained in 2022 covering the region of the CMZ extending from the Brick to the Sgr C molecular clouds (corresponding to a 1\degree\ $\times$ 0.75\degree\ region of the sky). We combine DR2 with the first FIREPLACE data release (DR1) to obtain full coverage of the CMZ (a 1.5\degree\ $\times$ 0.75\degree\ region of the sky). After applying total and polarized intensity significance cuts on the full FIREPLACE data set, we obtain $\rm\sim$64,000 Nyquist-sampled polarization pseudovectors. The distribution of polarization pseudovectors confirms a bimodal distribution in the CMZ magnetic field orientations, recovering field components that are oriented predominantly parallel or perpendicular to the Galactic plane. This distribution of orientations is similar to what was observed in DR1 and other studies. We also inspect the magnetic fields toward a set of prominent CMZ molecular clouds (the Brick, Three Little Pigs, 50 \kms\ and 20 \kms\ clouds, Circum-nuclear Disk, CO 0.02-0.02, and Sgr C), revealing spatially varying magnetic fields having orientations that generally trace the total intensity morphologies of the clouds. We find evidence that compression from stellar winds and shear from tidal forces are prominent mechanisms influencing the structure of the magnetic fields.

\end{abstract}


\section{INTRODUCTION} \label{sec:intro}

\begin{figure*}
    \centering
    \includegraphics[width=1.0\textwidth]{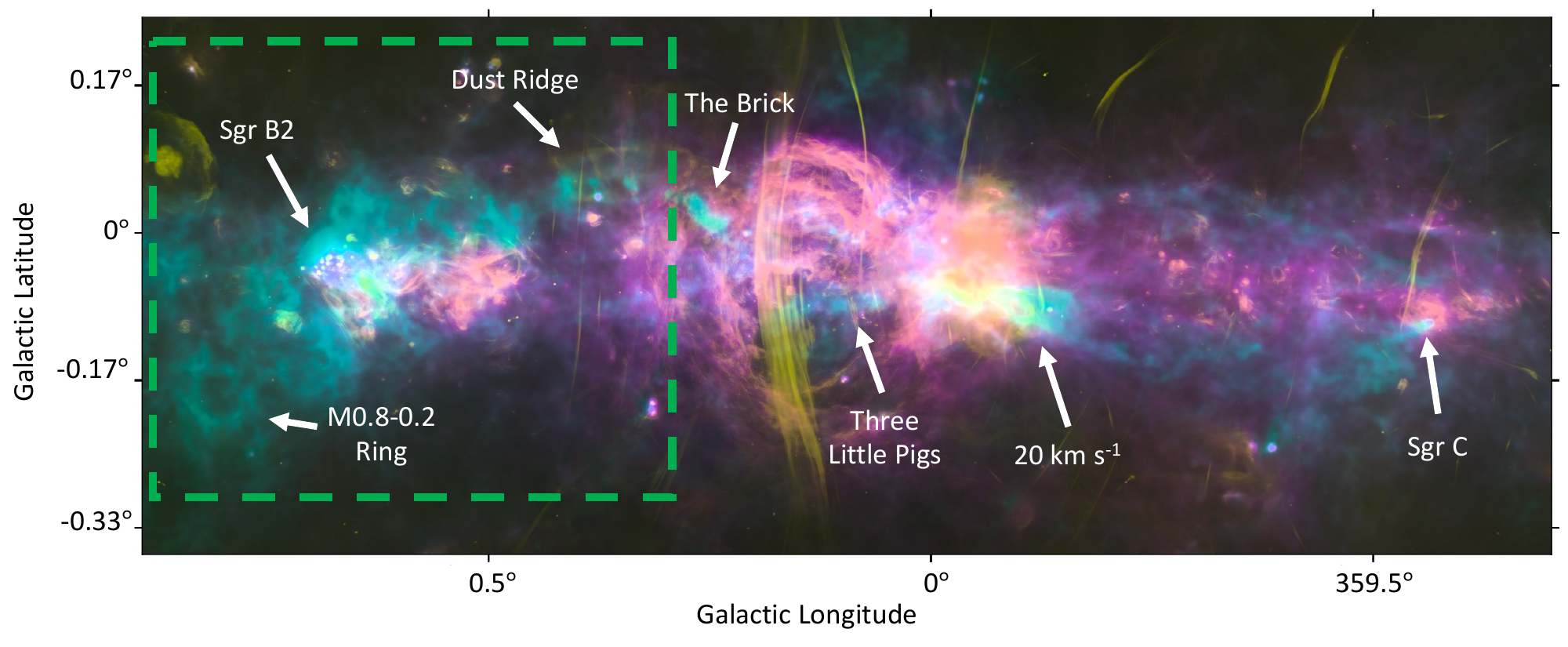}
    \caption{A 3-color view of the CMZ with 20 cm (1 GHz) MeerKAT radio emission in yellow \citep{Heywood2022} and 250 \micron{} cool dust and 70 \micron{} warm dust emission in cyan and magenta, respectively, both observed by Herschel \citep{Molinari2011}. The green dashed box roughly marks the region covered by the FIREPLACE DR1 observations presented in \citet{Butterfield2023}. Prominent CMZ molecular clouds are marked and labeled.}
    \label{fig:legend}
\end{figure*}

The Central Molecular Zone (CMZ) of the Galactic Center (GC) is an extreme region of the Milky Way, containing elevated molecular densities \citep[10$\rm^3$ -- 10$\rm^6$ cm$^{-3}$; \eg,][]{Mills2018} and magnetic field strengths \citep[100s of $\rm\mu$G -- mG,][]{Yusef-Zadeh1987a,Plante1995,Chuss2003a,Pillai2015,Hsieh2018,Mangilli2019,Guerra2023} compared to what is observed in the Galactic disk. A 3-color view of this region probed by radio and infrared observations is shown in Figure \ref{fig:legend}, where prominent molecular clouds throughout the CMZ have been labeled. The combination of its extreme properties and proximity to Earth compared to other galactic nuclear regions makes it a useful proxy for more distant gas-rich galactic nuclei throughout the universe.

The star formation rate (SFR) is much lower in the CMZ than in the Galactic disk \cite[e.g.][]{Longmore2013,Morris2023}. The low SFR is particularly unexpected given the amount of dense gas in the CMZ \citep[\eg{} 2-6$\rm\times$10$\rm^{7}$ \msun,][]{Morris1996b,Barnes2017}. There are multiple possible explanations for the low SFR in the CMZ, such as the strength and compressibility of turbulence in the region or that the CMZ could be in a period of inactivity between episodic bursts of star formation \citep{Krumholz2015}. Another factor that could contribute to the inhibition of star formation in the CMZ is the elevated magnetic field strengths observed in the region \citep[e.g.,][]{Morris1989iaus,Morris1993}.

The distribution of the magnetic field in the CMZ has been studied using both radio and infrared observations. Radio observations reveal a population of synchrotron structures known as non-thermal filaments (NTFs) that are understood to trace ordered magnetic fields. These NTFs are predominantly oriented perpendicular to the Galactic plane \citep{Morris1996a,Yusef-Zadeh2022,Heywood2022}. The general orientation of the NTF population has been interpreted to indicate a large-scale vertical (or possibly poloidal) magnetic field that pervades the GC \citep{Yusef-Zadeh2004,Morris2006sum}.

Conversely, the magnetic field derived from large-scale studies of dust polarization at sub-millimeter and infrared wavelengths reveals a distribution that seemingly conflicts with the vertical magnetic field traced by the NTFs. The magnetic field derived from the dust polarization observations is aligned parallel to the distribution of the molecular clouds within the CMZ -- an orientation that is largely parallel to the Galactic plane and consistent with a horizontal (or possibly toroidal) field \citep{Novak2000,Nishiyama2010,Mangilli2019,Guan2021}. 

One possible explanation for these different magnetic field orientations could be the distinct mechanisms responsible for the polarization (relativistic electrons illuminating the NTFs for the radio observations compared to dust-grain alignment for the infrared observations). However, polarimetric observations at 90 and 150 GHz reveal regions where the magnetic field becomes vertical rather than horizontal \citep{Guan2021}. These vertical magnetic field regions could either indicate connections between the fields traced by dust polarization and the NTFs or a spatially variable distribution comprised of multiple magnetic field components \citep{Guan2021}. The latter is likely given the sensitivity of these particular observations to both thermal dust and synchrotron emission.

The Far-InfraREd Polarimetric Large Area CMZ Exploration (FIREPLACE) survey was proposed to address the question of how the magnetic field at infrared wavelengths varies throughout the CMZ by utilizing  observations made by the High-resolution Airborne Wideband Camera plus (HAWC+) instrument on the Stratospheric Observatory for Infrared Astronomy (SOFIA). The first FIREPLACE data release (DR1) covered a region of the CMZ centered on Sgr B2 \citep[][hereafter reffered to as \citetalias{Butterfield2023}]{Butterfield2023}. The field of view of these first FIREPLACE observations is marked by the green dashed rectangle in Figure \ref{fig:legend}. The DR1 observations reveal a more spatially varying magnetic field at 19.6\arcsec\ resolution than is seen in previous CMZ-wide surveys \citep[$\rm\sim$1\arcmin,][]{Mangilli2019,Guan2021}. An in-depth study of the magnetic field system local to one of the molecular structures identified in DR1 indicates that the magnetic field probed by FIREPLACE traces smaller-scale molecular structures local to the CMZ than in these previous CMZ-wide surveys \citep[][hereafter referred to as \citetalias{Butterfield2024}]{Butterfield2024}.

In this paper we present the second FIREPLACE data release (DR2), which is comprised of SOFIA/HAWC+ observations covering the extent of the CMZ from the Brick to Sgr C (a roughly 1\degree\ $\times$ 0.75\degree\ region of the sky). We then combine the DR1 and DR2 FIREPLACE data sets to obtain a polarimetric map with full coverage of the CMZ (a 1.5\degree\ $\times$ 0.75\degree\ region). In Section \ref{sec:reduc} we describe the data reduction procedure applied to the DR2 FIREPLACE observations to make maps from the raw observations obtained from SOFIA/HAWC+. Section \ref{sec:qa} describes the data quality checks performed to verify the reduction procedure. The CMZ-wide FIREPLACE observations are then presented and discussed in Section \ref{sec:full}. In Section \ref{sec:cloud_mag} we study the detailed polarimetric properties of prominent molecular clouds in our DR2 data set and we present our conclusions in Section \ref{sec:conc}.

\section{OBSERVATIONS AND DATA REDUCTION} \label{sec:reduc}
\begin{figure*}
   \centering
  \includegraphics[width=1.0\textwidth]{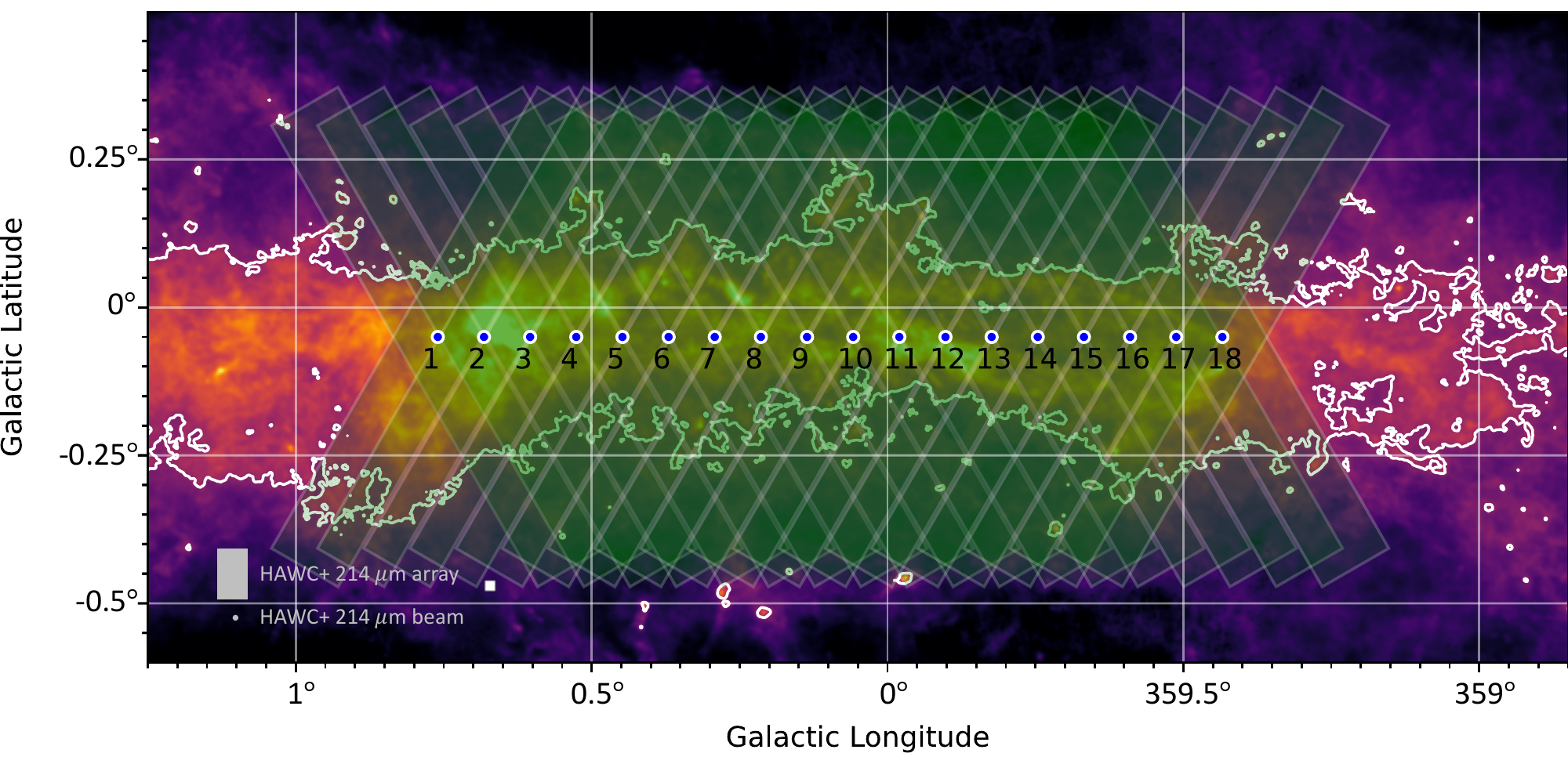}
   \caption{The observing strategy for the FIREPLACE survey is shown. Numbered dots show the centers of each field and semi-transparent rectangles show the approximate areas covered by the boresight Lissajous trajectories. The background color scale is Herschel 250 \micron\ data \citep{Molinari2011}, with the white contour level representing the targeted sensitivity of FIREPLACE (7189 MJy sr$\rm^{-1}$). The field numbers of the FIREPLACE observations discussed in the text are labelled in the figure. The HAWC+ array footprint and beam size at 214 \micron\ (E-band) are shown in the lower left corner of the figure. The fields shown in this figure are described in more detail in Appendix \ref{sec:cal_tab}.}
   \label{fig:scans}
\end{figure*}
\subsection{DR2 Observations: 2022 SOFIA Flights} \label{sec:obs}
FIREPLACE DR2 is an expansion of the FIREPLACE pilot program \citepalias{Butterfield2023}. The DR1 pilot program observed the eastern third of the CMZ (the greater Sgr B2 complex) during SOFIA flights F775 and F777 on August 31, 2021 and September 2 2021, respectively. The new DR2 release expands on this pilot program to include observations obtained from several 2022 SOFIA flights: four Southern hemisphere Christchurch-based flights in June and July 2022 (flights F890, F891, F893, and F895) and three Northern hemisphere Palmdale-based flights in September 2022 (flights F916, F917, and F918).

The full FIREPLACE survey covers the central 1.5\degree\ $\times$ 0.75\degree\ of the CMZ with a pixel size of 5.3\arcsec, beam size of 19.6\arcsec, central wavelength of 214 \micron, and a bandwidth of 175 \micron. This is the largest polarimetric map made by SOFIA. Given both the large map size and the extended flux in this region at 214 \micron, the standard Chop-Nod-Match \citep[CNM,][]{Harper2018} observing strategy was not practical.  Instead, FIREPLACE utilized the on-the-fly mapping (OTFMAP) mode. The OTFMAP mode involves rapid scanning of the HAWC+ array over the field to create time streams that are subsequently fit for correlated noise components and map values using the SOFIA data reduction pipeline, which is based on the \textit{CRUSH} algorithm \citep{Kovacs2008}. This is done for each of four angular positions of the half-wave plate (0\degree, 22.5\degree, 45\degree, and 67.5\degree), and the Stokes $I$, $Q$, and $U$ parameters are fit to the resulting images. A detailed description of the OTFMAP polarimetric observing mode using HAWC+ is presented in \citet{Lopez-Rodriguez2022}.

Even with the OTFMAP mode strategy, the full extent of the CMZ is larger than what can be feasibly observed with a single scan. Therefore we developed an observing plan utilizing  a set of fields that fully cover the CMZ as shown in Figure \ref{fig:scans}. The fields are reduced individually using the SOFIA OTFMAP reduction software as described below and then merged into a final map. This observation approach is novel, and there is significant correlated noise on short timescales. Therefore, care has to be taken planning the observations, in optimizing the reduction parameters, and verifying the data set. The reduction process is described in Section \ref{sec:red_meth} and the data verification methods are discussed in Section \ref{sec:qa}.

For FIREPLACE, each scan is designed to cross the Galactic plane at an angle of either $\rm\pm$30\degree\ relative to Galactic North. For each scan, we follow the same parameters as \citetalias{Butterfield2023} -- the boresight of the detector follows a Lissajous scan with amplitudes of 27\arcmin\ $\times$ 4.5\arcmin.  The ratio of the frequencies between the parametric sinusoidal curves of the Lissajous pattern is set to $\rm\sim\sqrt{2}$, with the higher frequency corresponding to the long axis.  The on-source time for each scan is 120 seconds and the scan rate is 300\arcsec\ s$\rm^{-1}$.
The planned boresight path for each field of view is shown in Figure \ref{fig:scans}. The numbering in the figure marks the center of each field and the area covered by each boresight trajectory is indicated by the semi-transparent rectangles. Each field was observed using two different scan alignments that are oriented $\rm\pm$30\degree\ with respect to Galactic North. The region targeted for polarimetry measurements is shown by the white contour (at a level of 7189 \Mjsr\ at 250 \micron). 

OTFMAP mode is inherently a differential image reconstruction technique that does not preserve the map intensity zero point. For this reason, each scan was designed with two considerations in mind. First, scans were required to have sufficient extent for the detector array to cross an effective ``zero reference'' common to other scans, where the instrument field of view encounters a low-intensity reference region that is subdominant to CMZ molecular cloud intensity. Second, an overlap of two-thirds is employed between adjacent fields for repeated measurements of any sky position in our target region. We used the same overlap for both scan orientations to ensure additional measurements and improve the cross-linking of our observations. This latter point enables sufficient cross-linking between consecutive scans for the simultaneous fitting of all of the time streams when generating the combined map.  

Although the current version of the SOFIA pipeline does not allow for simultaneous fitting over all observational scans, future data reduction efforts are being explored to take advantage of this aspect of the scan strategy for increased fidelity in recovering emission from larger-scale structures. This future analysis can incorporate CMB map-making strategies to reconstruct spatial structures larger than the instantaneous array field of view. Full sky implementations have been explored previously \citep[e.g.,][]{Wright1996,Tegmark1997}. More recently there are examples of implementations at smaller angular scales \citep{Aiola2020,Qu2023,Eimer2023,Li2023}.

Observations in the region spanning fields 1 - 5 were taken in 2021 and comprise the DR1 FIREPLACE observations presented in \citetalias{Butterfield2023}. Note that the 2021 fields were implemented differently than those shown in Figure \ref{fig:scans} due to setup issues during the 2021 observation runs (see Figure 2 of \citetalias{Butterfield2023} which shows the actual orientations of the DR1 observation fields). DR2 re-observed some of the fields covered in DR1 (fields 4 and 5), to increase the spatial overlap between the DR1 and DR2 observations. However, DR2 primarily focuses on fields 6-18 to complete the map of the CMZ (Figure \ref{fig:scans}). 

\subsection{Data Reduction} \label{sec:red_meth}
To reduce the raw observations obtained from the SOFIA Infrared Science Archive at IPAC (IRSA)\footnote{https://irsa.ipac.caltech.edu/frontpage/} we use the HAWC+ Data Reduction Pipeline (DRP). We used DRP version 2.7.0, which is a Python-wrapper built on a Java version of the \textit{CRUSH} reduction algorithm \citep{Kovacs2008}. Since the latest version of DRP (version 3.2.0) has not been verified for such large maps with extended emission we did not use this more recent version of DRP to reduce our HAWC+ observations. The use of DRP version 2.7.0 has the added benefit of maintaining consistency with DR1, which also used this DRP version.
 
Apart from \citetalias{Butterfield2023}, the OTFMAP polarimetry mode on SOFIA has previously been demonstrated only for sources that have spatial extents smaller than the instantaneous field-of-view of the HAWC+ detector \citep{Lopez-Rodriguez2022}. The abundance of diffuse, extended sources in the CMZ therefore presents a unique challenge, given that it is difficult for \textit{CRUSH} to differentiate between these extended structures and correlated noise. In the reduction of our 2022 observations we closely follow and build on the methods from \citetalias{Butterfield2023}.
 
\begin{figure*}
   \centering
  \includegraphics[width=1.0\textwidth]{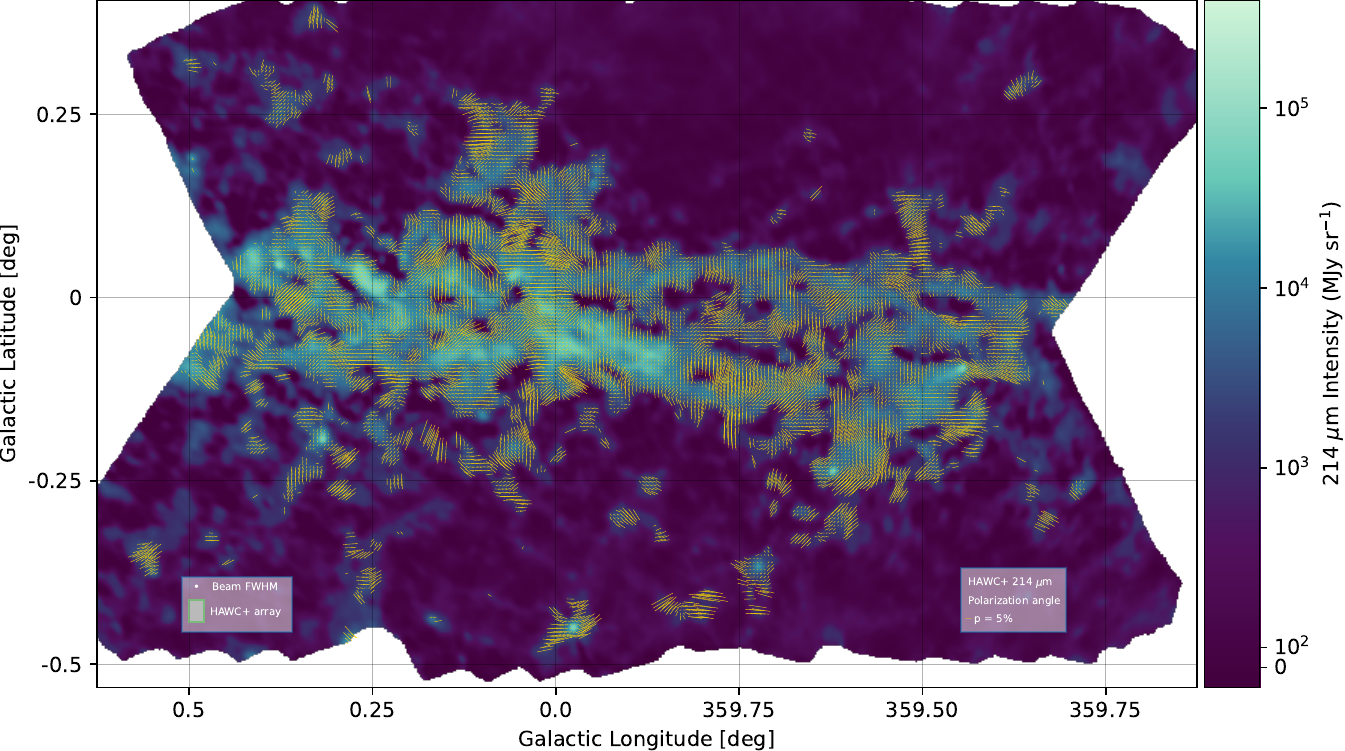}
   \caption{Total intensity distribution of merged 214 \micron\ FIREPLACE fields from the 2022 flights. Yellow pseudovectors reveal the polarization orientations derived from FIREPLACE. Inset boxes show a representative polarization angle, the beam size, and the instantaneous HAWC+ array size.}
  \label{fig:fire_2022}
\end{figure*}
 
To reduce the DR2 observations using \textit{CRUSH} we set the rounds parameter to ``\textit{-rounds=85},'' which increases the number of iterations \textit{CRUSH} performs when numerically modeling the correlated noise. We also set the ``\textit{-extended}'' option, which indicates that \textit{CRUSH} needs to account for extended emission during data reduction to better preserve large scale flux when identifying correlated noise. Furthermore, we specified ``\textit{-fixjumps}'' and ``\textit{-downsample=1}'' to remove the effects of quantum flux jumps in the SQUID amplifiers and to prevent spatial averaging, respectively. These \textit{CRUSH} settings match the parameters used for the FIREPLACE DR1 data release presented in \citetalias{Butterfield2023}. For each reduced field, DRP removes the $\sim$2\% instrumental polarization that \citet{Harper2018} found in all HAWC+ bands.

Previous polarimetric studies using SOFIA/HAWC+ observations have found that using the ``\textit{-extended}'' flag introduced artificial polarization in low-intensity regions \citep{Lopez-Rodriguez2022,LeGouellec2023}. We observed this same phenomenon when reducing the FIREPLACE observations using DRP version 3.2.0, but not when using DRP version 2.7.0. Figure \ref{fig:artcomp} in Appendix \ref{sec:artifacts} provides a representative example of the artifacts encountered when using DRP version 3.2.0 for a single FIREPLACE field. The data reductions presented in \citet{LeGouellec2023} and \citet{Lopez-Rodriguez2022} also describe an extensive ``zero-level background'' correction. In this work we use the filtered version of the FIREPLACE observations obtained from the \textit{CRUSH} output and do not correct for the intensity ``zero-point.''
 
Using the parameters specified above we reduced each field separately. For each field, all scans were analyzed simultaneously where possible (8 scans for most fields accounting for 2 sets of 4 HWP angles). In some cases, an observation consisted of only one set of 4 scans. In these cases, 4 scans were included in the \textit{CRUSH} step. Table \ref{tab:DR2} in Appendix \ref{sec:cal_tab} provides details on each of the scans observed and their mapping to the fields shown in Figure \ref{fig:scans}. Table \ref{tab:DR2} also indicates how many scans were obtained for each field. 

Due to systematically high correlated noise that was evident in the post-reduction scans for the relevant fields, data from flight F890 were omitted from the final reduction. Fortunately, the regions observed during this flight were also observed during other flights, and so we still obtain full CMZ coverage. For completeness, Table \ref{tab:DR2} includes information on the observations conducted during flight F890, \textit{but we emphasize that the results presented in this work do not incorporate the observations obtained from that flight.}
 
After reducing each field using \textit{CRUSH}, the resulting total intensity emission of the fields was cross-correlated with Herschel 250 \micron\ observations from \citet{Molinari2011} to identify pointing offsets in our fields. We found that the majority of fields corresponding to the 2022 observations had pointing offsets on the scales of $\rm\sim$10\arcsec (corresponding to offsets of 2 or more pixels in our map). These significant pointing corrections were applied to all fields in which an offset of more than 5\arcsec\ (angular size of 1 pixel in our map) was identified. Correcting the FIREPLACE pointings using the Herschel 250 \micron\ observations completely corrects these offsets. Offsets smaller than 5\arcsec\ are below the pixel size of our observations, so we did not implement offset corrections for these fields.
 
After applying the offset correction, the instrumental polarization is then removed from each field. The DRP has a built-in instrumental polarization correction which has been applied to the reduced FIREPLACE observations. The instrumental polarization is determined using images of the sky, and the polarization from these calibrated observations is subtracted from the $Q$ and $U$ FIREPLACE data sets. This correction is expected to be accurate within $Q$/$I$ and $U$/$I <$ 0.3\% according to the HAWC+ handbook\footnote{https://irsa.ipac.caltech.edu/data/SOFIA/docs/instruments/hawc/index.html}. 

The calibrated, offset-corrected, and instrumental polarization-corrected fields were then co-added into one combined data set using the DRP \textit{merge} algorithm. The \textit{merge} algorithm resamples the input fields onto a common pixel grid based on the input World Coordinate Systems of the individual fields. The input pixels are then weighted based on their distance from the nearest coordinate in the common grid. Figure \ref{fig:fire_2022} shows the total intensity distribution for the merged 2022 observations with polarization angle pseudovectors shown as yellow lines. These polarization angles are calculated using
\begin{equation}
   \phi = \frac{1}{2}\arctan\frac{U}{Q} ,\label{eq:phi}
\end{equation}
where $Q$ and $U$ are defined following the IAU convention where North corresponds to an angle of 0\degree{} with $\rm\phi$ increasing in the counterclockwise direction\footnote{The original reduction was done in equatorial coordinates. Thus ``North'' here refers to ``equatorial North.'' When switching to Galactic coordinates, the angle $\phi$ is rotated to be relative to ``Galactic North.''}.
 
The length of each pseudovector in Figure \ref{fig:fire_2022} is proportional to the debiased fractional polarization defined as
\begin{equation}
   p = \sqrt{p^2_m - \sigma^2_p} ,
\end{equation}
where $\sigma_p$ is the estimate of the uncertaintainty of the measured fractional polarization propagated from the fitted weights in the \textit{CRUSH} step, and $p_m$ is the measured fractional polarization. Pseudovectors are only derived for lines-of-sight having total intensity above a significance threshold of $I/\sigma_I > 200$, percentage polarizations less than 50\% ($p{}<$ 50\%), and significant polarized intensity ($p/\sigma_p > 3$), which are quality cuts consistent with standard SOFIA polarimetry practice \citep{Gordon2018}. The individual calibrated files and the merged map of the entire CMZ are publicly available and can be downloaded from the NASA Infrared Processing and Analysis Center (IPAC) Legacy Surveys webpage as FITS files\footnote{https://irsa.ipac.caltech.edu/data/SOFIA/docs/data/legacy-programs/two-color-polarimetric-survey-galactic-center-pilot-legacy-program/index.html}.

\section{DR2 Quality Checks} \label{sec:qa}
\begin{figure*}
    \centering
    \includegraphics[width=1.0\textwidth]{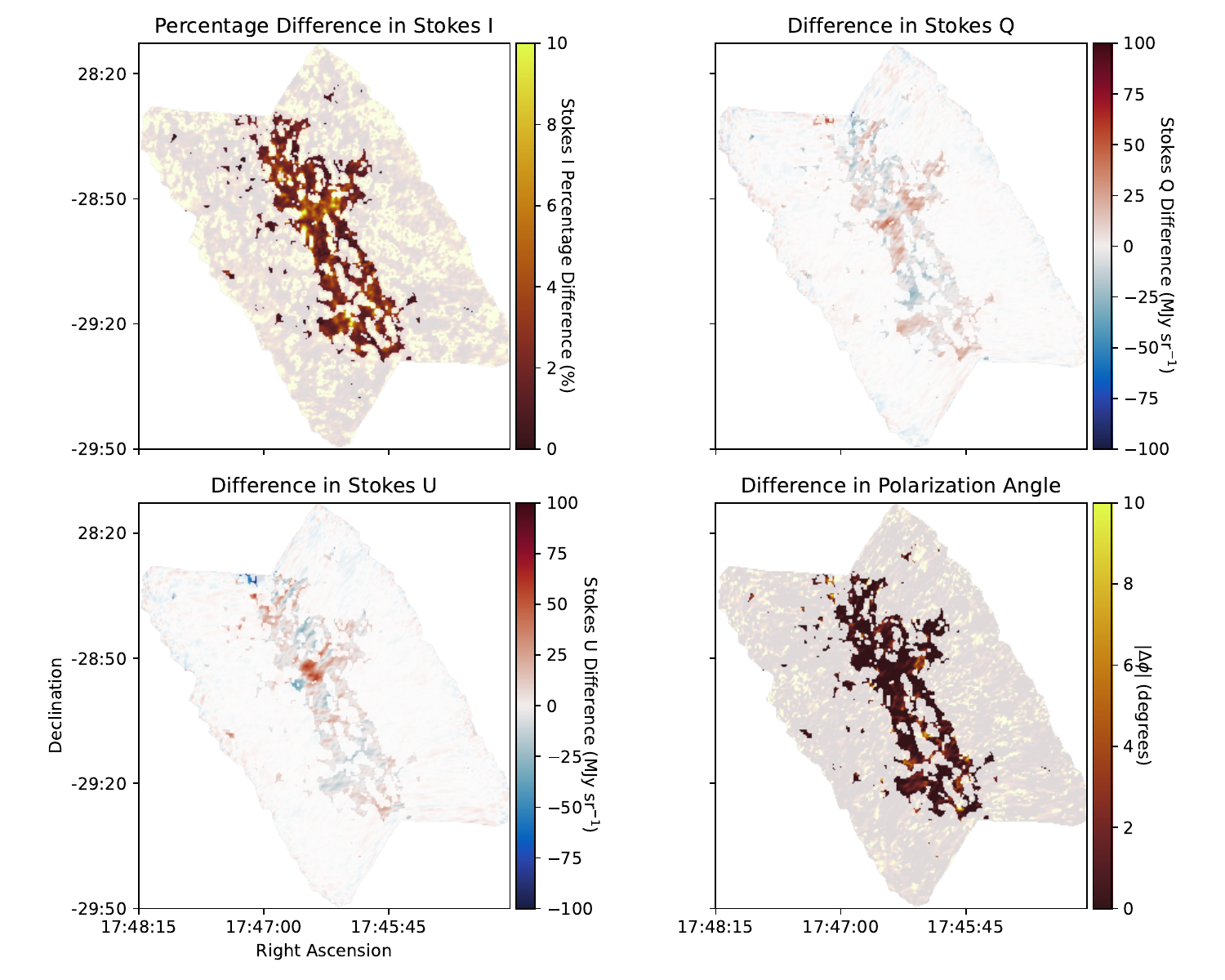}
    \caption{Maps of the differences in Stokes $I$ (top left), $Q$ (top right), $U$ (bottom left), and polarization angle (bottom right) between the 75 and 85 rounds reductions of our FIREPLACE observations. The highlighted portion of the distributions indicates the region of significant total and polarized intensity at which the polarization pseudovectors are determined using the significance cuts discussed in Section \ref{sec:red_meth}.}
    \label{fig:convergence}
\end{figure*}
\subsection{Testing Convergence of \textit{CRUSH} Reduction} 
The large amount of extended emission present in our observations of the CMZ requires that we run \textit{CRUSH} for more than the default 15 rounds of fitting iterations. To confirm that the \textit{CRUSH} fits are converging using this larger number of rounds, we performed a second reduction of our HAWC+ observations that used the same \textit{CRUSH} parameters (e.g. using the ``\textit{-extended}'' flag and setting ``\textit{-downsample=1}'') but changing the number of rounds to 75 instead of 85 (``\textit{-rounds=75}'' instead of ``\textit{-rounds=85}''), matching the reduction verification conducted in \citetalias{Butterfield2023}. We then created difference maps of Stokes $I$, $Q$, $U$, and of the polarization angle ($\phi$) between the 85 and 75 rounds reductions to assess whether the different numbers of iterations are converging to similar solutions. The difference maps for the merged 2022 flights are shown in Figure \ref{fig:convergence}.

The top left panel of Figure \ref{fig:convergence} shows the difference in Stokes $I$ between the different number of rounds as a percentage difference. These differences are generally $\rm\lesssim$1\% in the region satisfying the significance cuts described in Section \ref{sec:red_meth}. A percentage difference of 1\% corresponds to a magnitude difference of 1000 \Mjsr\ in Stokes $I$. The magnitudes of the differences in the Stokes $Q$ (upper right panel of Figure \ref{fig:convergence}) and $U$ (lower left panel of Figure \ref{fig:convergence}) distributions are on the order of $\sim$100 \Mjsr, corresponding to $\sim$1\% the $Q$ and $U$ intensities observed in the significance region. Because Stokes $Q$ and $U$ are angle-based parameters, we express the differences in $Q$ and $U$ as absolute differences rather than percentage differences in Figure \ref{fig:convergence}. The polarization angle differences shown in the lower right panel of Figure \ref{fig:convergence} are  generally $\rm<$10\degree. Since a 10\degree\ uncertainty in polarization angle corresponds to $p$/$\sigma_p = 3$, the variation resulting from the different numbers of rounds is smaller than the statistical noise resulting from the significance cuts employed in this paper. We therefore conclude that changing the number of iterations from 75 to 85 rounds is not significantly changing the results obtained from the \textit{CRUSH} algorithm. 

\subsection{Variance Within DR2 Observations} \label{sec:variance}
\begin{figure*}
    \centering
    \includegraphics[width=1.0\textwidth]{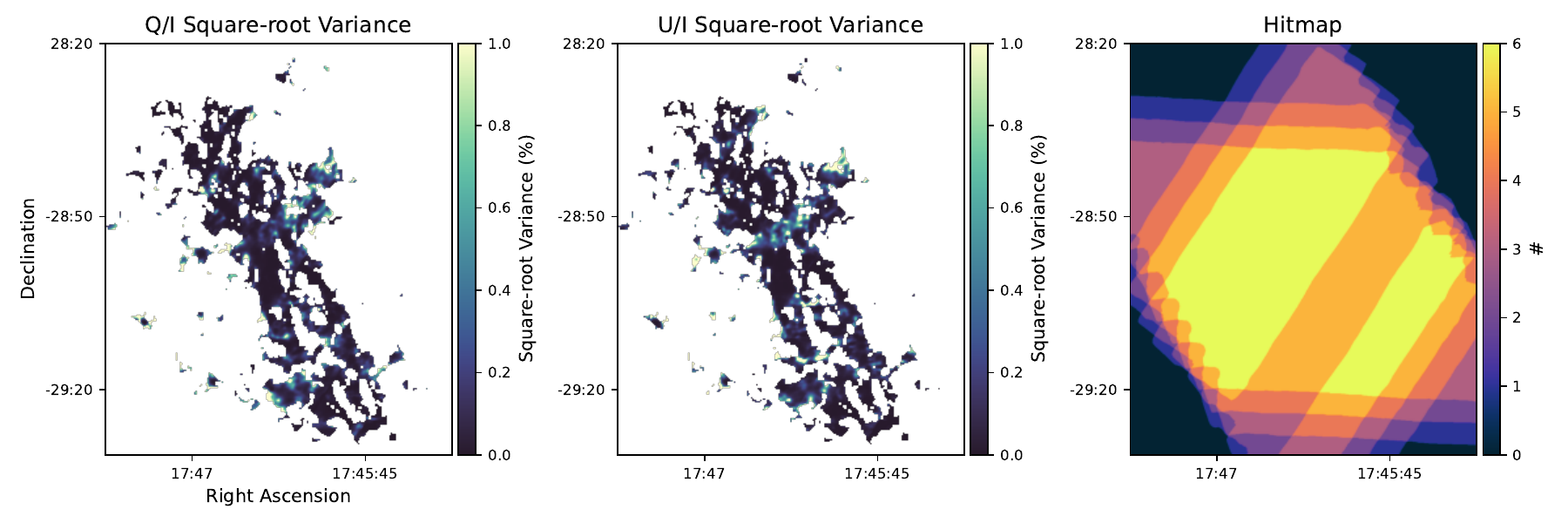}
    \caption{Square-root of the variances obtained for the fractional $q$ (right) and $u$ (middle) distributions from our 2022 observations in regions of significant polarization using the standard SOFIA total and polarimetric cuts. The right panel shows the hitmap indicating how many submerges overlap at each pixel location in the full 2022 data set.}
    \label{fig:variance}
\end{figure*}
We further test the convergence of the \textit{CRUSH} reduction by computing the square-root of the variance for the fractional $q$ and $u$ parameters across the entire CMZ. The fractional Stokes parameters (in percentage) are defined such that: $q = 100\times(Q/I); u = 100\times(U/I)$. We are able to compute variance maps of our observations because our observing strategy ensures that each pixel is sampled in multiple fields. The overlap between each field is such that an image made using every third field would have contiguous, non-redundant coverage over most of the CMZ. We therefore create sub-merges using every third field and for both of the $\rm\pm$30\degree\ scan orientations, resulting in 6 sub-merges of the FIREPLACE observations that cover the extent of the CMZ observed in DR2. We compute the variance for each line-of-sight in our observations by treating the measurement from the full merged data set at that sky location as the ``model'' value, with each of our six data subsets being the different measurements of this ``model'' value. 

The square-root of the variance values obtained for the fractional $q$ and $u$ distributions are shown in the left and middle panels of Figure \ref{fig:variance}. The right panel of Figure \ref{fig:variance} displays the hitmap indicating how many of the six subsets of observations overlap with each pixel evaluated in this procedure. The central region of this hitmap peaks at an overlap of six. The number of overlapping subsets decreases towards the edge of the map because of how the fields are distributed and oriented on the sky (Figure \ref{fig:scans}).

We compute the variance only for the lines-of-sight that pass the significance thresholds discussed at the end of Section \ref{sec:red_meth}. In these regions of significant total and polarized intensity, the square root of the variance is generally $<$1\%. The variance is also generally quite flat, though there are some regions where the variance is $\rm\sim1\%$. The small percentage variances observed indicate that the Stokes $I$, $Q$, and $U$ emission is generally consistent across the different sub-portions of the combined 2022 data set. Furthermore, these variances are of the order of the reported instrumental polarization of the HAWC+ instrument ($\sim$2\%), indicating the impact of polarization leakage in high $\rm{}I_{214}$ emission regions may be negligible. Some areas of the map indicate the possible presence of systematic errors, however, as can be seen where the $q$ and $u$ square root variance approaches 1\%.

\subsection{CNM Observing Mode Comparison} \label{sec:chop}
To further validate the quality of our reduction, we follow the procedure employed in \citetalias{Butterfield2023} to compare our observations with archival CNM mode observations for 3 prominent molecular clouds covered by FIREPLACE DR2: the Brick, the Circum-nuclear Disk (CND), and Sgr C. These CNM mode data sets are available on the SOFIA:IRSA data archive. The CNM mode data set for the Brick, consisting of 5 files, was obtained on flight 394 on May 12, 2017 by Principal Investigator T. Pillai (proposal ID: 05\_0206). The CNM mode data for the CND and Sgr C regions were obtained on flight 397 on May 18, 2017 by Principal Investigator D. Chuss (proposal ID: 05\_0018), consisting of 4 files for the CND and 3 files for Sgr C.

\begin{figure*}
    \centering
    \includegraphics[width=0.95\textwidth]{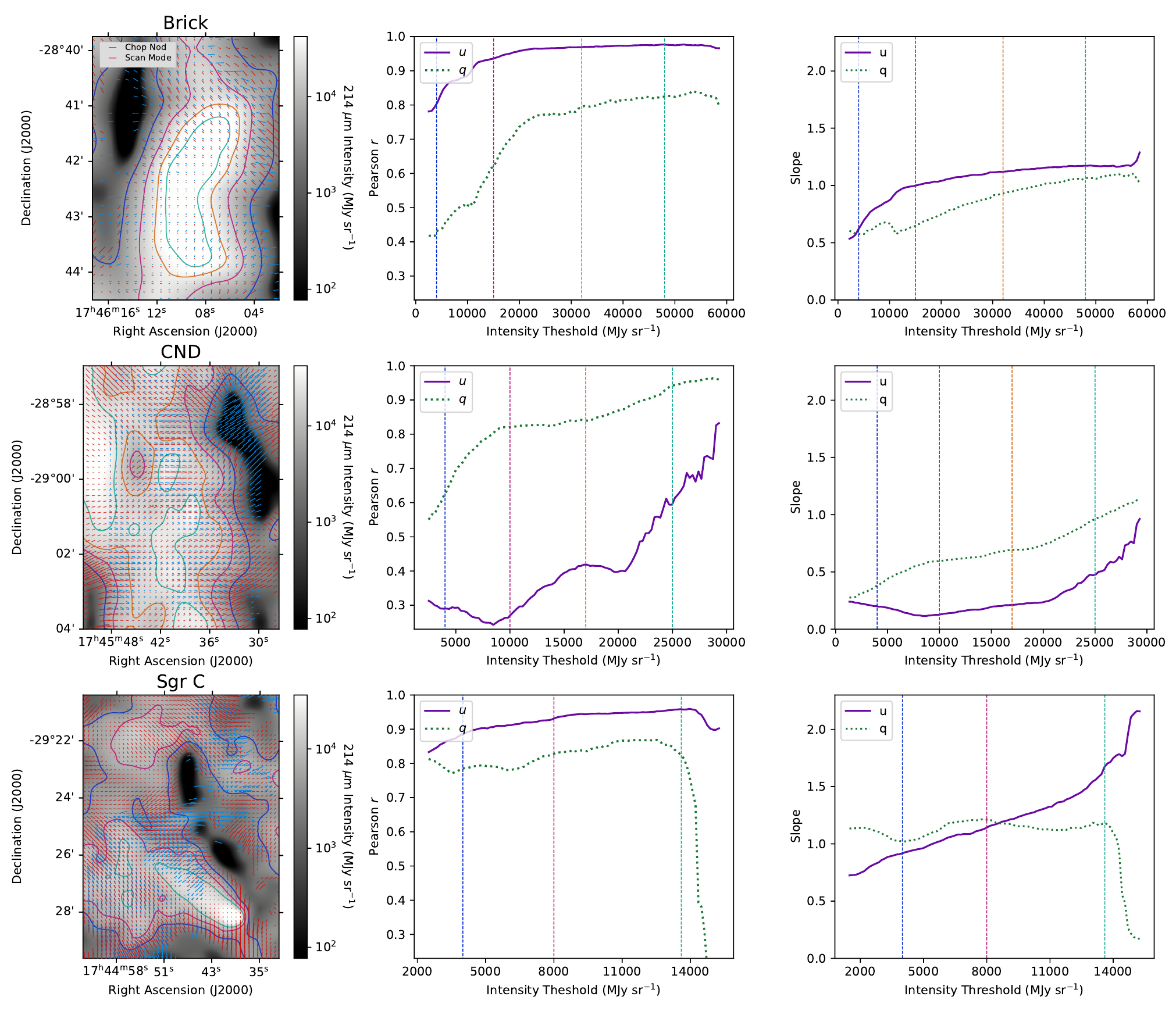}
    \caption{(Left Column) Comparisons between the CNM mode (blue) and OTFMAP mode (red) polarization pseudovectors in the Brick, CND, and Sgr C. OTFMAP mode and CNM mode pseudovectors are spatially shifted from one another to allow for better comparison of vector orientation. (Middle Column) For each cloud we show the Pearson $r$ coefficient between the normalized fractional Stokes parameters, $u$ (solid purple) and $q$ (dotted green) of the CNM mode and OTFMAP mode data as a function of intensity threshold. (Right Column) slope of $q$ and $u$ vs intensity threshold, where slope is determined by fitting the CNM mode emission to the OTFMAP mode emission using a linear model (e.g. $I_{CNM} = m\times{}I_{OTFMAP} + b$. The contours shown in the left column correspond to the intensity levels marked by the vertical lines of the same colors in the middle and right columns. These intensity levels were chosen to indicate the extent and intensity levels of the targeted molecular clouds.} 
    \label{fig:cn}
\end{figure*}

Figure ~\ref{fig:cn} (left column) shows the inferred polarization pseudovectors of the CNM mode (blue pseudovectors) and OTFMAP mode (red pseudovectors) data in the Brick (top), the CND (middle), and Sgr C (bottom). The pseudovectors in these maps have been shifted by 1 pixel ($\sim$5\arcsec) from one another for ease of comparison. We applied the same significance cuts to the CNM mode observations that were applied to our OTFMAP mode observations discussed in Section \ref{sec:red_meth} of $p/\sigma_p>3$, $p<$50\%, and $I/\sigma_I>200$ to preserve consistency between the two observing modes. The polarization angles of the CNM mode and OTFMAP mode observations generally agree for intensities greater than the red dashed vertical line in the middle and right columns of Figure \ref{fig:cn}. Less agreement is seen around low intensity regions for all three clouds where the reference beam contamination for the CNM mode observations is potentially more significant (i.e. the spatial filtering differences between the two observing methods is most impactful).

The middle column of Figure \ref{fig:cn} quantifies the comparison through calculation of the Pearson $r$ correlation between the normalized Stokes $q = Q/I$ and $u = U/I$ parameters of the CNM mode and OTFMAP mode data as a function of a total intensity threshold. The Pearson-$r$ value is calculated only between points where the total intensity exceeds the intensity threshold, as determined from the OTFMAP mode total intensity map. The maximum intensity value was chosen to ensure that a minimum of 100 pixels (approximately 6 independent beams) are included in each correlation.

For intensities less than the red dashed vertical line in the right panels of Figure \ref{fig:cn} the correlations are observed to drop. This drop is likely due to differences in spatial filtering between the OTFMAP and CNM mode observations, since spatial filtering is more significant in lower total intensity regions. At higher total intensities there are discrepancies between the Stokes parameters where one of the normalized Stokes parameters has a higher Pearson $r$ correlation than the other. For example, the correlations derived for the CND reveal a systematically higher correlation for normalized fractional Stokes $q$ than for normalized fractional Stokes $u$. This discrepancy is caused by different signal-to-noise levels for the different normalized parameters where in the case for the CND there is generally much higher signal in $q$ than in $u$. In general, for all three clouds there is an increase in the Pearson $r$ correlation with increasing intensity. This trend is expected, since at higher total intensities contamination in the reference beam is less significant.

The right column of Figure \ref{fig:cn} displays the slope of the $q$ and $u$ parameters vs intensity threshold. For all three clouds we observe the slope approach a value of 1. The slope of 1 indicates we are finding similar polarization fractions between the CNM and OTFMAP mode observations. Such a result indicates that the differences we observe between the CNM and OFTMAP mode observations are not a result of a systematic offset between the different observations, but rather differences in spatial filtering or potential reference beam contamination in the CNM mode observations. The trend of increasing correlation with total intensity and the $q$ and $u$ slopes of 1 support the validity of the data reduction pipeline employed for our FIREPLACE observations.

\subsection{Comparison With Herschel 250 \micron} \label{sec:Herschel}
\begin{figure}
    \centering
    \includegraphics[width=0.45\textwidth]{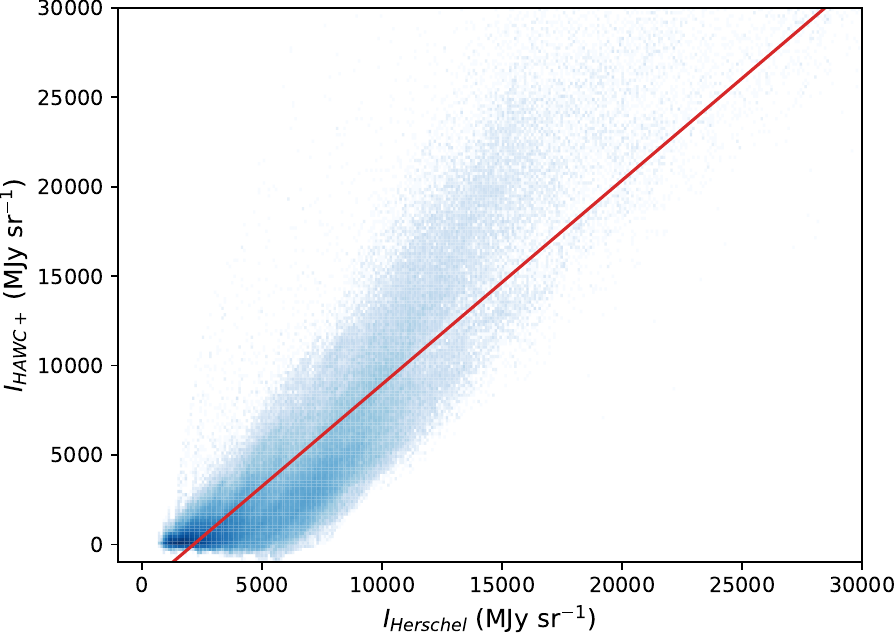}
    \caption{A 2-D histogram showing the correlation between Herschel 250 \micron\ from \citet{Molinari2011} and the SOFIA/HAWC+ 214 \micron\ emission from FIREPLACE DR2. The red line shows the best fit between the two and has a slope of 1.14 and an intercept of -2438 MJy\,sr$^{-1}$.}
    \label{fig:Herschel}
\end{figure}

As an additional check of the consistency of our new FIREPLACE observations, we plot the FIREPLACE DR2 Stokes $I$ (214 \micron\ intensity) against the Herschel 250 \micron\ intensity \citep{Molinari2011} in Figure \ref{fig:Herschel}. The Pearson $r$ value is found to be 0.89, showing good correlation between the two data sets.

There are instances where the 214 \micron\ intensity from FIREPLACE is $\rm<$0 as can be seen in Figure \ref{fig:Herschel}. The negative emission indicates that the SOFIA/HAWC+ observations are insensitive to large-scale emission associated with very extended structures. Though there is variation in dust temperature throughout the CMZ, we note that the slope of 1.14 found is consistent with dust having a temperature of 17.5 K with a spectral index of $\beta=1.5$. This temperature is in good agreement for the dust in the GC, which exhibits an average temperature of 20 K \citep{Molinari2011}. 

The SOFIA/HAWC+ Stokes $I_{214}$ map is possibly less sensitive to extended emission than the Herschel 250 \micron. This loss in sensitivity could explain the negative 214 \micron\ intensities observed in Figure \ref{fig:Herschel}. Fitting only the positive $I_{214}$ values in Figure \ref{fig:Herschel} yields essentially the same values for Pearson r and slope, indicating that the negative values are not significantly impacting our correlation with the Herschel 250 \micron\ observations.

\subsection{Validation Summary} \label{sec:val_sum}
As mentioned previously, the large area and prevalence of extended emission in the CMZ makes the polarimetric data reduction of our observations quite challenging. We have shown that the DRP reduction for this region is generally robust and the results are in good agreement with brighter molecular CMZ structures that have existing CNM mode observations at 214 \micron. The reduced FIREPLACE data set presented here still likely contains residual systematics, and individual measurements in lower 214 \micron\ intensity regions should be viewed with caution. 

An additional challenge for the GC is the large amount of magnetized dust along the line-of-sight. By virtue of our reduction strategy, the spatial filtering of the FIREPLACE observations limits our sensitivity to features smaller than our beam size ($\rm\sim$19.6\arcsec). Future work that incorporates improved data reduction methods may yield a better understanding of the 3D magnetic field structure towards the GC.

\section{The Polarimetric CMZ at 214 \micron} \label{sec:full}
\begin{figure*}
    \centering
    \includegraphics[width=1.0\textwidth]{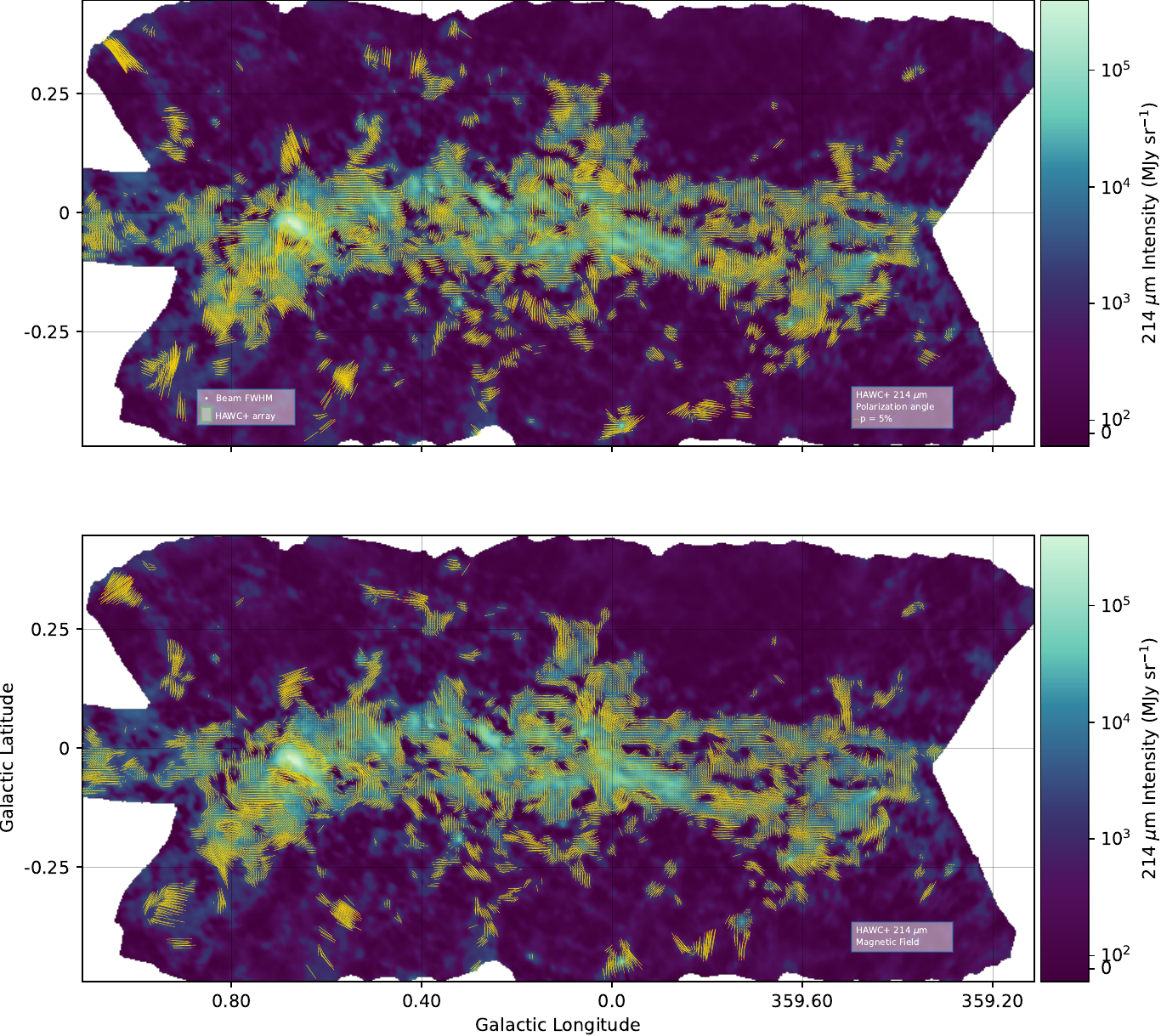}
    \caption{Distribution of the combined DR1+DR2 214 \micron\ FIREPLACE observations. Upper panel: the polarization angle distribution (electric field orientation) plotted over the $I_{214}$ emission. Inset boxes show a representative polarization angle, the beam size, and the instantaneous SOFIA/HAWC+ array size at 214 \micron. Lower panel: the derived magnetic field pseudovectors from the FIREPLACE observations obtained by rotating the polarization angle pseudovectors by 90\degree.}
    \label{fig:fire_full}
\end{figure*}

\begin{figure*}
    \centering
    \includegraphics[width=1.0\textwidth]{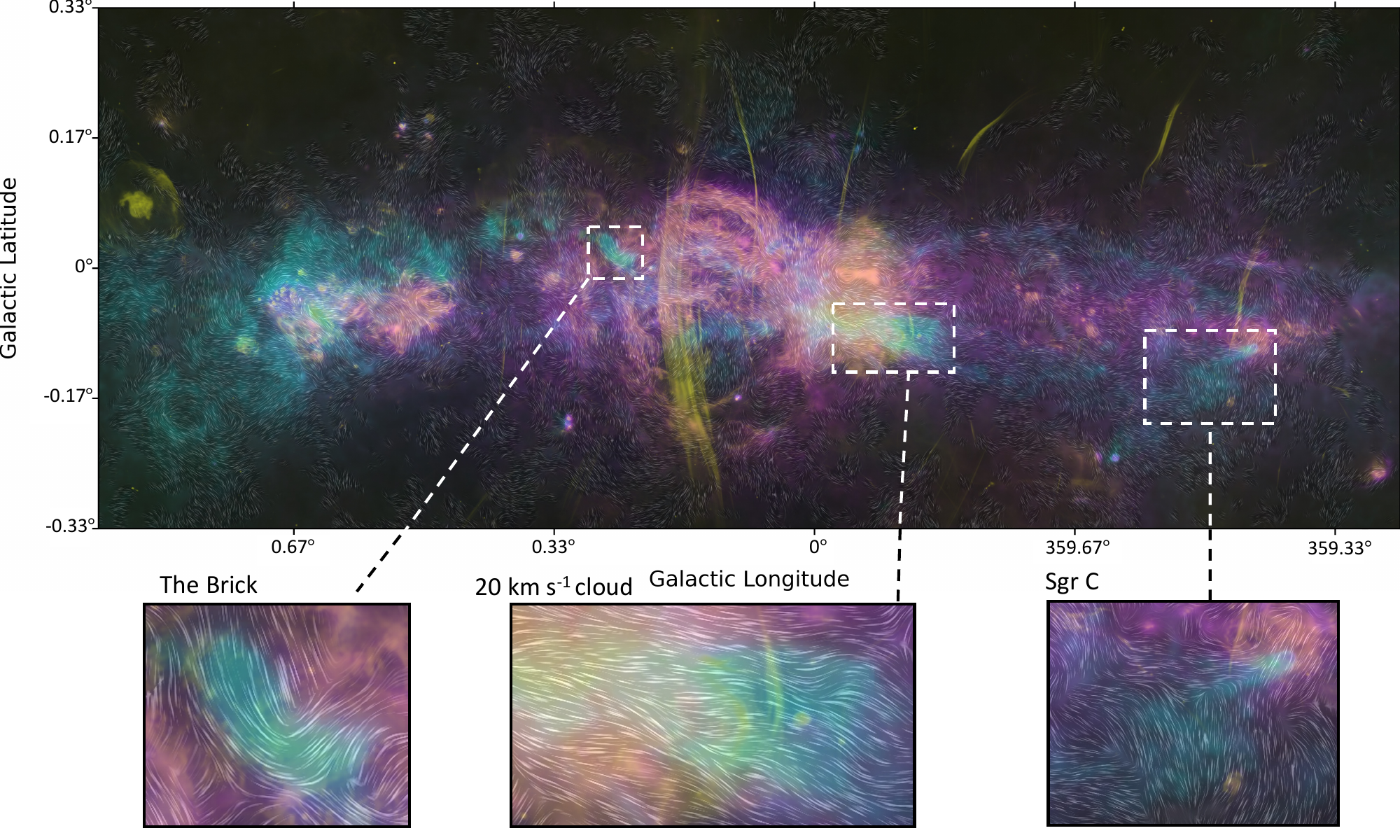}
    \caption{The LIC representation of the magnetic field orientations derived from FIREPLACE for the entire CMZ made using a kernel of 1\arcmin.  Three color background is the same as in Figure \ref{fig:legend}. Inset panels show zoom-in views of specific molecular clouds in the CMZ emphasizing the detailed view FIREPLACE provides of the magnetic fields coincident with these clouds.}
    \label{fig:LIC}
\end{figure*}
Having verified our DR2 reduction, we combine these DR2 data with the DR1 observations presented in \citetalias{Butterfield2023} to obtain a full map of the CMZ. This combination is straightforward since the DR1 and DR2 fields were reduced using identical \textit{CRUSH} parameters. We therefore co-added all of the individual DR1 and DR2 fields using the DRP \textit{merge} algorithm. The combined fields consist of the DR2 fields presented previously in Table \ref{tab:DR2} and the set of DR1 fields shown in Table \ref{tab:DR1}.

The upper panel of Figure \ref{fig:fire_full} reveals the polarization pseudovectors (yellow pseudovectors) obtained from our combined FIREPLACE DR1 and DR2 observations that cover the entire CMZ. Polarization orientations are only displayed for lines-of-sight that pass the selection criteria described in Section \ref{sec:red_meth}. Even with these significance cuts we obtain $\rm\sim$64,000 Nyquist-sampled pseudovectors from our full 214 \micron\ map of the CMZ, a significant increase from the 24,000 pseudovectors obtained in DR1 \citepalias{Butterfield2023}. 

Following the approach of \citetalias{Butterfield2023} we assume that the dust grain alignment mechanism responsible for the polarized emission results from the grains preferentially aligning perpendicular to the magnetic field orientation. The dust grains spin preferentially about their axis of greatest moment of inertia. This scenario implies that the inferred direction of the magnetic field in the plane of the sky is perpendicular to the emergent polarization (or electric field) direction. This is consistent with $B$-RAT alignment theory \citep{Lazarian07,Andersson2015}.  We note that ultimately this interpretation of polarization will rely on a more complete theoretical understanding of the physics of grain alignment than currently exists. 

The lower panel of Figure \ref{fig:fire_full} shows the derived magnetic field pseudovectors obtained by rotating the polarization angle pseudovectors shown in the upper panel of Figure \ref{fig:fire_full} by 90\degree\ under the assumption of $B$-RAT alignment. These magnetic field pseudovectors are shown overlayed on the FIREPLACE $I_{214}$ emission.

We display the Line Integral Contour (LIC) representation of the FIREPLACE magnetic field pseudovectors in Figure \ref{fig:LIC}, made using a kernel size of 1\arcmin. The LIC is a method of illustrating the magnetic field pseudovectors as streamlines \citep{Cabral1993}. To do so, we filter the input distribution of magnetic field pseudovectors along local streamlines to produce an output image that represents the `flow' direction of the magnetic field orientations. This visualization technique leads to a representation of the magnetic field that emphasizes the field geometry on the scales of molecular clouds in the CMZ.

The magnetic field orientations obtained from FIREPLACE are seen to generally trace the morphology of individual molecular clouds, as evident in the inset panels of Figure \ref{fig:LIC}. The fact that the magnetic field is impacted by the morphology of the CMZ molecular clouds indicates that the field derived from FIREPLACE is likely local to the CMZ rather than some other magnetic field component encountered along the line-of-sight. This conclusion is consistent with the idea that polarized thermal emission is column density weighted within the volume of the beam \citep[e.g.,][]{Reissl2020}.
\begin{figure*}
    \centering
    \includegraphics[width=1.0\textwidth]{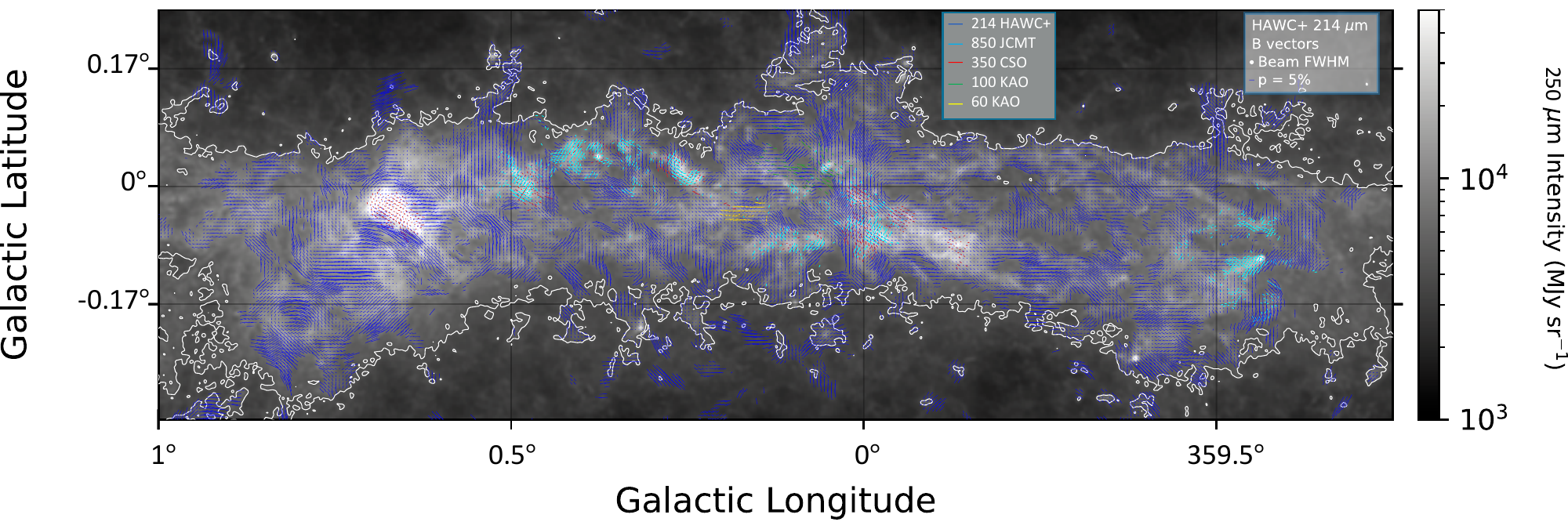}
    \caption{FIREPLACE magnetic field orientations are shown as blue pseudovectors with yellow and green pseudovectors being magnetic field orientations from 60 \micron\ and 100 \micron\ observations, respectively, from the KAO \citep[][]{Dotson2000}. Red pseudovectors are magnetic field orientations from 350 \micron\ observations from the CSO \citep[][]{Dotson2010} and cyan psudovectors are from 850 \micron\ SCUBA/POL2 observations made using the JCMT \citep[][]{Lu2023}. Greyscale background is 250 \micron\ Herschel emission of the CMZ \citep{Molinari2011}. Zoom-ins of prominent molecular clouds are shown and discussed in Section \ref{sec:cloud_mag} (Figures \ref{fig:brick} -- \ref{fig:SgrC}).}
    \label{fig:B-field_comp}
\end{figure*}

A comparison of the FIREPLACE pseudovectors to previous infrared polarimetric data sets of individual molecular clouds within the CMZ made using the Kuiper Airborne Observatory (KAO), Caltech Submillimeter Observatory (CSO), and James Clerk Maxwell Telescope (JCMT) is shown in Figure \ref{fig:B-field_comp}. The KAO and CSO pseudovectors are scaled in a manner similar to the FIREPLACE pseudovectors: proportional to fractional polarization. The JCMT pseudovectors, conversely, are not scaled, which matches their presentation in \citet{Lu2023}. It is clear from Figure \ref{fig:B-field_comp} that the FIREPLACE survey provides extensive polarimetric coverage of the CMZ, yielding a large sample of pseudovectors that will be used for statistical analysis. Figure \ref{fig:B-field_comp} clearly reveals the ability of the FIREPLACE observations to measure the magnetic field in lower intensity regimes. In particular, the FIREPLACE observations recover pseudovectors corresponding to fainter dust emission regions further from the Galactic plane and between the prominent molecular clouds in the CMZ compared to these other observations. This increased sensitivity is a result of our observing strategy utilizing OTFMAP mode polarimetry and overlapping fields. The FIREPLACE magnetic field orientations are in general agreement with those of previous infrared polarimetric observations of prominent CMZ molecular clouds, as can be seen in Figure \ref{fig:B-field_comp}; however, there are examples where the field derived from FIREPLACE is systematically different than previous observations. Comparisons between the pseudovectors derived from the FIREPLACE observations and previous individual cloud observations are discussed in more detail in Section \ref{sec:cloud_mag}.

\subsection{The CMZ Magnetic Field on Multiple Spatial Scales} \label{sec:surv_comp}
\begin{figure*}
    \centering
    \includegraphics[width=1.0\textwidth]{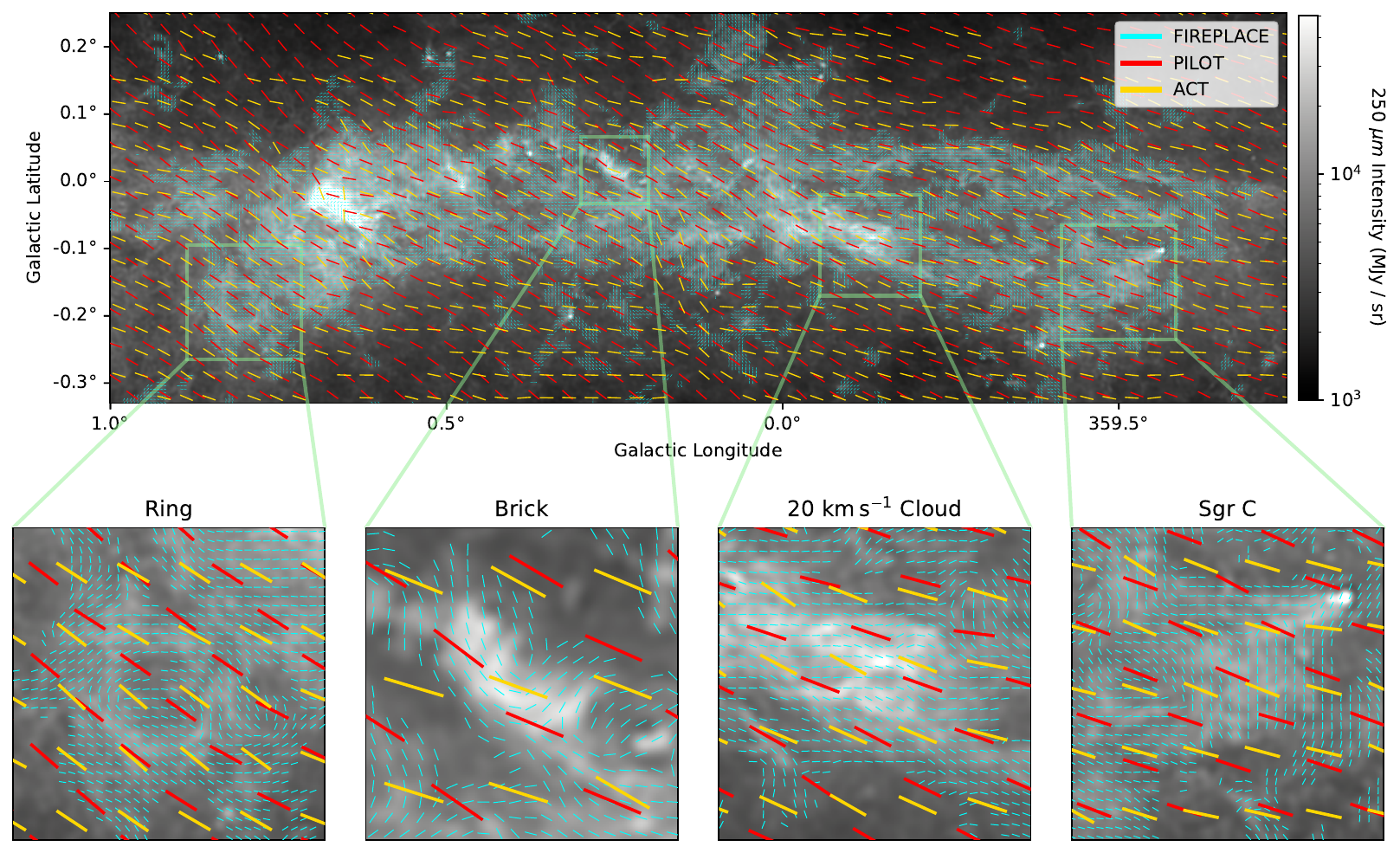}
    \caption{Comparison of magnetic field orientations seen towards the CMZ over multiple angular scales. Background colorscale indicates the $I_{250}$ emission from Herschel \citep{Molinari2011}, with the cyan pseudovectors indicating the FIREPLACE magnetic field orientations at a resolution of 19.6\arcsec. Red pseudovectors are magnetic field orientations from PILOT at 240 \micron\ smoothed to 2\arcmin\ and gold pseudovectors are magnetic field orientations from ACTPol at 220 GHz \citep[1.36 mm, also 2\arcmin,][]{Mangilli2019,Guan2021}.}
    \label{fig:surv_comp}
\end{figure*}
We first compare the magnetic field orientations obtained from FIREPLACE (19.6\arcsec) with the magnetic fields derived from ACTPol smoothed to 2\arcmin\ \citep{Guan2021} and PILOT \citep[also 2 \arcmin,][]{Mangilli2019}. This comparison is shown in Figure \ref{fig:surv_comp} where the magnetic fields are superposed on the 250 \micron\ total intensity ($I_{250}$) CMZ emission from Herschel \citep{Molinari2011}. In this figure, the pseudovectors have not been scaled by fractional polarization to make it easier to compare the orientations observed in the different observations. In general, the magnetic fields from PILOT and ACTPol have little variation between neighboring pseudovectors and are parallel to one another, indicating that they are tracing a similar large-scale, spatially uniform, magnetic field component. The magnetic field traced by FIREPLACE shows more spatial variation than either the PILOT or ACTPol distributions. The FIREPLACE magnetic field tends to trace the total intensity morphology of prominent CMZ molecular clouds as seen in high $I_{250}$ emissivity regions in Figure \ref{fig:surv_comp} and previously in Figure \ref{fig:LIC}. In lower $I_{250}$ emissivity regions the magnetic field does not exhibit an apparent preferential orientation, ranging from being parallel to the Galactic plane to perpendicular to the Galactic plane.

Because of the higher sensitivity of ACTPol and PILOT to large-scale polarization patterns relative to FIREPLACE, the magnetic fields derived from these surveys likely include a magnetic field contribution originating from emission external to the CMZ. In other words, the increased spatial filtering and higher resolution of FIREPLACE, relative to PILOT and ACTPol, likely results in measurements of polarization local to the dust structures in the CMZ, whereas the PILOT and ACTPol measurements include contributions from a larger part of the central kpc of the Galaxy. Different dust temperatures encountered along the line-of-sight for the surveys is an alternative explanation for the different magnetic field orientations. However, the similarity of the field orientation observed between PILOT at 240 \micron\ and ACTPol at 1.36 mm potentially indicates that different dust temperatures is not a significant factor in the different magnetic field orientations. Careful modeling of potential line-of-sight magnetic field components for these data sets will enable an analysis of the magnetic field configuration along the line-of-sight. For example, comparison of IR emission from stars on the near and far side of the CMZ indicates that this large-scale field component is connected to material that is within the central 1$-$2 kpc of the Galaxy \citep{Nishiyama2009}. 

\subsection{The Orientation of the Magnetic Field Throughout the CMZ} \label{sec:b_orient}
\begin{figure*}
    \centering
    \includegraphics[width=1.0\textwidth]{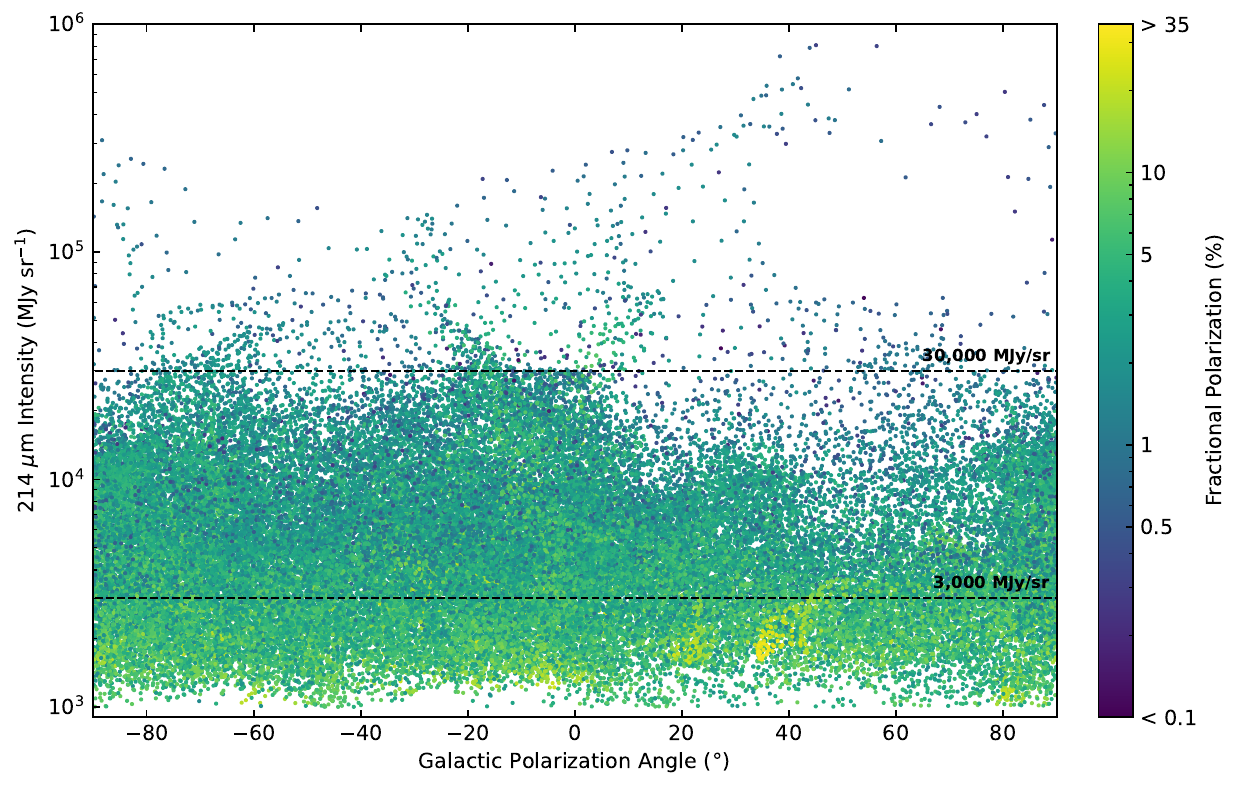}
    \caption{Total intensity plotted versus polarization angle for regions of significant total and polarized intensity throughout the CMZ. Data points are colored by polarization percentage. Black dashed lines demarcate different total intensity regimes separating high, middle, and low total intensity regions in the CMZ as described in the text and used later in Figure \ref{fig:Hist}.}
    \label{fig:pol_vec_pix}
\end{figure*}
\begin{figure*}
    \centering
    \includegraphics[width=1.0\textwidth]{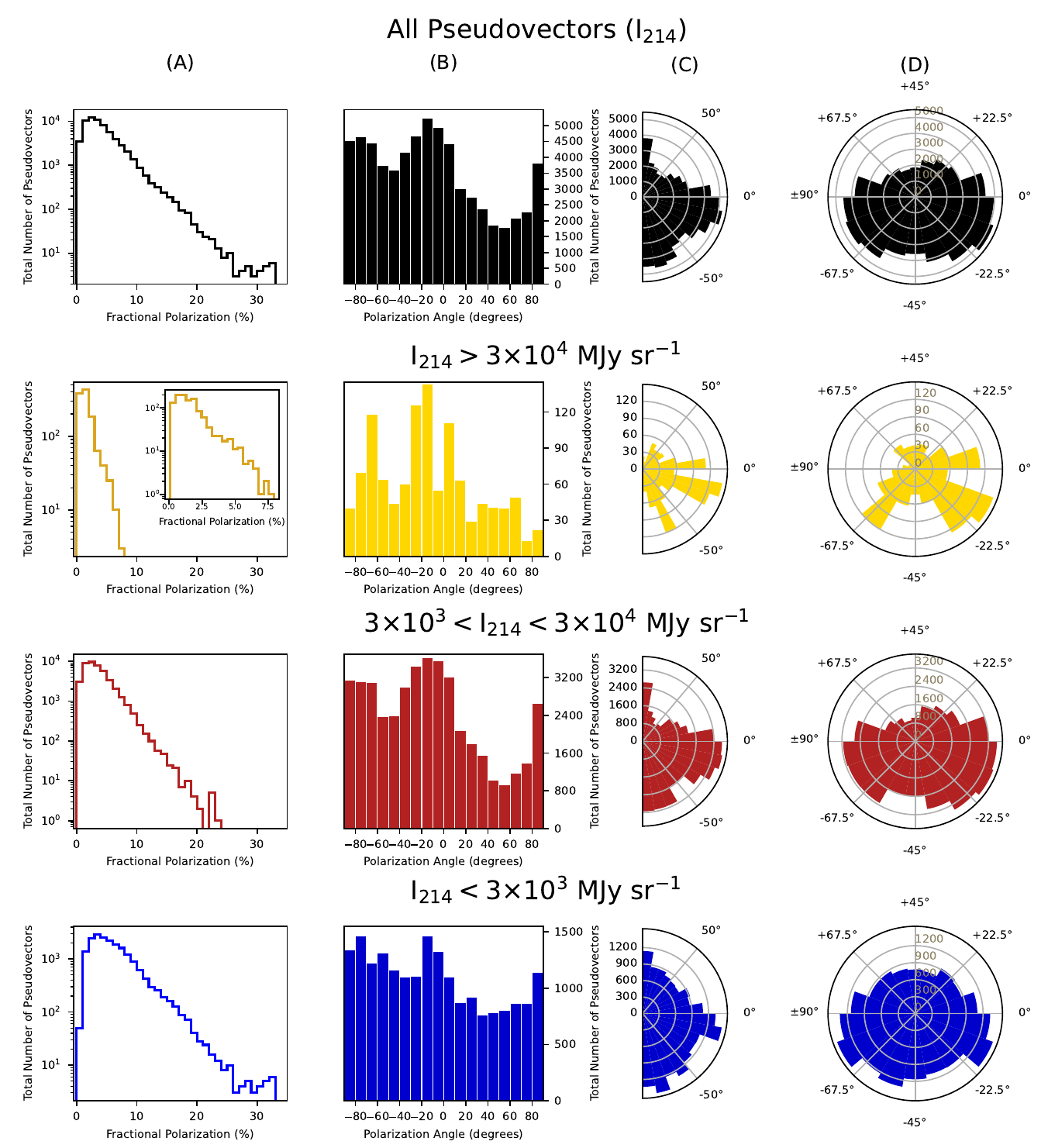}
    \caption{Distribution of the FIREPLACE polarization pseudovectors for the entire CMZ (top row; black histograms; 64,276 pseudovectors) compared with three different intensity regimes, as titled: $I_{214}>$30,000 \Mjsr\ (yellow; 1134 pseudovectors); 3000$<I_{214}<$30,000 \Mjsr\ (red; 43,560 pseudovectors); $I_{214}<$3000 \Mjsr\ (blue; 19,582 pseudovectors). These intensity ranges correspond with the dashed lines in Figure \ref{fig:pol_vec_pix}. The columns show the following information: (A) distribution of fractional polarization, summed over 1\% bins. (B) histograms showing the distribution of polarization angles in Galactic coordinates.  (C) Same distribution as Column B, but rotated to illustrate the plane-of-sky orientation. (D) Same distribution as Column C but ``wrapped" so that $\pm$90\degree\ is connected on the left side of the plot.}
    \label{fig:Hist}
\end{figure*}
Building on the comparison of magnetic field spatial scales presented in Section \ref{sec:surv_comp}, we study the distribution of polarization pseudovectors to determine systematic trends in the polarization angle (and by extension the magnetic field) throughout the CMZ. The objective here is to quantify the extent to which the magnetic field derived from FIREPLACE connects to the large-scale magnetic field observed by PILOT and ACTPol. We plot the total intensity versus polarization angle in Figure \ref{fig:pol_vec_pix}, where the individual data points are colored based on the polarization percentage ($p$/$I\times$100) for each line-of-sight. The polarization angle distribution covers the full range of angles from -90\degree\ to +90\degree. Furthermore, an inspection of the observed polarization percentage of the observations reveals that polarization levels $\rm>$20\% are rare ($\rm\leq$1\% of our total pseudovectors).

The horizontal dashed lines divide the plot into different total intensity regimes. The $I_{214}\leq$3000 \Mjsr\ regime corresponds to the $I_{214}$ emission level of the peripheries of the molecular clouds in the CMZ, whereas $I_{214} \geq$30,000 \Mjsr\ corresponds to the brightest molecular emission in the CMZ of the most prominent CMZ clouds (e.g., Sgr B2, Sgr C, and the Brick molecular clouds).

The higher polarization percentages of $\rm>$20\% are only present in the fainter $I_{214}$ emission regime, corresponding to only $\sim$100 lines-of-sight. These high polarization measurements connect to the previously observed trend that the polarization percentage is generally higher in the fainter peripheries of CMZ molecular clouds. These high values could also indicate large-scale, low-surface brightness emission is missing in our observations due to the OTFMAP mode.

At the highest total intensities ($I_{214}\geq$30,000 MJy sr$\rm^{-1}$, demarcated by the upper horizontal dashed line in Figure \ref{fig:pol_vec_pix}) the number of polarization angles obtained is more sparse. The polarizations are generally $\rm<$5\% in this regime. The generally lower polarization percentages in these denser regions relative to the percentages in the fainter cloud peripheries agrees with previous polarimetric results for CMZ molecular clouds (e.g. \citealt{Chuss2003a,Chuss2003b,Pillai2015,Lu2023}; \citetalias{Butterfield2023}).

The large number of individual lines-of-sight in Figure \ref{fig:pol_vec_pix} ($\rm\sim$64,000 Nyquist-sampled pseudovectors) makes it difficult to assess whether there are any preferential polarization angle orientations in the CMZ. To obtain better insight into CMZ-wide polarization angle trends we sort the individual polarization angles shown in Figure \ref{fig:pol_vec_pix} into 10\degree\ bins. The histogram representation of the binned polarization angles is shown in Figure \ref{fig:Hist} (similar to what is done in \citetalias{Butterfield2023}). In these histograms, a polarization angle of 0\degree\ indicates a magnetic field that is aligned with the Galactic plane, with an angle of $\rm\pm$90\degree\ indicating a magnetic field that is oriented perpendicular to the Galactic plane.

The top row of Figure \ref{fig:Hist} reveals the polarization angle distribution for all polarization pseudovectors from the CMZ (corresponding to the entire polarization angle distribution shown in Figure \ref{fig:pol_vec_pix}). The remaining rows reveal the polarization angle distributions for the different total intensity regimes marked in Figure \ref{fig:pol_vec_pix}.

Column A illustrates the fractional polarization distribution for pseudovectors of each intensity range. We observe a steeper slope for the higher total intensities (where $I_{214}>$30,000 MJy sr$\rm^{-1}$). This steeper  distribution was also observed in \citetalias{Butterfield2023} and could indicate increased depolarization towards the brighter (higher density) regions. This increased depolarization could be a result of decreased grain alignment efficiency in the interiors of molecular clouds \citep{Santos2019}, superpositions of multiple field components along the line-of-sight, or field tangling on scales smaller than the 19.6\arcsec\ SOFIA/HAWC+ beam size of these observations \citep[e.g.][]{Fissel2016}. Higher resolution observations are needed to determine which of these mechanisms is responsible for the increased depolarization observed in our results. We see an increasingly shallow slope as we move to fainter intensity regimes as marked in the third and fourth rows of Figure \ref{fig:Hist}. This trend agrees with what was observed in DR1 \citepalias{Butterfield2023}.

The histograms presented in columns B - D reveal preferential polarization angle orientations. The histogram corresponding to all of the polarization pseudovectors (top row) reveals a bimodal distribution in the polarization angle, where the histogram peaks at an angle of approximately -20\degree\ and $\rm\pm$90\degree. This is a similar bimodal distribution to the one observed in \citetalias{Butterfield2023}. We observe the same polarization angle enhancements in the middle and low $I_{214}$ emission regimes as seen in rows 3 and 4 of Figure \ref{fig:Hist}. In the high $I_{214}$ regime presented in the second row of Figure \ref{fig:Hist} there is no $\rm\pm$90\degree\ enhancement and there is instead an enhancement at 0\degree.

The peak at $0$\degree\ only appears in the brightest intensity regime corresponding to the emission originating from the brightest and densest regions of the prominent CMZ molecular clouds. We therefore find it is unlikely that the orientation at $0$\degree\ is a magnetic field component external to the CMZ.

We also bin the 2022 polarization angle distribution shown in Figure \ref{fig:fire_2022} to assess whether the same polarization angle enhancements are recovered in the DR2 portion of the FIREPLACE observations. These histograms are presented in Appendix \ref{sec:2022_Hist} and shown in Figure \ref{fig:2022_Hist}. We find a similar bimodal polarization angle enhancement at angles of $\rm\sim$ -20\degree\ and $\rm\sim$ -70\degree\ in all 2022 pseudovectors and in the middle and low $I_{214}$ emission regimes. These are similar bimodal enhancements to those observed in the full FIREPLACE data set shown in Figure \ref{fig:Hist} for these intensity regimes. However, in the high $I_{214}$ emission regime of only the 2022 portion of our FIREPLACE data set we do not detect the 0\degree\ enhancement and instead observe an enhancement at -60\degree, as seen in Figure \ref{fig:2022_Hist}. Therefore, the 0\degree\ component seen in the $I_{214}$ emission regime of the full data set can be associated with Sgr B2. Since this high intensity regime corresponds to the cores of the brightest and most dense molecular clouds, the different enhancements observed in this regime indicate that the magnetic field is preferentially aligned along the morphology of these clouds, further corroborating the conclusion that the magnetic field in this regime is local to the CMZ molecular clouds.

\begin{figure*}[t]
    \centering
    \includegraphics[width=1.0\textwidth]{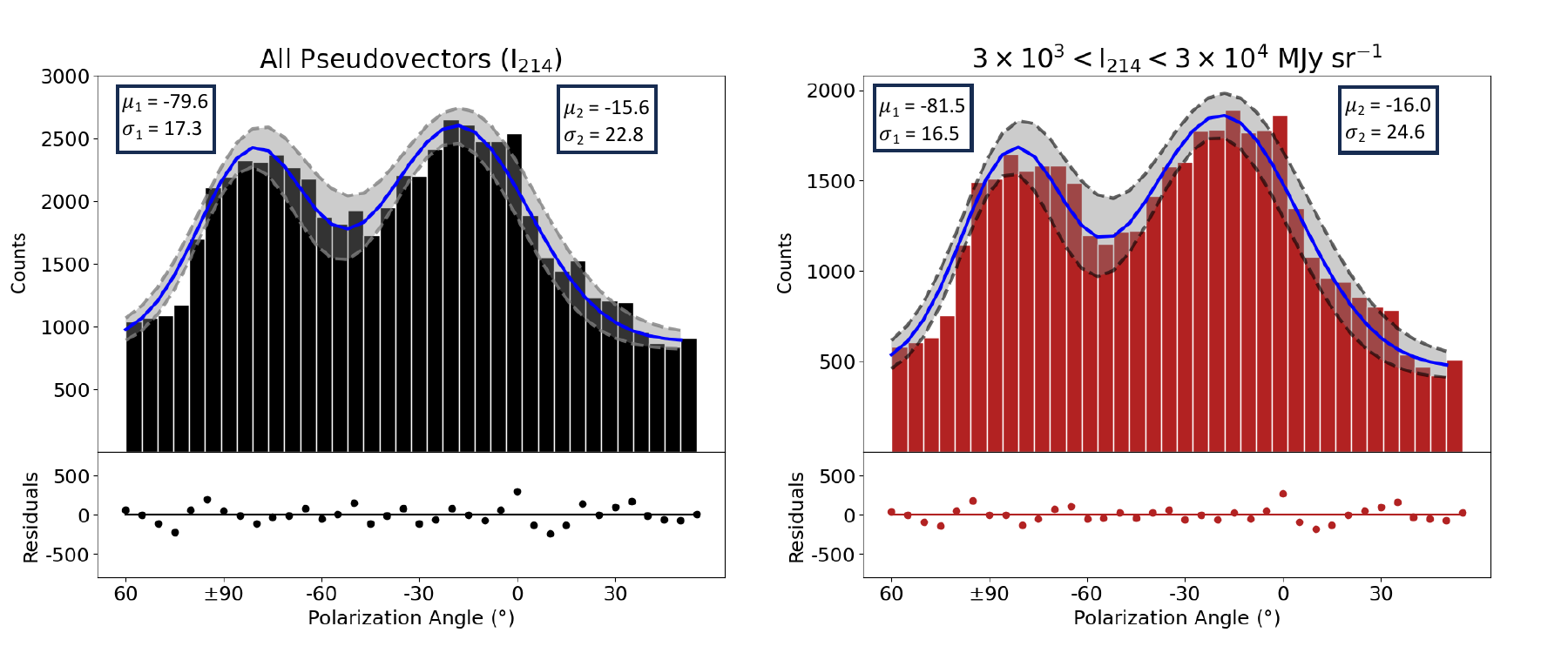}
    \caption{Distribution of polarization angle orientations in 5\degree\ bins with best-fit two-Gaussian model. Left: all significant polarization pseudovectors, like the top row of Figure \ref{fig:Hist}. Right: polarization pseudovectors in the middle $I_{214}$ intensity range where 3,000$<I_{214}<$30,000 \Mjsr, like the third row of Figure \ref{fig:Hist}. The best-fit model to these histograms is shown as a blue line in both panels, with best-fit centroid and width parameters of the model Gaussian components indicated in the figure. Shadowed regions represent the $\pm$1$\sigma$ dispersion of the best-fit models.}
    \label{fig:mod_fit}
\end{figure*}
A more comprehensive analysis of the magnetic field alignment will be conducted in a further work utilizing the Histogram of Relative Orientation \citep[HRO,][]{Soler2013} and Projected Rayleigh Statistic \citep[PRS,][]{Jow2018} methods. For now, to further quantify the bimodal distribution from the full FIREPLACE data set, we fit the histogram of polarization pseudovectors obtained for the full FIREPLACE data set with a two-Gaussian model. To perform this fitting, we assume the uncertainties on the counts in each histogram bin is the square-root of the counts in that bin (ranging from 31$-$51 counts). These fits are shown in Figure \ref{fig:mod_fit}, where the polarization angles derived from FIREPLACE have been sorted into 5\degree\ bins for the full set of polarization angles and the middle intensity regime. The histogram and model have been shifted by 60\degree\ in the figure to better show the bimodal distribution in the histogram and resulting fit. We note that though the range of angles is technically wrapped such that 60\degree\ connects to 59\degree, the model fit was performed on a linear representation of our polarization angles where 60\degree\ and 59\degree\ are on opposite sides of the histogram as shown in Figure \ref{fig:mod_fit}. This disconnect in linear space is why the model fit is not continuous when wrapped. The blue line is the best-fit result obtained from fitting a two-Gaussian model to the histogram of the polarization angle data with the shadowed envelope showing the $\pm$1$\sigma$ dispersion of the model fit. The equation of this two-Gaussian model is defined as:
\begin{equation}
    y = H + A_1e^{\frac{-(x-\mu_1)^2}{2\sigma_1^2}} + A_2e^{\frac{-(x-\mu_2)^2}{2\sigma_2^2}}
    \label{eq:mod}
\end{equation}
where $H$ is the offset above zero for the Gaussian components (in counts), $A_1$ and $A_2$ are the amplitudes of each Gaussian (in counts), $x$ is the set of polarization angles from -90\degree\ to +90\degree\ that comprise the histogram, $\mu_1$ and $\mu_2$ are the centroids of each Gaussian (in \degree), and $\sigma_1$ and $\sigma_2$ are the standard deviations of each Gaussian (in \degree). This results in a seven-parameter model, with all parameters allowed to vary to determine the best-fit to the underlying histogram. The resulting best fit parameters for the full polarization vector distribution and the middle intensity regime are shown in Table \ref{tab:comp} with 1$\sigma$ uncertainties.

The peaks of the modelled Gaussian components for the full set of pseudovectors are located at polarization angles of$-15.6\pm1.1$\degree\ and $-79.6\pm1.2$\degree. The residuals are shown below the model fit, and range from $\sim$10 to a few hundred counts. These residuals are fairly large compared to the uncertainties in the histogram bins, but this two-Gaussian model fit is simple, and is likely not fully characterizing the structure of the histogram.

The histogram and resulting best-fit model  for the middle intensity regime corresponding to the third row of Figure \ref{fig:Hist} is shown in the right panel of Figure \ref{fig:mod_fit}. The fit returns similar parameters that are generally consistent within the error bars of the Gaussian model components found for the full pseudovector distribution. The Gaussian peaks for the middle intensity regime are located at $-16.0\pm1.2$\degree\ and $-81.5\pm1.2$\degree. As with the full polarization angle model fit, the residuals corresponding to the best-fit model are implying that the model is under-fitting the structure of the histogram. More detailed characterization of these histograms would require a model consisting of additional model components. Theoretical modeling is needed to estimate possible additional model components.

The polarization angle peak from the model fit at approximately $-16$\degree\ corresponds to a magnetic field orientation that is aligned with the CMZ-wide magnetic field observed by the PILOT and ACTPol programs \citep{Mangilli2019,Guan2021}. As discussed this field is connected to material in the inner 1$-$2 kpc of the Galaxy \citep{Nishiyama2009}. This enhancement could therefore be tracing the orientation of the CMZ molecular clouds or it could be a magnetic field component external to the CMZ. The Gaussian component centered at approximately $\pm{}90$\degree\ relates, conversely, to a vertical magnetic field and provides evidence of a connection between the magnetic fields in the dust to those traced by the radio NTF population. We do not fit a two-Gaussian model to the high-intensity regime because of the lower number of polarization angles per histogram bin in this regime as can be seen in Figure \ref{fig:Hist} -- $\rm\sim$100 counts compared to the full distribution with $\rm\sim$1000 counts.

The high-total intensity regime cutoff of 3$\times$10$\rm^4$ \Mjsr\ is lower than was used in \citetalias{Butterfield2023}, where a threshold of 5$\times$10$\rm^4$ \Mjsr\ was used to isolate the magnetic field orientation obtained from only the brightest regions of Sgr B2. As a further verification of the combined FIREPLACE data set we analyze the histogram of polarization angles obtained from using the DR1 upper total intensity regime of $I_{214}>$5$\times$10$\rm^4$ \Mjsr\ to isolate the polarization angles from Sgr B2. We would expect to obtain the same distribution as shown in \citetalias{Butterfield2023}. We find the same number of polarization pseudovectors to what was found in DR1 (359) corresponding to this regime where $I_{214}>$5$\times$10$\rm^4$. We find essentially the same histogram distribution as reported in Figure 8 of \citetalias{Butterfield2023}. There is some variation where certain 10\degree\ bins are one count higher or lower in our combined observations compared to the histogram presented in \citetalias{Butterfield2023}. Such slight differences in individual histogram bins are likely a result of slight rounding differences where polarization angles near the 10\degree\ bin edges are placed in different bins.

\begin{figure*}
    \centering
    \includegraphics[width=1.0\textwidth]{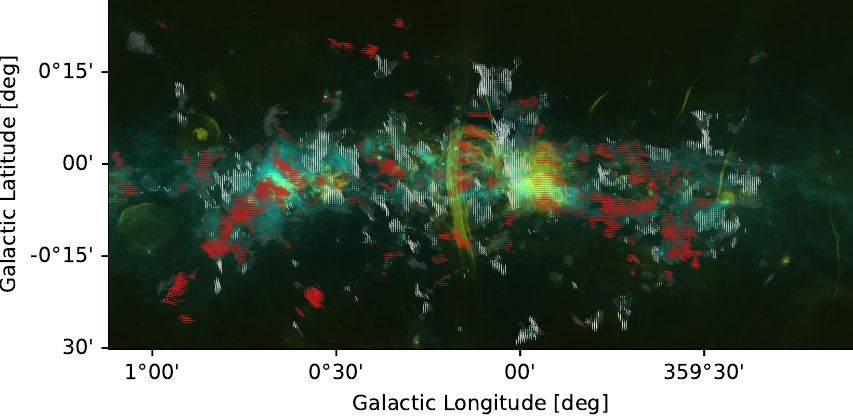}
    \caption{FIREPLACE magnetic field pseudovectors with orientations within $\pm$1$\sigma$ of the centroids of the Gaussian fit components shown in the left column of Table \ref{tab:comp} found for all polarization pseudovectors (2 Gaussian model fit shown in the left panel of Figure \ref{fig:mod_fit}). Pseudovectors corresponding to the -15.6\degree\ component are shown in red and pseudovectors corresponding to the -78.8\degree\ component are shown in white. Grey pseudovectors indicate all magnetic field pseudovectors from FIREPLACE that are oriented outside of these angle ranges. 2-color background shows 250 \micron\ Herschel emission in cyan \citep{Molinari2011} and 20 cm (1 GHz) MeerKAT emission in yellow \citep{Heywood2022}.}
    \label{fig:paraperp}
\end{figure*}

The orientation of the magnetic field enhancements helps to clarify a picture of the magnetic fields across the central degree of the Galaxy. In all histograms, we see excess measurements at an angle consistent with that of the large-scale component seen in the PILOT and ACTPol data (the $\sim$ -20\degree\ enhancement shown in Figures \ref{fig:Hist}, \ref{fig:mod_fit}, \ref{fig:2022_Hist}). 

In the high intensity regime (the yellow histogram of Figure \ref{fig:Hist}), the polarization angle enhancements observed at -20\degree\ splits into two sub-components: one at -20\degree\ and one at 0\degree. However, in the 2022 high intensity portion shown in Figure \ref{fig:2022_Hist} the enhancements are at -60\degree\ and -20\degree. The 0\degree\ enhancement is only observed in the brightest portion of Sgr B2 is almost certainly associated with the core of the Sgr B2 region, since we do not recover this component in the 2022 observations (Figure \ref{fig:2022_Hist}). The -20\degree\ component is detected in all our intensity regimes. In the high intensity regime the -20\degree\ enhancement is likely tracing a local CMZ field. For the lower intensity regimes, it could be at least partially an external magnetic field component given the large spatial extent over which this field is coherent at arcminute resolutions \citep[e.g. ACTPol,][]{Guan2021}. The enhancement at $\rm\pm$90\degree\ is only detected in the middle and low intensity regimes. The detection of this enhancement in only these regimes is consistent with the picture of a connection between the vertical field traced by the NTFs and a horizontal field that is a result of the field being sheared in denser, kinematically-dominated regions.

Previously, it has been posited that an initially vertical (or poloidal) field could be sheared into a horizontal configuration in denser molecular regions \citep{Morris1996b}. Our findings here support early submillimeter data \citep{Novak2003b, Chuss2003a} and near-infrared polarimetry data \citep{Nishiyama2010} that have suggested a transition occurs from a sheared field in dense regions to a vertical field in regions having lower kinetic energy density. The two scenarios that have been considered are the above-mentioned evolution from an initially vertical field and the creation of a vertical field from the sheared field via winds and outflows.

In producing these histograms, we have utilized polarimetry data across the range of intensities down to $\sim$1000 MJy sr$^{-1}$. If systematic errors remain after our data cuts, they are unlikely to produce the histogram peaks observed.  The scan angles ($\pm30^\circ$) are sufficiently different from the maxima of the histograms that a systematic scan-synchronous effect is unlikely to produce the results seen here. 

To further quantify the locations of the pseudovectors corresponding to the Gaussian peaks of the model fit shown in the left panel of Figure \ref{fig:mod_fit} we overlay the vectors that are within $\pm$1$\sigma$ of the peaks of the two Gaussian components identified for the full pseudovector histogram distribution (left panel of Figure \ref{fig:mod_fit}). The pseudovectors corresponding to the -15.6\degree\ and -78.8\degree\ model fit components are shown as red and white pseudovectors, respectively, in Figure \ref{fig:paraperp}. The locations of the red pseudovectors in the CMZ are observed to largely coincide with molecular clouds (shown in cyan in this figure) and are generally within 0.25\degree\ in Galactic latitude of the Galactic plane. In particular, the set of red pseudovectors observed at a Galactic Longitude of $\sim$359.7\degree\ traces and coincides with the molecular gas of the ``twisted ring'' \citep{Molinari2011}. Conversely, the white pseudovectors are generally spread over a larger latitude range.

The tendency of the red pseudovectors to coincide with the molecular clouds supports the idea that a pervasive initially vertical field could have been sheared by the motion of the higher density CMZ clouds. The larger latitude spread of the white pseudovectors supports the idea of a ubiquitous vertical field. The possibility of a ubiquitous field of this nature is further strengthened by the fact that these vertical pseudovectors are not uniquely spatially coincident with the NTFs that are shown in yellow in Figure \ref{fig:paraperp}. This finding suggests that the NTFs are illuminated portions of a larger vertical field, rather than being local magnetic field enhancements. A more quantitative analysis of the CMZ-wide magnetic field distribution will be performed in a future work.

\begin{table*}
\centering
\caption{Best Fit 2-Gaussian Model Parameters}
\begin{tabular}{ccc}
\hline
Parameter & All Pseudovectors ($I_{214}$) & $3\times10^3 < I_{214} < 3\times10^4$ \Mjsr \\ \hline\hline
$H$       & 877$\pm$62  & 459$\pm$56 \\
$A_1$      & 1517$\pm$78  & 1184$\pm$66 \\
$A_2$     & 1722$\pm$75  & 1403$\pm$64 \\
$\mu_1$   & -79.6\degree$\pm$1.2\degree & -81.5\degree$\pm$1.2\degree \\
$\mu_2$   & -15.6\degree$\pm$1.1\degree & -16.0\degree$\pm$1.2\degree \\
$\sigma_1$ & 17.3\degree$\pm$1.2\degree  & 16.5\degree$\pm$1.2\degree \\
$\sigma_2$ & 22.8\degree$\pm$1.5\degree  & 24.6\degree$\pm$1.6\degree \\\hline
\end{tabular}
\label{tab:comp}
\end{table*}
\section{THE MAGNETIC FIELDS OF PROMINENT CMZ MOLECULAR CLOUDS} \label{sec:cloud_mag}
\begin{figure*}
    \centering
    \includegraphics[width=1.0\textwidth]{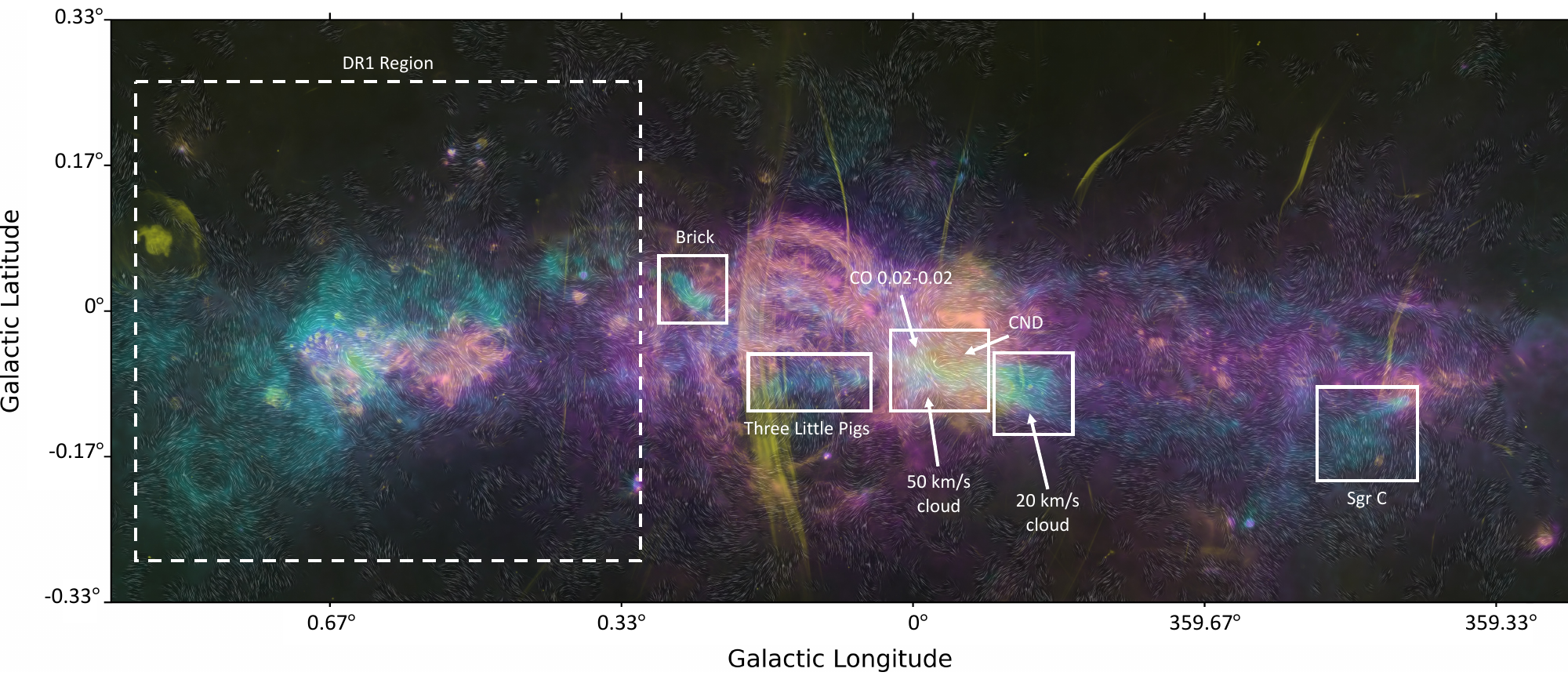}
    \caption{Annotated version of the LIC representation of the FIREPLACE magnetic field distribution as shown in Figure \ref{fig:LIC}, indicating the locations of prominent molecular clouds in the CMZ studied in this section. Solid white rectangles indicate the fields-of-view of Figures \ref{fig:brick} - \ref{fig:SgrC} that are used to discuss prominent CMZ molecular clouds. The dashed white rectangle indicates the region of the CMZ observed in \citetalias{Butterfield2023}. Clouds in the region of the CMZ observed in \citetalias{Butterfield2023} are marked in that paper and not indicated here.}
    \label{fig:LIC_anno}
\end{figure*}
We have argued the FIREPLACE results predominantly reveal the magnetic field local to the CMZ, especially in bright $I_{214}$ emission regions (Section \ref{sec:b_orient}). In this section, we compare the FIREPLACE 214 \micron\ results with previous polarimetric observations of individual molecular clouds. Furthermore, the increased sensitivity of the FIREPLACE survey enables a better determination of how the magnetic field local to the CMZ varies in the peripheries of these clouds than previous studies. Here we present an overview of the magnetic field configurations obtained for a set of prominent CMZ molecular clouds, as indicated in Figure \ref{fig:LIC_anno}.

In the following discussion, Figures \ref{fig:brick} - \ref{fig:SgrC} are used to motivate the discussion of the magnetic fields local to the distinct clouds. These figures show how the FIREPLACE magnetic field pseudovectors compare to previous polarimetric observations of these clouds. In general, our new 214 \micron\ observations find a magnetic field that is consistent with these previous studies. Instances where this is not the case are discussed in the relevant subsections.

\subsection{M0.253+0.016 (The Brick)} \label{sec:brick}
\begin{figure*}
    \centering
    \includegraphics[width=1.0\textwidth]{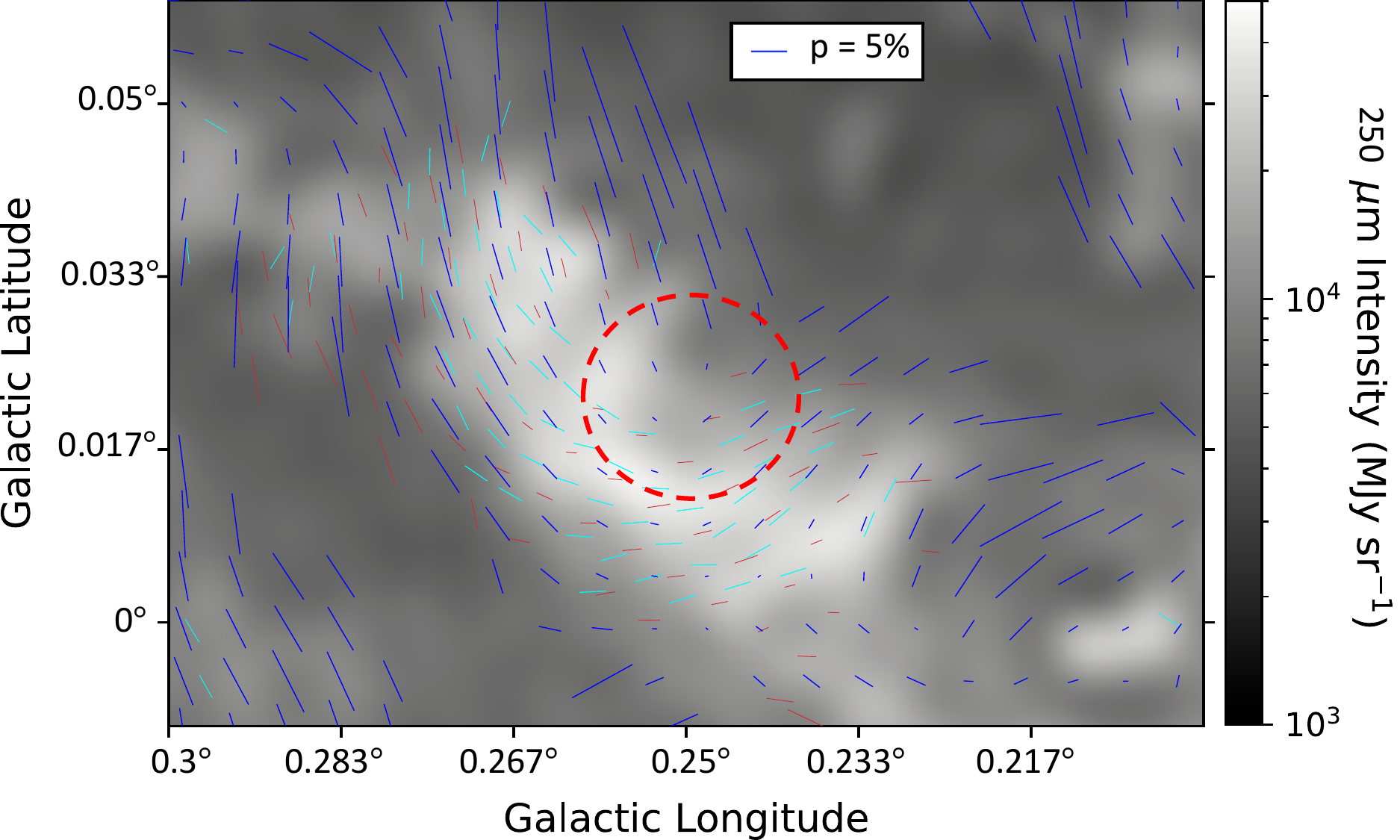}
    \caption{Zoom-in of Figure \ref{fig:B-field_comp} on the Brick molecular cloud. Total intensity background is from 250 \micron\ Herschel observations \citep{Molinari2011}. Blue pseudovectors are the derived magnetic field orientations from the FIREPLACE survey at 214 \micron\ and red pseudovectors are from CSO observations at 350 \micron\ \citep{Dotson2010}. Cyan pseudovectors are from recent JCMT observations at 850 \micron\ \citep{Lu2023}. The red circle indicates the extent of the wind-blown bubble identified in \citet{Henshaw2022}. A 5\% polarization pseudovector is also shown for reference.}
    \label{fig:brick}
\end{figure*}
M0.253+0.016, known colloquially as the ``Brick,'' lies to the Galactic Northeast of Sgr A$^*$ and is near the arched filaments that are proximal to the Radio Arc NTFs \citep{Yusef-Zadeh1987,Lang1999b}. The location of the Brick in the GC can be observed in Figure \ref{fig:LIC_anno}. A zoom-in view of the Brick's total intensity emission obtained from Herschel at 250 \micron\ is shown in Figure \ref{fig:brick}. Overlaid on this total intensity distribution are the magnetic field pseudovectors derived from our FIREPLACE observations (blue line segments), the magnetic field pseudovectors derived from 350 \micron\ CSO observations \citep[red line segments,][]{Dotson2010,Pillai2015}, and the magnetic field pseudovectors derived from 850 \micron\ JCMT observations \citep[cyan line segments,][]{Lu2023}. We obtain $\rm>$50 magnetic field pseudovectors corresponding to the Brick, extending into fainter $I_{250}$ emission regions than the previous CSO or JCMT observations. The FIREPLACE pseudovectors are observed to trace the total intensity morphology of the Brick, generally following the extended curvature of the cloud. Our FIREPLACE magnetic field orientations also agree with those derived from previous HAWC+ CNM mode observations of this cloud, as seen in Figure \ref{fig:cn}.

The ordered magnetic field obtained from the CSO observations led \citet{Pillai2015} to postulate that the magnetic pressure exceeds the turbulent pressure in this cloud. If true, this indicates that the magnetic field is more dynamically important than turbulence and plays a more significant role in shaping the morphology of the Brick.  The similarly ordered magnetic field derived from our FIREPLACE observations support this possibility, and we furthermore observe that the ordered field extends into fainter regions of the Brick than was measured by \citet{Pillai2015}. Additionally, HCO$^+$ absorption filaments observed within the Brick also appear to trace the magnetic field orientation within the cloud \citep{Bally2014}, which indicates that the density structures within the cloud could be shaped by the internal magnetic field geometry. 

\citet{Pillai2015} argue that the curvature of the magnetic field seen towards the Brick implies the presence of a shock. A recent study of the kinematics and structure of the Brick implies that the cloud is likely a conglomeration of distinct velocity components along the line-of-sight \citep{Henshaw2019,Henshaw2022}. The Brick could therefore be multiple distinct molecular structures that, although superimposed along the line-of-sight, are at different locations in the gravitational potential of the CMZ.

There is also evidence of feedback in the Brick given the presence of an expanding bubble likely driven by the stellar wind of a high mass star \citep{Henshaw2022}. The expanding bubble could be driving the shock observed in \citet{Pillai2015}. The extent of this wind-blown bubble \citep[a radius of 1.3 pc,][]{Henshaw2022} is marked with a red dashed circle in Figure \ref{fig:brick}. Feedback shocks resulting from the expansion of this wind-blown bubble could generate the shocked gas and explain the ordered and curved magnetic field of the Brick. In fact, inspection of the FIREPLACE magnetic field pseudovectors that coincide with the periphery of the bubble reveals that they largely trace the periphery of the bubble, particularly in high $I_{250}$ emission regions coincident with the core of the Brick molecular cloud.

The magnetic field external to the periphery of the wind-blown bubble parallels the largely azimuthal magnetic field geometry seen for the M0.8-0.2 Ring near Sgr B2 that is studied in \citetalias{Butterfield2024}. The M0.8-0.2 ring is likely a supernova remnant, as shown from the multi-wavelength results discussed in \citetalias{Butterfield2024}. They report that the magnetic field in M0.8-0.2 becomes enhanced as a result of compression of the magnetic field due to the expansion of the shell. Similar physics could be occurring with the wind-blown bubble within the Brick.

The fractional polarizations North of the Brick are generally $\rm\geq$5\%, whereas the fractional polarizations coinciding with the Brick range from 5\% to $<$1\%. The fractional polarization within the Brick is observed to steadily decrease from North to South in Galactic Latitude. Outside of the Brick we observe fractional polarizations that are generally $\rm\geq$5\% (like for the pseudovectors observed to the Southwest of the Brick). The large fractional polarization observed external to the Brick, coupled with the generally lower fractional polarizations internal to the cloud, is consistent with the CMZ-wide trend of increased depolarization with higher $I_{214}$ emission as shown in Figure \ref{fig:Hist}.

\subsection{M0.084-0.081 (The Three Little Pigs)} \label{sec:tlp}
\begin{figure*}
    \centering
    \includegraphics[width=1.0\textwidth]{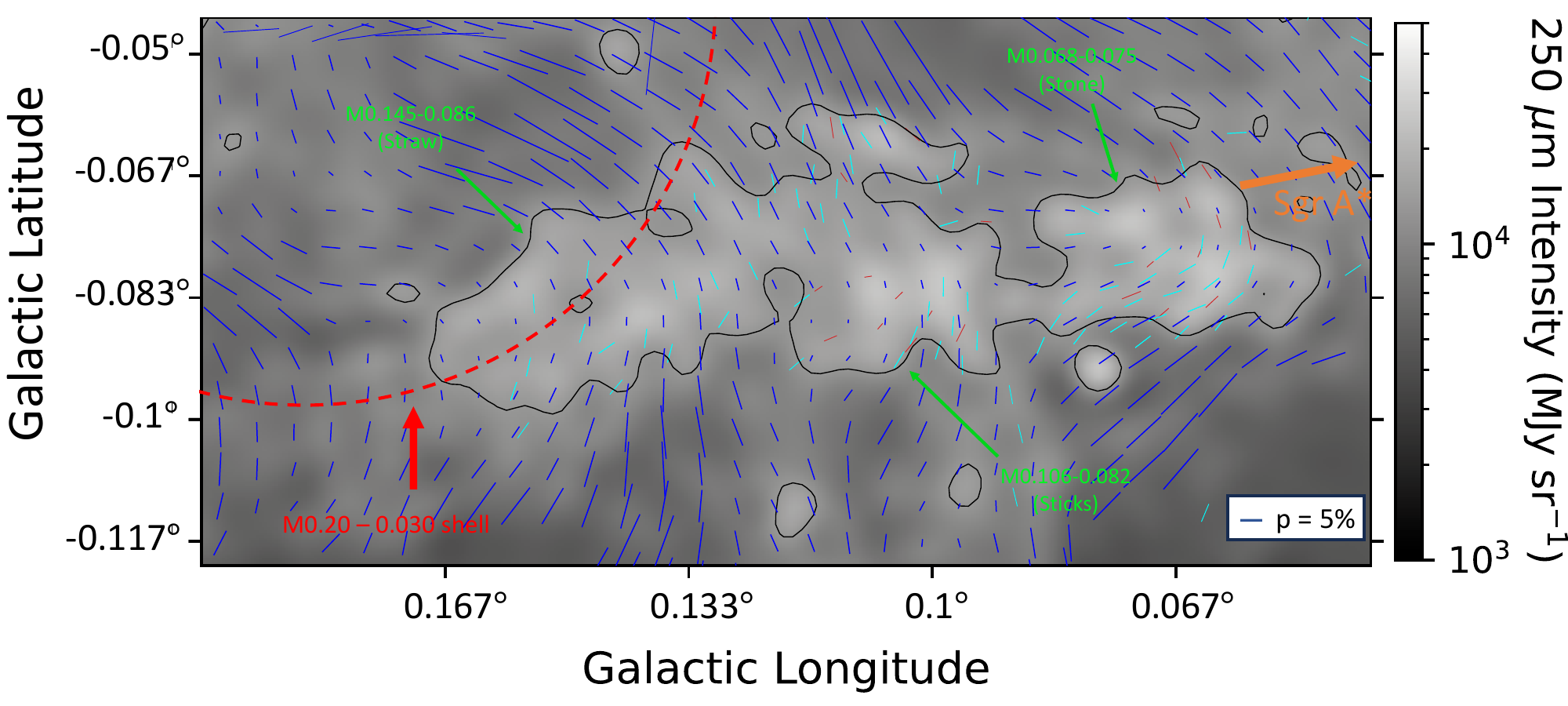}
    \caption{Zoom-in of Figure \ref{fig:B-field_comp} on the Three Little Pigs cloud complex. Total intensity background is from 250 \micron\ Herschel observations \citep{Molinari2011}. Pseudovector colors are the same as for Figure \ref{fig:brick}. The red dashed circle marks an expanding shell inferred from molecular line emission \citep{Butterfield2018,Butterfield2022}. The orange arrow indicates the direction towards Sgr A$\rm^*$. The contour level indicates the extents of the TLP clouds.}
    \label{fig:TLP}
\end{figure*}
The M0.145-0.086 (Straw), M0.106-0.082 (Sticks), and M0.068-0.075 (Stone) comprise the cloud complex known as the ``Three Little Pigs'' \citep[TLP,][]{Battersby2020}. These clouds are labeled and shown in Figure \ref{fig:TLP}. The extent of this complex is oriented nearly parallel to the Galactic plane. The clouds all have similar velocities \citep[$\sim$50 \kms\ e.g.,][]{Kruijssen2015,Butterfield2018}. This indicates they are likely dynamically associated with each other and not simply close in projection. The TLP is located $\rm\sim$10\arcmin\ in projection from Sgr A$\rm^*$ as can be seen in Figure \ref{fig:LIC_anno}.

Previous polarimetric observations of the TLP clouds yielded preliminary evidence of a curved magnetic field within the Stone cloud in the TLP. This curvature occurs on the periphery of this cloud closest to Sgr A$\rm^*$ \citep{Chuss2003a,Chuss2003b,Lu2023}. The Straw and the Sticks clouds, conversely, were observed to have magnetic fields that are oriented largely perpendicular to the Galactic plane and do not exhibit as much warping as seen in the Stone cloud. The perpendicular orientation of the magnetic field near the Straw cloud is particularly interesting since this cloud coincides with the Radio Arc NTF \citep[e.g.][]{YMC1984,YM1987,Pare2019}, as can be seen in Figure \ref{fig:LIC_anno}. This perpendicular field is parallel to the orientation of the filaments in the Radio Arc, which could indicate a connection between the field inferred from dust polarization and the field traced by the synchrotron emission of the Radio Arc filaments.

The difference in magnetic field curvature between the TLP clouds was previously observed in the arcminute-resolution ACTPol results presented in \citet{Guan2021}. Our results are consistent with these previous observations but show more detail thanks to our 19.6\arcsec\ resolution. Our new FIREPLACE observations provide over 100 magnetic field pseudovectors in the vicinity of the TLP, including most of the extended, lower intensity regions. In the cores of the Straw, the Sticks, and the Stone clouds the FIREPLACE pseudovectors have similar orientations to those derived from the CSO and JCMT observations.

The FIREPLACE magnetic field to the South of the Straw cloud is perpendicular to the Galactic plane at lower Galactic latitudes (-0.1\degree\ and below in the figure). Conversely, the magnetic field above the cloud is oriented almost parallel to the Galactic plane. The field above and below the Sticks cloud is oriented perpendicular to the Galactic plane and only becomes warped for lines-of-sight where the $I_{250}$ emission exceeds $\rm\sim3\times{}10^4$ \Mjsr. The field surrounding the Stone cloud exhibits the same curvature as is observed in the core of the cloud from the previous CSO and JCMT observations.

The systematic differences in the magnetic field orientation seen above and below the Straw cloud coincide with the low-latitude outer extent of the M0.20-0.033 shell likely originating from the Quintuplet cluster \citep{Butterfield2018,Butterfield2022} as marked in Figure \ref{fig:TLP}. They argue that the TLP clouds could be interacting with the M0.20-0.033 shell. The magnetic field pseudovectors from FIREPLACE do appear to be impacted by the presence of the shell: changing from a systematically perpendicular (with respect to the Galactic plane) magnetic field orientation outside the shell to parallel or rotated within the shell.

The impact of the shell on the magnetic field pseudovectors is also observed in higher $I_{250}$ emission regions of the Straw cloud, where the core of the cloud coincides with the outer extent of the shell. This result corroborates the argument of \citet{Butterfield2018,Butterfield2022} of an interaction between the shell and the TLP complex; indicating that at least the Straw cloud is likely impacted by the presence of the shell. A similar argument is made in \citet{Lu2023} based on their JCMT observations of this cloud.

Neither the Sticks nor the Stone clouds show any evidence of interaction with the M0.20-0.033 shell in our observations. However, the field coinciding with and surrounding the Stone cloud is noticeably more warped than that of the Sticks cloud. The Stone cloud in particular is too far in projection from the M0.20-0.033 shell for the curvature of the magnetic field in that region to be explained by interaction with that shell. Instead, the curvature could be the deformation of an initially vertical magnetic field induced by the orbital motion of the TLP cloud complex as it moves through the field.


An explanation for the differences in the magnetic field geometries observed between the Straw, Sticks, and Stone clouds could be due to differences in column density, as can be seen from the Herschel column density maps of the GC \citep{Molinari2011}, with the Straw cloud being the least dense and the Stone being the most dense. This column density gradient could indicate changes in the magnetic critical parameter which quantifies the relative importance of the magnetic field and self-gravity within these clouds \citep[e.g.,][]{Nakano1978}.  

A difference in the magnetic critical parameter between the clouds could indicate changes in the tendency of the clouds to contract and form stars \citep{Nakano1978}. This possibility is consistent with recent star formation rate (SFR) estimates obtained by the CMZoom survey.  They found an SFR of 6.4$\rm\pm$3.8 10$^{-3}$\msun\ yr$^{-1}$ for the Stone cloud compared to 0.4$\rm\pm$0.3 10$^{-3}$\msun\ yr$\rm^{-1}$ for the Straw cloud \citep{Hatchfield2024}. The higher SFR of the Stone cloud increases the likelihood of there being outflowing winds within this cloud. However, it remains unclear what the cause for the different magnetic field distribution could be. Follow-up work is necessary to determine whether it is the orbital motion of the TLP or differences in the presence of stellar winds between the clouds that best explains the differences in the magnetic field seen for the Sticks and the Stone cloud.

As with the Brick, we observe increased polarization fractions towards the periphery of the TLP cloud complex. The highest polarization fractions observed for this cloud are $\rm\sim$5\%, with regions in the center of the cloud having generally lower polarization fractions ($\rm<$1\%) than what is observed in the fainter peripheries of the cloud complex. 

\subsection{Clouds Surrounding Sgr A$^*$} \label{sec:sgrA}
\begin{figure*}
    \centering
    \includegraphics[width=1.0\textwidth]{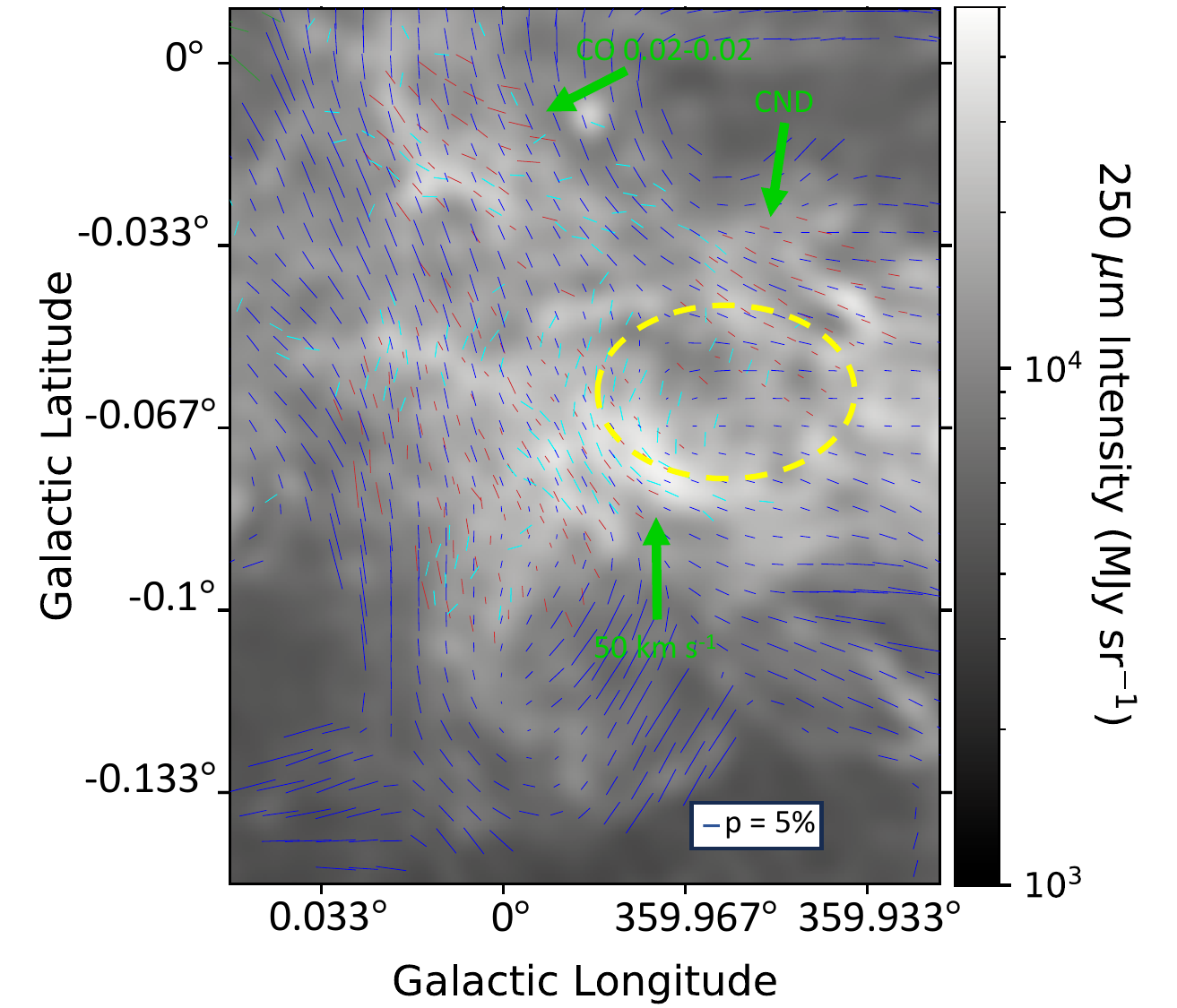}
    \caption{Zoom-in of Figure \ref{fig:B-field_comp} on the 50 \kms, Circum-nuclear Disk, and CO 0.02-0.02 clouds. Total intensity background is from 250 \micron\ Herschel observations \citep{Molinari2011}. Pseudovector colors are the same as for Figure \ref{fig:brick}. The yellow ellipse marks the extent of the Sgr A East supernova remnant discussed in the text \citep{Ehlerova2022}.}
    \label{fig:50kms}
\end{figure*}
There are multiple clouds that are within 5\arcmin\ in projection from Sgr A$\rm^*$ (Figure \ref{fig:LIC_anno}). These prominent clouds are M0.001-0.058 (50 \kms\ cloud), M359.948-0.052 (Circum-nuclear Disk), CO 0.02-0.02 (as labeled in \citealt{Lu2023}), and M359.889-0.093 (20 \kms\ cloud). Zoom-in views of these clouds are presented in Figures \ref{fig:50kms} and \ref{fig:20kms}.

\subsubsection{M0.001-0.058 (50 \kms\ Cloud)} \label{sec:50kms_foc}
The 50 \kms{} cloud is located to the Galactic Southeast of Sgr A$\rm^*$ and is contained within the greater Sgr A complex (Figure \ref{fig:LIC_anno}).

We obtain $\rm\sim$50 magnetic field pseudovectors from our 214 \micron\ observations coinciding with the 50 \kms\ cloud as seen in Figure \ref{fig:50kms}. These pseudovectors are largely oriented perpendicular to the Galactic plane. However, there are deviations from this perpendicular field orientation. For example, there is a dense dust ridge within the 50 \kms\ cloud at $l$ = 359.967\degree, $b$ = -0.067\degree. The magnetic field coincident with this dust ridge traces the orientation of this ridge, which is rotated by 45\degree\ relative to the Galactic plane. This dust ridge in the 50 \kms\ cloud is thought to be compressed by the nearby Sgr A East supernova remnant, the extent of which is marked by the dashed yellow circle in Figure \ref{fig:50kms}  \citep{Serabyn1992,Novak2000,Ehlerova2022}, which can explain why the magnetic field is oriented along the ridge. The dust ridge traces the edge of the supernova remnant, meaning the FIREPLACE magnetic field is largely azimuthal with respect to the supernova remnant. A similar, largely azimuthal, magnetic field morphology is observed in the periphery of the Ring structure studied in \citetalias{Butterfield2024}.

The magnetic field observed in the 50 \kms\ cloud changes direction at Southern Galactic latitudes where the intensity decreases and the previous CSO and JCMT observations lack sensitivity.  This is particularly noticeable at $l$ = 359.97\degree, $b$ = -0.117\degree\ where the field subtends an angle of 60\degree\ with respect to the Galactic plane. The fractional polarizations are also quite large in this region (generally $>$5\%). The magnetic field orientation also systematically changes from being generally oriented perpendicular to the Galactic plane to the East of the 50 \kms\ cloud to being parallel to the Galactic plane to the West of the 50 \kms\ cloud. 

\subsubsection{M359.948-0.052 (Circum-nuclear Disk)} \label{sec:CND}
The Circum-nuclear Disk (CND) is located within 5 pc of the super massive black hole Sgr A$\rm^*$ (labeled in Figure \ref{fig:50kms}). The CND has an inclination of $\sim$70\degree\ relative to the plane of the sky and an inner radius of $\sim1.5$ pc \citep{Lau2013,Guerra2023}. 

Images by SOFIA/FORCAST at 19.7, 31.5, and 37.1, and SOFIA/HAWC+ images at 53 \micron\ show a clear indication that the CND is experiencing gravitational shear \citep{Lau2013,Guerra2023}.  At far-infrared wavelengths \citep[250 \micron,][]{Molinari2011}, the structure of the CND is only faintly apparent, with the emission being dominated by more extended, cooler  CMZ dust, which is located external to the central 5 pc. The FIREPLACE 214 \micron\ Stokes $I$ maps are consistent with these images. 

The FIREPLACE dataset contains $\rm\sim$50 Nyquist-sampled magnetic field pseudovectors in the vicinity of the CND. The FIREPLACE 214 \micron\ data is consistent with the geometry measured at 350 \micron\ at 20\arcsec\ resolution \citep[red pseudovectors in Figure \ref{fig:50kms},][]{Novak2000}. Parts of the CND were included in the SCUBA-2/Pol-2 map at 850 \micron\ and 20\arcsec\ resolution \citep[cyan pseudovectors in Figure \ref{fig:50kms};][]{Lu2023}. There is general agreement with these vectors in the Northeastern part of the CND region; however, to the South, the 850 \micron\ vectors do not agree with the shorter wavelengths. This disagreement could be an intrinsic variation in the source polarization at these different wavelengths. Alternatively, it could be a result of differences in spatial filtering from the different observing modes (OTFMAP mode for this study compared to CNM mode for the CSO and Daisy mode for the JCMT \citep{Lu2023}). However, our FIREPLACE results agree with the HAWC+ CNM mode observations of this cloud as seen in Figure \ref{fig:cn}. Therefore the difference is likely a result of changes in the polarization over the different wavelengths. Previous 850 \micron\ polarization measurements have been made of the CND region \citep{Hsieh2018}, in which the authors claimed to detect a sheared geometry of the field. 

The claim of a sheared field was supported by more recent observations of the CND \citep{Guerra2023}. The pseudovectors from these observations show a field that is highly aligned with the dust streamer morphology in \citet{Guerra2023}, who argue for a steady-state shear flow leading to a magnetic field that traces the morphology of the CND arms \citep[alternatively referred to as ``wings'' in other CND studies e.g.,][]{Zhao2016}.  They estimate a $\sim$mG magnetic field strength in the CND using a modified version of the Davis-Chandrasekhar-Fermi technique that accounts for shear \citep{Davis1951, CF1953}. 

The differences in morphology and polarization orientation from 53 to 214 \micron\ (and longer wavelengths) are most likely due to two cloud components superposed along the line-of-sight: the warm dust local to the streamers of the CND and the cooler dust component in the wider CMZ.  The former contains magnetic fields that are dynamically connected to the sheared flow of the inner 5 pc; the latter are likely shaped by the dynamics of the greater CMZ.

\subsubsection{CO 0.02-0.02} \label{sec:CO22}
Cloud CO 0.02-0.02 is only 4\arcmin\ away from the CND in projection to the Northeast of the CND as can be seen in Figure \ref{fig:50kms}. The FIREPLACE observations reveal $\rm\sim$20 magnetic field pseudovectors coinciding with the CO 0.02-0.02 cloud. These pseudovectors are generally oriented perpendicular to the Galactic plane; however, this orientation does not agree with what is observed in previous CSO and JCMT observations of this cloud \citep{Lu2023}. These pseudovectors differ by a $\rm\sim$30\degree\ angle with respect to the FIREPLACE pseudovectors. In addition, there are instances where the CSO and JCMT observations disagree, which could indicate magnetic field complexity in this region of the cloud.

The disagreement in magnetic field orientation could be a result of different spatial filtering between our FIREPLACE observations and that of the CSO and JCMT observations. In addition, we note that the CO 0.02-0.02 cloud falls within the region of elevated $q$ and $u$ variance observed in Figure \ref{fig:variance} from our FIREPLACE observations. The angle differences observed for this cloud could therefore indicate remaining systematic uncertainties in our FIREPLACE observations. This explanation is further supported by the fact that the CSO and JCMT observations are in general agreement over most of this cloud.

\subsubsection{M359.889-0.093 (20 \kms\ Cloud)} \label{sec:20kms}
\begin{figure*}
    \centering
    \includegraphics[width=1.0\textwidth]{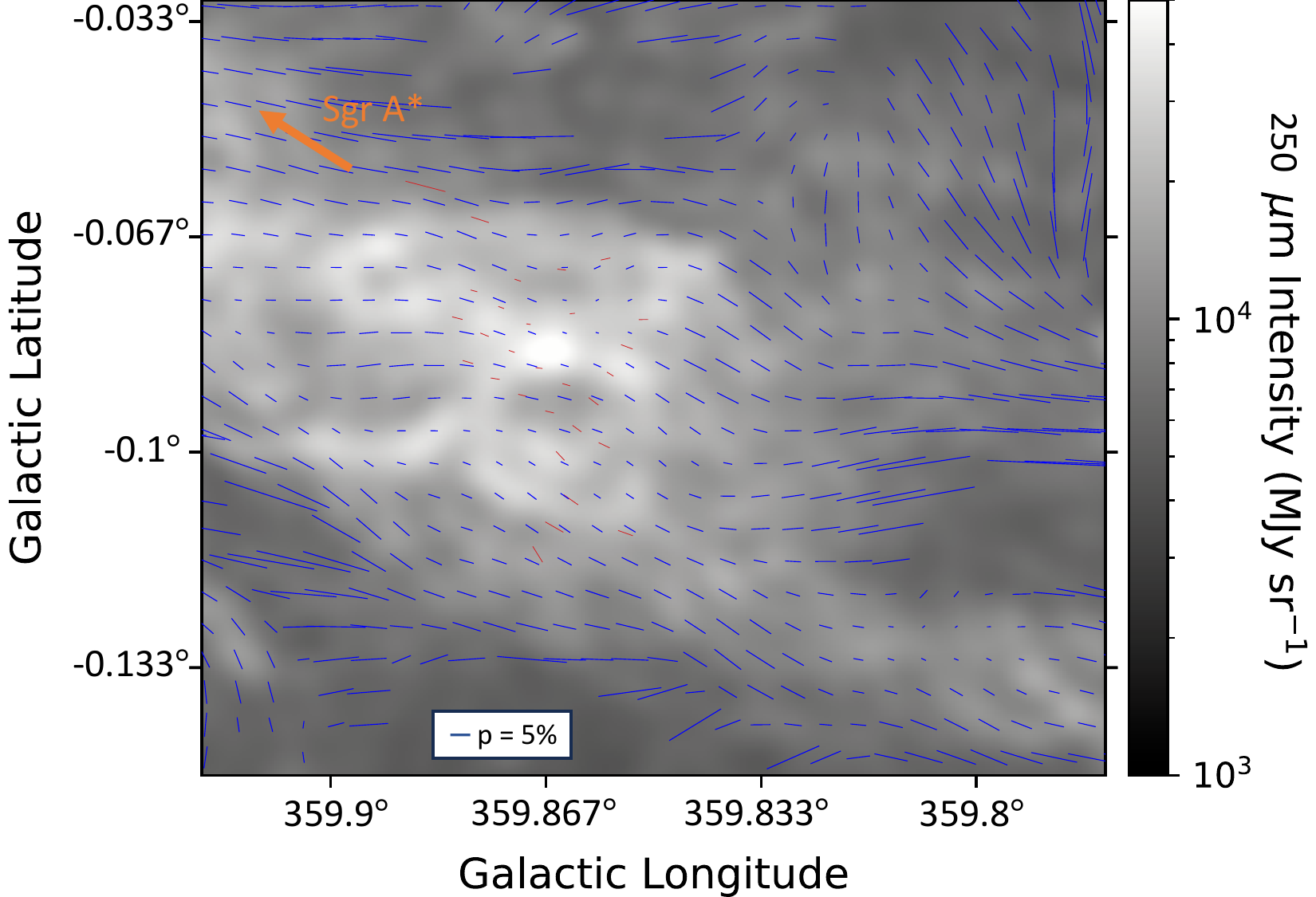}
    \caption{Zoom-in of Figure \ref{fig:B-field_comp} on the 20 \kms\ cloud. Total intensity background is from 250 \micron\ Herschel observations \citep{Molinari2011}. Pseudovector colors are the same as for Figure \ref{fig:brick}. The orange arrow indicates the direction towards Sgr A$\rm^*$.}
    \label{fig:20kms}
\end{figure*}
M359.889-0.093, known colloquially as the 20 \kms\ cloud, is located 10\arcmin\ in projection from Sgr A$\rm^{*}$ to the Southwest as shown in Figure \ref{fig:LIC_anno}. A zoom-in view of the 20 \kms\ cloud is shown in Figure \ref{fig:20kms}. The orange arrow in Figure \ref{fig:20kms} indicates the direction to Sgr A$\rm^*$.

Our SOFIA/HAWC+ observations yield $\rm\sim$100 distinct pseudovectors that span the full extent of the 20 \kms\ cloud. The field within the cloud is predominantly oriented parallel to the Galactic plane. The field above and below the cloud in Galactic latitude is also largely oriented parallel to the Galactic plane. 

The core of the 20 \kms\ cloud is comprised of a set of filamentary structures in the dust continuum that appear to coalesce at the brightest 250 \micron\ clump in the cloud, located at $l$=359.867\degree, $b$=-0.083\degree.
At this location we observe low fractional polarization values ($\ll$1\%), suggesting that depolarization in this region may be present. This depolarization could be caused by the overlapping of field orientations along our line-of-sight from these multiple filamentary structures. Alternative explanations for depolarization in this region could be caused by loss of grain alignment in the densest regions of the cloud or multiple field components within the size of our beam. Depolarization in these denser regions has been observed in several of CMZ clouds including Sgr B2 and Dust Ridge cloud E/F \citepalias{Butterfield2023}. 
Previous studies using combined SMA and VLA observations find that the star formation of the 20 \kms\ cloud is likely in an early evolutionary stage \citep{Lu2015}. Sgr B2 and the Dust Ridge cloud E/F, which also show depolarization in their cores show signs of early and on-going star formation - indicating that the 20 \kms\ cloud may be at a similar evolutionary stage. 



\subsection{M359.484-0.132 (Sgr C)} \label{sec:sgrc}
\begin{figure*}
    \centering
    \includegraphics[width=1.0\textwidth]{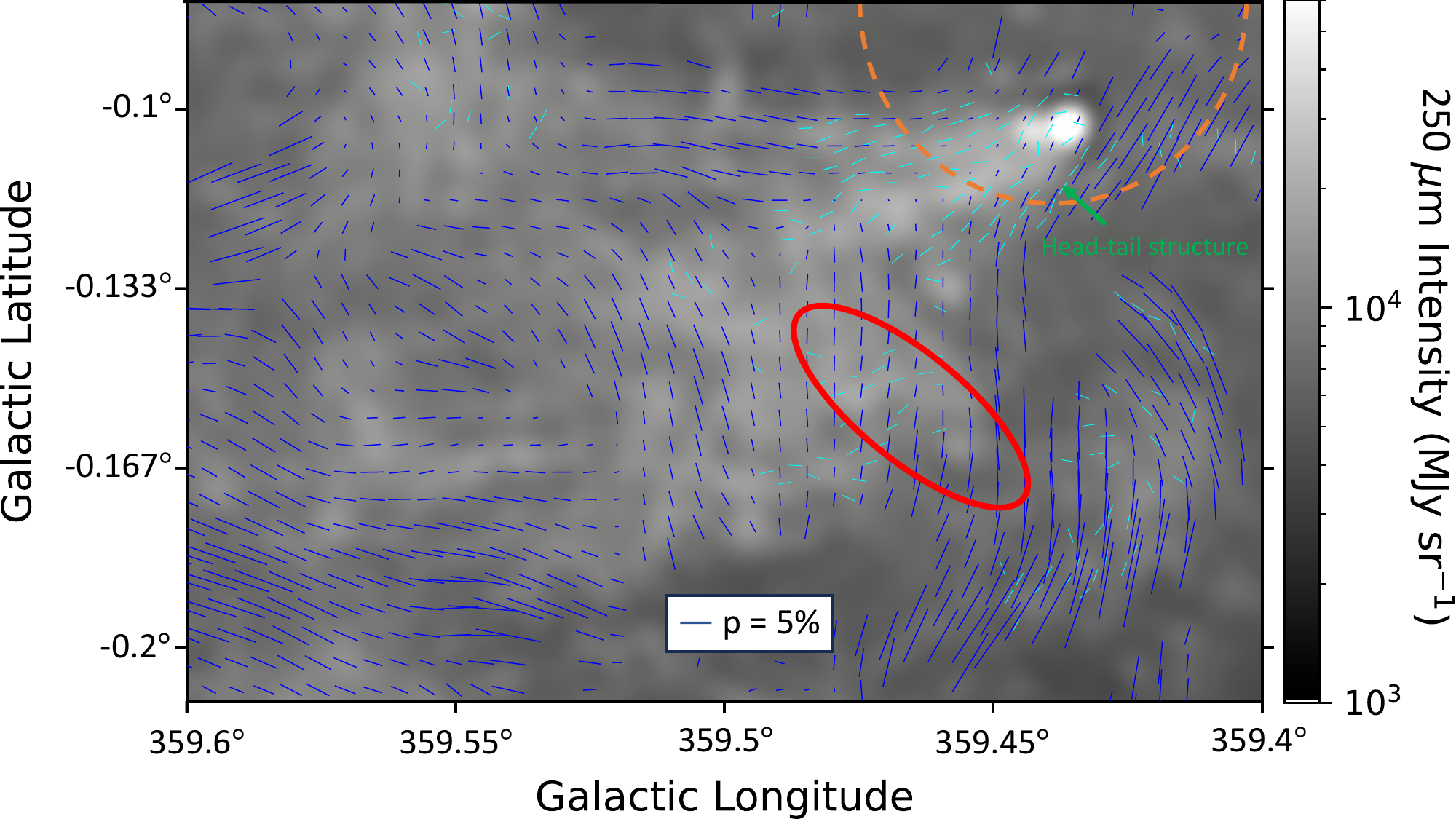}
    \caption{Zoom-in of Figure \ref{fig:B-field_comp} on  Sgr C. Total intensity background is from 250 \micron\ Herschel observations \citep{Molinari2011}. Pseudovector colors are the same as for Figure \ref{fig:brick}. The dashed orange circle marks the extent of the prominent \hii\ region discussed in the text. The red ellipse marks the extent of the AGAL 359.474-0.152 clump from the ATLASGAL catalog discussed in the text \citep{Contreras2013}. The head-tail structure discussed in the text is marked with the green arrow.}
    \label{fig:SgrC}
\end{figure*}
Sgr C is located $\rm\sim$0.5\degree{} in projection from the 20 \kms\ cloud and is on the Western periphery of  the CMZ as shown in Figure \ref{fig:LIC_anno}. The cloud manifests a mix of diffuse and compact emission coupled with a broad range of $I_{250}$ emissivities as shown in Figure \ref{fig:SgrC}. This cloud was recently observed polarimetrically using the JCMT \citep{Lu2023} and the CNM mode observations made using SOFIA/HAWC+ that are shown in Figure \ref{fig:cn}. Prior to these observations Sgr C had not been studied polarimetrically at a resolution less than 1\arcmin. The FIREPLACE observations of Sgr C yield $\rm>$100 magnetic field pseudovectors within the periphery of the cloud. These observations largely agree with the orientations derived from the SOFIA/HAWC+ CNM observations of this cloud as seen in Figure \ref{fig:cn}. 

The $I_{250}$ emission from Sgr C obtained by Herschel is generally diffuse, but there is a region of compact, high intensity emission at $l$ = 359.433\degree, $b$ = -0.1\degree. There is a ridge or tail of emission emanating from this compact emission towards the Southeast, which we hereafter refer to as the ``head-tail structure'' (labeled in Figure \ref{fig:SgrC}). The magnetic field derived from our 214 \micron\ observations follows the morphology of the head-tail structure and is largely consistent with the magnetic field orientation derived from the 850 \micron\ JCMT observations of \citet{Lu2023}. \citet{Lu2023} argue that the magnetic field orientation in the head-tail structure could result from interaction of the Sgr C molecular cloud with a nearby prominent \hii\ region, the extent of which has been marked with an orange circle in Figure \ref{fig:SgrC} \citep{Lang2010,Hankins2020}. Evidence of this  interaction between Sgr C and the \hii\ region has previously been presented in e.g. \citet{Lu2019}. The similar magnetic field orientations derived from FIREPLACE bolster the argument that winds from the prominent \hii\ region create molecular pillars leading to a compressed magnetic field that follows the orientation of the head-tail structure. Magnetohydrodynamic simulations of expanding \hii\ regions reveal similar molecular and magnetic field morphologies to what is seen in the head-tail structure of Sgr C \citep{Arthur2011,Mackey2011}.

Our results corroborate the simulation results obtained from these magnetohydrodynamic models of expanding \hii\ regions. Furthermore, the field geometry seen in the high $I_{250}$ emission is preserved in lower intensities as measured by our FIREPLACE observations. \citet{Arthur2011} and \citet{Mackey2011} argue that the magnetic field is dragged along the head-tail structure by the expanding wind from the \hii\ region.

The magnetic field revealed by our SOFIA/HAWC+ observations Southeast of the head-tail structure at $l$ = 359.467\degree, $b$ = -0.15\degree\ is largely oriented perpendicular to the Galactic plane. The JCMT pseudovectors in this Southern portion of Sgr C generally do not agree with those obtained in the FIREPLACE survey and are more spatially varying. To help understand the difference in polarization angle between the FIREPLACE and JCMT observations, we note that the significance of the JCMT polarized intensity in this region is reported as $p/\sigma_p > 3$, whereas the significance of the JCMT polarized intensity in the head-tail structure in Sgr C is $p/\sigma_p > 5$ \citep{Lu2023}.  $p/\sigma_p = 3$ is equivalent to a 10\degree\ uncertainty in the polarization angle, and the JCMT orientations disagree with those obtained from FIREPLACE by 70-90\degree\ in this Southern region. 

To the Southwest of the head-tail structure our SOFIA/HAWC+ observations reveal a curved magnetic field structure at $l$ = 359.417\degree, $b$ = -0.167\degree. There are also JCMT pseudovectors at this location which are again highly spatially variable. The curvature in this region is similar to the field structures seen for the Brick and the TLP clouds that were attributed to out-flowing winds.A possible source of these winds is a compact source identified in the ATLASGAL survey \citep[AGAL 359.474-0.152;][]{Contreras2013}, whose location is marked with a red ellipse in Figure \ref{fig:SgrC}. This source has an integrated flux of 168 Jy and a total gas mass of 5 $\times$ 10$\rm^4$ \msun\ \citep{Kendrew2013}. These properties make AGAL359.474-0.152 favorable for forming massive stars \citep{Lada2012}.

Therefore, out-flowing winds from star formation in Sgr C, likely originating from AGAL 359.474-0.152 is a reasonable origin for the structure of the magnetic field in this region of Sgr C, especially in light of the ample evidence of star formation in Sgr C \citep[e.g.][]{Yusef-Zadeh2009,Lu2019}. However, there are alternative mechanisms that could result in the magnetic field distribution seen in this region, such as shear from orbital motion of the structure.

As with the previously discussed clouds, the fractional polarizations derived from our 214 \micron\ SOFIA/HAWC+ observations increase towards the periphery of Sgr C in lower $I_{250}$ emission regions. In the region of diffuse emission to the South of the head-tail structure fractional polarizations range from 4 -- 5\%. In the head-tail structure, conversely, the fractional polarizations are $\rm<$1\%, similar to what is observed in the peak $I_{250}$ emission regions of other prominent CMZ clouds.

\section{CONCLUSIONS} \label{sec:conc}
In this work, the full FIREPLACE survey covering the entire CMZ polarimetrically at 214 \micron\ is presented. First, the reduction method performed on the observations obtained in 2022 is detailed, comprising the second FIREPLACE data release (DR2) and observing the region of the CMZ extending from the Brick to Sgr C (corresponding to a 1\degree\ $\times$ 0.75\degree\ region of the sky). The reduced DR2 observations are then combined with the DR1 observations presented in \citetalias{Butterfield2023} to obtain a polarimetric map of the entire CMZ at 214 \micron\ (a 1.5\degree\ $\times$ 0.75\degree\ region of the sky). CMZ-wide trends of the magnetic field are then inferred and analyzed, with the fields local to a set of prominent molecular clouds inspected in more detail. The following summarizes the results from the inspection of the full FIREPLACE survey:

\begin{itemize}
    \item Our reduction method, using the SOFIA DRP software, results in a robust data set that is well verified and consistent with other SOFIA/HAWC+ CMZ polarimetric observations made using the CNM observing mode. The combination of the convergence, variance, and CNM mode comparison presented in Section \ref{sec:qa} convincingly shows that our reduction is reliably removing the correlated noise while accurately recovering the source emission.
    \item The sensitivity of our HAWC+ observations makes it possible to derive the polarization angle and magnetic field orientations in fainter molecular structures in the CMZ at a resolution of 19.6\arcsec\ than ever before. The distribution of $\rm\sim$64,000 Nyquist-sampled magnetic field pseudovectors from the full FIREPLACE survey is shown in Figure \ref{fig:B-field_comp}.
    \item A bimodal distribution in the CMZ-wide polarization angle orientations is identified from the full FIREPLACE survey. This bimodal distribution shows enhancements in directions parallel (approximately -16\degree) and perpendicular (approximately -80\degree) to the Galactic plane. This is a similar bimodal distribution to the one recovered in \citetalias{Butterfield2023} for the Sgr B2 region. The component that is parallel to the Galactic plane could either indicate a horizontal magnetic field component in the CMZ \citep{Mangilli2019,Guan2021} or a component external to the CMZ. The bimodal component that is perpendicular to the Galactic plane could be related to the vertical magnetic field distribution in the GC \citep[e.g.][]{Morris2006sum}, possibly revealing connections between the magnetic fields traced by infrared dust polarization and radio synchrotron emission.
    \item Regions of vertical magnetic field (as shown in Figure \ref{fig:paraperp}) are not confined to the locations of detected NTFs, whereas regions where the magnetic field is parallel to the Galactic plane tend to be coincident with molecular clouds. This result supports the theory that there is a vertical magnetic field that pervades the GC and is sheared into a horizontal field in the higher density CMZ molecular clouds.
    \item The distributions of the magnetic fields of a set of prominent molecular clouds in the CMZ are also analyzed: the Brick, Three Little Pigs (TLP) cloud complex, 50 \kms\ cloud, CND, CO 0.02-0.02, 20 \kms\ cloud, and Sgr C. The magnetic field derived from FIREPLACE generally traces the morphology of the molecular clouds. In addition, there is evidence that expanding shock fronts from SNRs, advancing ionization fronts from \hii\ regions, and tidal forces from orbital motion might cause compression and shearing leading to the organization of the magnetic fields in the Brick, TLP cloud complex, 50 \kms\ cloud, and Sgr C. However, not all of the prominent clouds studied in this work exhibit signatures of compression or winds (the 20 \kms\ cloud, for example). These results nonetheless indicate that the compression from winds and shear from tidal forces are important mechanisms for ordering the magnetic field distributions within CMZ clouds.
\end{itemize}
The FIREPLACE observations presented in this work have the potential to enhance our understanding of the role magnetic fields play in the star formation process in the complex CMZ. Furthermore, the magnetic field distribution derived from these observations will improve our ability to develop a unified picture of the CMZ-wide magnetic field throughout the GC region. The reduced observations presented in this paper are publicly available on the NASA IPAC Legacy Surveys webpage as FITS files \footnote{https://irsa.ipac.caltech.edu/data/SOFIA/docs/data/legacy-programs/two-color-polarimetric-survey-galactic-center-pilot-legacy-program/index.html}.

\begin{acknowledgements}
    This work is based on observations made with the NASA/DLR Stratospheric Observatory for Infrared Astronomy (SOFIA). SOFIA was jointly operated by the Universities Space Research Association, Inc. (USRA), under NASA contract NNA17BF53C, and the Deutsches SOFIA Institut (DSI) under DLR contract 50 OK 2002 to the University of Stuttgart. Financial support for this work was provided by NASA through award \#09-0054 issued by USRA to Villanova University.

    We would like to thank Dr. Simon Coude, Dr. Sachin Shenoy, Dr. Peter Ashton, Dr. Sarah Eftekharzadeh, Dr. Ryan Arneson, and the rest of the SOFIA team for their help with the observations and data reduction, including providing the 2.7.0 HAWC+ DRP software used for the reduction. We would like to thank Dr. Thushara Pillai and Dr. Jens Kauffmann for their assistance and helpful discussion on calibration. We would like to thank Joe Michail (Northwestern University) for his LIC python code used to create Figure  \ref{fig:LIC} (based on IDL code by Diego Falceta-Gonçalves). We would like to thank the anonymous referee for their helpful comments on this work.
\end{acknowledgements}

\facility{
    SOFIA, 
    MeerKAT, 
    Herschel,
    CSO,
    KAO,
    WISE
    }

\software{
    SOFIA-USRA Data Reduction Pipeline,
    \textit{CRUSH} \citep{Kovacs2008},
    Astropy \citep{Greenfield2014},
    Matplotlib \citep{Hunter2007}
    }

\appendix
\counterwithin{figure}{section}
\counterwithin{table}{section}

\renewcommand{\thesection}{A.\arabic{section}}
\renewcommand{\thefigure}{A.\arabic{figure}}
\renewcommand{\thetable}{A.\arabic{table}}
\addtocounter{table}{-1}

\section{Extended Polarization Artifacts in Different DRP Versions} \label{sec:artifacts}
\begin{figure*}
    \centering
    \includegraphics[width=1.0\textwidth]{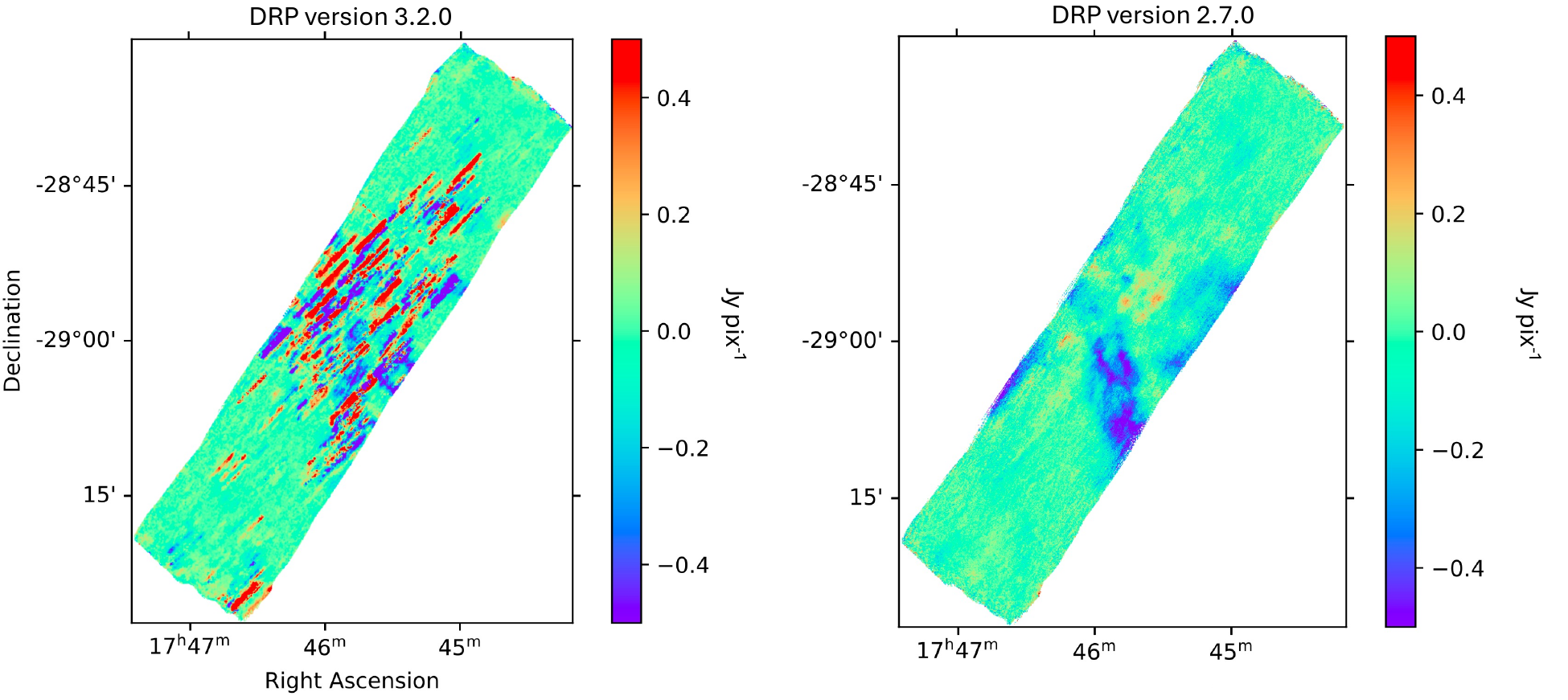}
    \caption{An example of the reduced Stokes $Q$ distributions from a single field of our FIREPLACE observations  (field 7 as numbered in Figure \ref{fig:scans}) reduced using (left) DRP version 3.2.0 and (right) DRP version 2.7.0. These fields were reduced using identical \textit{CRUSH} parameters (\textit{``-extended,-downsample=1,-fixjumps,-rounds=85''}), with the only difference being the DRP version.}
    \label{fig:artcomp}
\end{figure*}
\begin{figure*}
    \centering
    \includegraphics[width=1.0\textwidth]{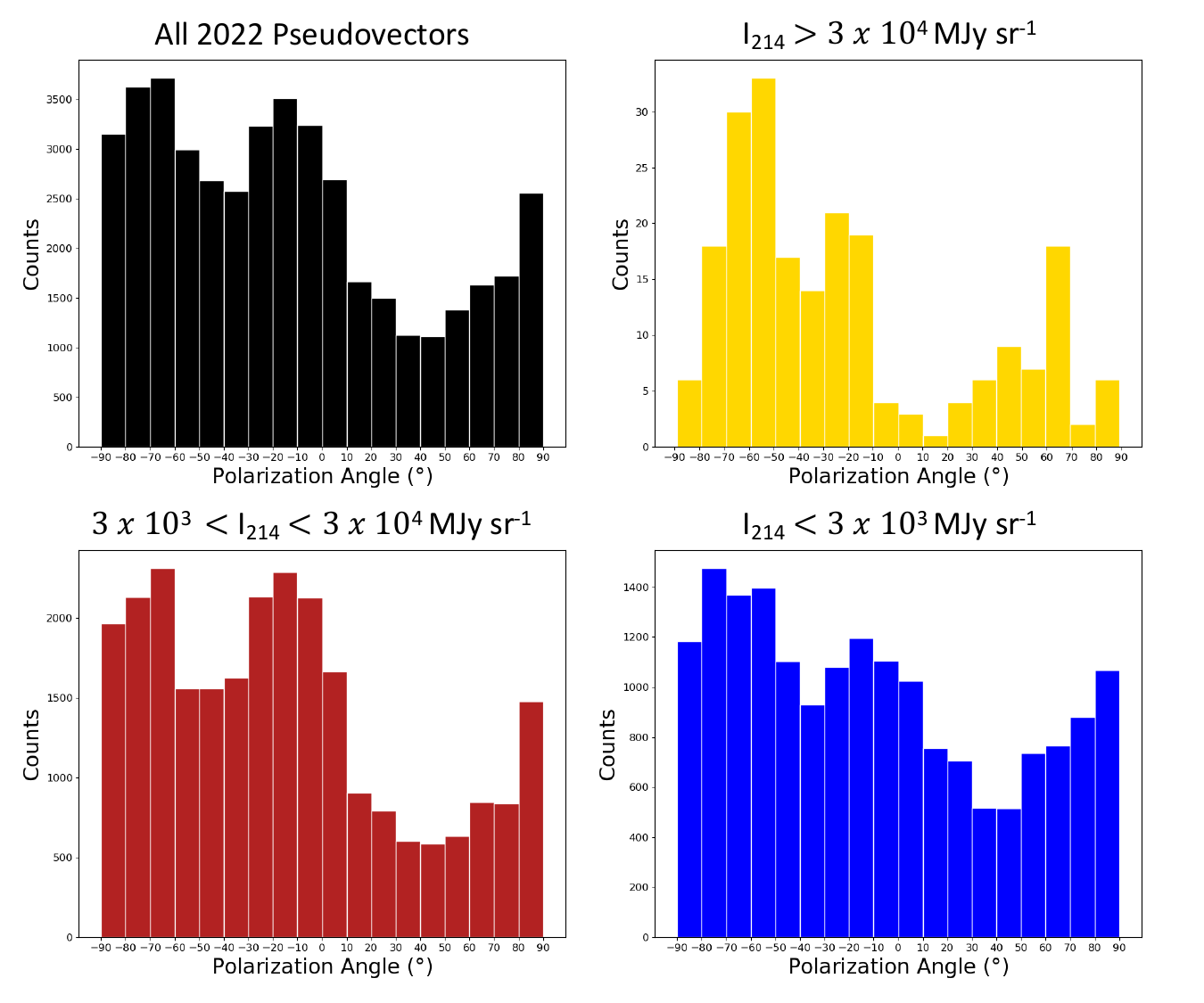}
    \caption{Distribution of the FIREPLACE polarization pseudovectors corresponding to the merged DR2 fields (upper left: black histogram; 44,090 pseudovectors) compared with three different intensity regimes, as titled: $I_{214}>$30,000 \Mjsr\ (yellow; 218 pseudovectors); 3000$<I_{214}<$30,000 \Mjsr\ (red; 26,057 pseudovectors); $I_{214}<$3000 \Mjsr\ (blue; 17,815 pseudovectors). These intensity ranges correspond with the dashed lines in Figure \ref{fig:pol_vec_pix}.}
    \label{fig:2022_Hist}
\end{figure*}
We show an example of the polarization artifacts occurring in DRP version 3.2.0 but not in DRP version 2.7.0 for our FIREPLACE observations in Figure \ref{fig:artcomp}. These artifacts appear in DRP version 2.7.0 when using the \textit{``-extended''} flag, but do not appear without this flag\footnote{This difference could be a result of the updated image reconstruction SOFIA developed for version 3.2.0. The details of this update are outside of the scope of this paper but are discussed in the HAWC+ data reduction manuals: https://irsa.ipac.caltech.edu/data/SOFIA/docs/data/data-pipelines/}. The FIREPLACE field shown in Figure \ref{fig:artcomp} was reduced using identical reduction parameters (as indicated in the caption of Figure \ref{fig:artcomp}), with the only difference being the DRP version used to perform the reduction.

\section{Guide to Individual Fields and CAL Files} \label{sec:cal_tab}
We tabulate the fields and calibrated files used for the DR2 FIREPLACE data set in Table \ref{tab:DR2}. The files in Table \ref{tab:DR2} contain calibrated data with pointing corrections applied. Field numbers refer to the locations in Figure~\ref{fig:scans}. ``CCW'' and ``CW'' refer to the rotation direction of the major scan direction relative to perpendicular to Galactic North ($\rm\pm$30\degree\ as discussed in Section \ref{sec:obs}). For completeness, files from Flight 890 are included in this table, though as discussed in the main paper these files are not used in the DR2 reduction. Horizontal lines are included to separate observations from different flights.

Table \ref{tab:DR1} tabulates the FIREPLACE DR1 files that are combined with the DR2 files shown in Table \ref{tab:DR2} to produce the full FIREPLACE data release covering the entire CMZ. The format of Table \ref{tab:DR1} is identical to that of Table \ref{tab:DR2} except that orientation (``CCW'' or ``CW'') is not specified since the orientation of the DR1 observations was more variant, as discussed in Section \ref{sec:obs}.

\section{Histograms of the 2022 Polarization Pseudovectors} \label{sec:2022_Hist}
We produce histograms of the polarization pseudovectors of only the FIREPLACE DR2 fields to determine whether any of the magnetic field enhancements observed in Figure \ref{fig:Hist} are confined to only the DR1 or DR2 portions of the full FIREPLACE data set.

To do so we sort the Nyquist-sampled polarization angles shown in Figure \ref{fig:fire_2022} into 10\degree\ bins. The histogram representation of the binned polarization angles for the DR2 observations is shown in Figure \ref{fig:2022_Hist}. As with the histogram of the full FIREPLACE release, an angle of 0\degree\ indicates a magnetic field that is aligned with the Galactic plane, with an angle of $\rm\pm$90\degree\ indicating a magnetic field that is oriented perpendicular to the Galactic plane.

\begin{table}
\centering
\caption{Summary of calibrated files used for the DR2 maps}
\begin{tabular}{llll}
\hline
File Name & Field & Orientation & Scans \\ \hline\hline
F0890\_HA\_POL\_09005471\_HAWEHWPE\_CAL\_014-021.fits & 4 & CCW & 8\\
F0890\_HA\_POL\_09005472\_HAWEHWPE\_CAL\_006-013.fits & 4 & CW & 8\\
F0890\_HA\_POL\_09005473\_HAWEHWPE\_CAL\_031-038.fits & 5 & CCW & 8\\
F0890\_HA\_POL\_09005474\_HAWEHWPE\_CAL\_023-030.fits & 5 & CCW & 8\\
F0890\_HA\_POL\_09005475\_HAWEHWPE\_CAL\_058-061.fits & 6 & CW & 4\\
F0890\_HA\_POL\_09005475\_HAWEHWPE\_CAL\_064-067.fits & 6 & CW & 4\\
F0890\_HA\_POL\_09005476\_HAWEHWPE\_CAL\_068-071.fits & 6 & CCW & 4\\\hline
F0891\_HA\_POL\_09005476\_HAWEHWPE\_CAL\_004-007.fits & 6 & CCW & 4\\
F0891\_HA\_POL\_09005477\_HAWEHWPE\_CAL\_008-011.fits & 7 & CW & 4\\
F0891\_HA\_POL\_09005478\_HAWEHWPE\_CAL\_016-023.fits & 7 & CCW & 8\\
F0891\_HA\_POL\_09005479\_HAWEHWPE\_CAL\_024-031.fits & 8 & CW & 8 \\\hline
F0893\_HA\_POL\_09005479\_HAWEHWPE\_CAL\_009-016.fits & 8 & CW & 8 \\
F0893\_HA\_POL\_09005480\_HAWEHWPE\_CAL\_017-024.fits & 8 & CCW  & 8 \\
F0893\_HA\_POL\_09005481\_HAWEHWPE\_CAL\_029-032.fits & 9 & CW & 4 \\
F0893\_HA\_POL\_09005482\_HAWEHWPE\_CAL\_033-040.fits & 9 & CCW & 8\\
F0893\_HA\_POL\_09005483\_HAWEHWPE\_CAL\_041-044.fits & 10 & CW & 4 \\\hline
F0895\_HA\_POL\_09005483\_HAWEHWPE\_CAL\_003-006.fits & 10 & CW & 4 \\
F0895\_HA\_POL\_09005484\_HAWEHWPE\_CAL\_007-014.fits & 10 & CCW & 8 \\
F0895\_HA\_POL\_09005485\_HAWEHWPE\_CAL\_015-018.fits & 11 & CW & 4 \\
F0895\_HA\_POL\_09005485\_HAWEHWPE\_CAL\_020-023.fits & 11 & CW & 4 \\
F0895\_HA\_POL\_09005486\_HAWEHWPE\_CAL\_024-031.fits & 11 & CCW & 8 \\
F0895\_HA\_POL\_09005487\_HAWEHWPE\_CAL\_032-039.fits & 12 & CW & 8 \\
F0895\_HA\_POL\_09005488\_HAWEHWPE\_CAL\_040-043.fits & 12 & CCW & 4 \\
F0895\_HA\_POL\_09005488\_HAWEHWPE\_CAL\_047-050.fits & 12 & CCW & 4 \\
F0895\_HA\_POL\_09005489\_HAWEHWPE\_CAL\_051-054.fits & 13 & CW & 4 \\\hline
F0916\_HA\_POL\_09005489\_HAWEHWPE\_CAL\_004-011.fits & 13 & CW & 8 \\
F0916\_HA\_POL\_09005490\_HAWEHWPE\_CAL\_013-020.fits & 13 & CCW & 8 \\
F0916\_HA\_POL\_09005490\_HAWEHWPE\_CAL\_022-025.fits & 13 & CCW & 4 \\
F0916\_HA\_POL\_09005493\_HAWEHWPE\_CAL\_026-033.fits & 15 & CW & 8 \\ \hline
F0917\_HA\_POL\_09005494\_HAWEHWPE\_CAL\_002-005.fits & 15 & CCW & 4 \\
F0917\_HA\_POL\_09005497\_HAWEHWPE\_CAL\_007-014.fits & 17& CW & 8\\
F0917\_HA\_POL\_09005498\_HAWEHWPE\_CAL\_015-022.fits & 17 & CCW & 8\\\hline
F0918\_HA\_POL\_09005499\_HAWEHWPE\_CAL\_004-011.fits & 18 & CW & 8 \\
F0918\_HA\_POL\_090054100\_HAWEHWPE\_CAL\_012-019.fits& 18 & CCW & 8 \\
F0918\_HA\_POL\_09005496\_HAWEHWPE\_CAL\_020-027.fits & 16 & CCW & 8 \\\hline
\end{tabular}
\label{tab:DR2}
\end{table}

\begin{table}
\centering
\caption{Summary of DR1 calibrated files}
\begin{tabular}{lll}
\hline
File Name & Field & Scans \\ \hline\hline
F0775\_HA\_POL\_09005465\_HAWEHWPE\_CAL\_004-011.fits & 1 & 8\\
F0775\_HA\_POL\_09005466\_HAWEHWPE\_CAL\_012-019.fits & 1 & 8\\
F0775\_HA\_POL\_09005466\_HAWEHWPE\_CAL\_028-035.fits & 1 & 8\\
F0775\_HA\_POL\_09005467\_HAWEHWPE\_CAL\_020-027.fits & 2 & 8\\
F0775\_HA\_POL\_090054111\_HAWEHWPE\_CAL\_036-039.fits & 2 & 4\\\hline
F0777\_HA\_POL\_09005468\_HAWEHWPE\_CAL\_005-012.fits & 2 & 8\\
F0777\_HA\_POL\_09005469\_HAWEHWPE\_CAL\_013-020.fits & 3 & 8\\
F0777\_HA\_POL\_09005470\_HAWEHWPE\_CAL\_021-028.fits & 3 & 8\\
F0777\_HA\_POL\_09005471\_HAWEHWPE\_CAL\_029-036.fits & 4 & 8\\
F0777\_HA\_POL\_09005473\_HAWEHWPE\_CAL\_042-049.fits & 5 & 8\\\hline
\end{tabular}
\label{tab:DR1}
\end{table}

\bibliography{FIREPLACE_III}{}

\begin{thebibliography}{}
\expandafter\ifx\csname natexlab\endcsname\relax\def\natexlab#1{#1}\fi
\providecommand{\url}[1]{\href{#1}{#1}}
\providecommand{\dodoi}[1]{doi:~\href{http://doi.org/#1}{\nolinkurl{#1}}}
\providecommand{\doeprint}[1]{\href{http://ascl.net/#1}{\nolinkurl{http://ascl.net/#1}}}
\providecommand{\doarXiv}[1]{\href{https://arxiv.org/abs/#1}{\nolinkurl{https://arxiv.org/abs/#1}}}

\bibitem[{{Aiola} {et~al.}(2020){Aiola}, {Calabrese}, {Maurin}, {Naess},
  {Schmitt}, {Abitbol}, {Addison}, {Ade}, {Alonso}, {Amiri}, {Amodeo},
  {Angile}, {Austermann}, {Baildon}, {Battaglia}, {Beall}, {Bean}, {Becker},
  {Bond}, {Bruno}, {Calafut}, {Campusano}, {Carrero}, {Chesmore}, {Cho},
  {Choi}, {Clark}, {Cothard}, {Crichton}, {Crowley}, {Darwish}, {Datta},
  {Denison}, {Devlin}, {Duell}, {Duff}, {Duivenvoorden}, {Dunkley},
  {D{\"u}nner}, {Essinger-Hileman}, {Fankhanel}, {Ferraro}, {Fox}, {Fuzia},
  {Gallardo}, {Gluscevic}, {Golec}, {Grace}, {Gralla}, {Guan}, {Hall},
  {Halpern}, {Han}, {Hargrave}, {Hasselfield}, {Helton}, {Henderson},
  {Hensley}, {Hill}, {Hilton}, {Hilton}, {Hincks}, {Hlo{\v{z}}ek}, {Ho},
  {Hubmayr}, {Huffenberger}, {Hughes}, {Infante}, {Irwin}, {Jackson}, {Klein},
  {Knowles}, {Koopman}, {Kosowsky}, {Lakey}, {Li}, {Li}, {Li}, {Lokken},
  {Louis}, {Lungu}, {MacInnis}, {Madhavacheril}, {Maldonado}, {Mallaby-Kay},
  {Marsden}, {McMahon}, {Menanteau}, {Moodley}, {Morton}, {Namikawa}, {Nati},
  {Newburgh}, {Nibarger}, {Nicola}, {Niemack}, {Nolta}, {Orlowski-Sherer},
  {Page}, {Pappas}, {Partridge}, {Phakathi}, {Pisano}, {Prince}, {Puddu}, {Qu},
  {Rivera}, {Robertson}, {Rojas}, {Salatino}, {Schaan}, {Schillaci}, {Sehgal},
  {Sherwin}, {Sierra}, {Sievers}, {Sifon}, {Sikhosana}, {Simon}, {Spergel},
  {Staggs}, {Stevens}, {Storer}, {Sunder}, {Switzer}, {Thorne}, {Thornton},
  {Trac}, {Treu}, {Tucker}, {Vale}, {Van Engelen}, {Van Lanen}, {Vavagiakis},
  {Wagoner}, {Wang}, {Ward}, {Wollack}, {Xu}, {Zago}, \& {Zhu}}]{Aiola2020}
{Aiola}, S., {Calabrese}, E., {Maurin}, L., {et~al.} 2020, \jcap, 2020, 047,
  \dodoi{10.1088/1475-7516/2020/12/047}

\bibitem[{{Andersson} {et~al.}(2015){Andersson}, {Lazarian}, \&
  {Vaillancourt}}]{Andersson2015}
{Andersson}, B.~G., {Lazarian}, A., \& {Vaillancourt}, J.~E. 2015, \araa, 53,
  501, \dodoi{10.1146/annurev-astro-082214-122414}

\bibitem[{{Arthur} {et~al.}(2011){Arthur}, {Henney}, {Mellema}, {de Colle}, \&
  {V{\'a}zquez-Semadeni}}]{Arthur2011}
{Arthur}, S.~J., {Henney}, W.~J., {Mellema}, G., {de Colle}, F., \&
  {V{\'a}zquez-Semadeni}, E. 2011, \mnras, 414, 1747,
  \dodoi{10.1111/j.1365-2966.2011.18507.x}

\bibitem[{{Bally} {et~al.}(2014){Bally}, {Rathborne}, {Longmore}, {Jackson},
  {Alves}, {Bressert}, {Contreras}, {Foster}, {Garay}, {Ginsburg}, {Johnston},
  {Kruijssen}, {Testi}, \& {Walsh}}]{Bally2014}
{Bally}, J., {Rathborne}, J.~M., {Longmore}, S.~N., {et~al.} 2014, \apj, 795,
  28, \dodoi{10.1088/0004-637X/795/1/28}

\bibitem[{{Barnes} {et~al.}(2017){Barnes}, {Longmore}, {Battersby}, {Bally},
  {Kruijssen}, {Henshaw}, \& {Walker}}]{Barnes2017}
{Barnes}, A.~T., {Longmore}, S.~N., {Battersby}, C., {et~al.} 2017, \mnras,
  469, 2263, \dodoi{10.1093/mnras/stx941}

\bibitem[{{Battersby} {et~al.}(2020){Battersby}, {Keto}, {Walker}, {Barnes},
  {Callanan}, {Ginsburg}, {Hatchfield}, {Henshaw}, {Kauffmann}, {Kruijssen},
  {Longmore}, {Lu}, {Mills}, {Pillai}, {Zhang}, {Bally}, {Butterfield},
  {Contreras}, {Ho}, {Ott}, {Patel}, \& {Tolls}}]{Battersby2020}
{Battersby}, C., {Keto}, E., {Walker}, D., {et~al.} 2020, \apjs, 249, 35,
  \dodoi{10.3847/1538-4365/aba18e}

\bibitem[{{Butterfield} {et~al.}(2018){Butterfield}, {Lang}, {Morris}, {Mills},
  \& {Ott}}]{Butterfield2018}
{Butterfield}, N., {Lang}, C.~C., {Morris}, M., {Mills}, E.~A.~C., \& {Ott}, J.
  2018, \apj, 852, 11, \dodoi{10.3847/1538-4357/aa886e}

\bibitem[{{Butterfield} {et~al.}(2022){Butterfield}, {Lang}, {Ginsburg},
  {Morris}, {Ott}, \& {Ludovici}}]{Butterfield2022}
{Butterfield}, N.~O., {Lang}, C.~C., {Ginsburg}, A., {et~al.} 2022, \apj, 936,
  186, \dodoi{10.3847/1538-4357/ac887c}

\bibitem[{{Butterfield} {et~al.}(2024{\natexlab{a}}){Butterfield}, {Chuss},
  {Guerra}, {Morris}, {Par{\'e}}, {Wollack}, {Dowell}, {Hankins}, {Karpovich},
  {Siah}, {Staguhn}, \& {Zweibel}}]{Butterfield2023}
[FIREPLACE I], {Butterfield}, N.~O., {Chuss}, D.~T., {Guerra}, J.~A., {et~al.}
  2024{\natexlab{a}}, \apj, 963, 130, \dodoi{10.3847/1538-4357/ad12b9}

\bibitem[{{Butterfield} {et~al.}(2024{\natexlab{b}}){Butterfield}, {Guerra},
  {Chuss}, {Morris}, {Pare}, {Wollack}, {Costa}, {Hankins}, {Staguhn}, \&
  {Zweibel}}]{Butterfield2024}
[FIREPLACE II], {Butterfield}, N.~O., {Guerra}, J.~A., {Chuss}, D.~T., {et~al.}
  2024{\natexlab{b}}, arXiv e-prints, arXiv:2401.01983.
\newblock \doarXiv{2401.01983}

\bibitem[{Cabral \& Leedom(1993)}]{Cabral1993}
Cabral, B., \& Leedom, L.~C. 1993, in Proceedings of the 20th annual conference
  on Computer graphics and interactive techniques, ACM, 263--270

\bibitem[{{Chandrasekhar} \& {Fermi}(1953)}]{CF1953}
{Chandrasekhar}, S., \& {Fermi}, E. 1953, \apj, 118, 116,
  \dodoi{10.1086/145732}

\bibitem[{{Chuss} {et~al.}(2003{\natexlab{a}}){Chuss}, {Davidson}, {Dotson},
  {Dowell}, {Hildebrand}, {Novak}, \& {Vaillancourt}}]{Chuss2003a}
{Chuss}, D.~T., {Davidson}, J.~A., {Dotson}, J.~L., {et~al.}
  2003{\natexlab{a}}, \apj, 599, 1116, \dodoi{10.1086/379538}

\bibitem[{{Chuss} {et~al.}(2003{\natexlab{b}}){Chuss}, {Novak}, {Davidson},
  {Dotson}, {Dowell}, {Hildebrand}, \& {Vaillancourt}}]{Chuss2003b}
{Chuss}, D.~T., {Novak}, G., {Davidson}, J.~A., {et~al.} 2003{\natexlab{b}},
  Astronomische Nachrichten Supplement, 324, 173,
  \dodoi{10.1002/asna.200385099}

\bibitem[{{Contreras} {et~al.}(2013){Contreras}, {Schuller}, {Urquhart},
  {Csengeri}, {Wyrowski}, {Beuther}, {Bontemps}, {Bronfman}, {Henning},
  {Menten}, {Schilke}, {Walmsley}, {Wienen}, {Tackenberg}, \&
  {Linz}}]{Contreras2013}
{Contreras}, Y., {Schuller}, F., {Urquhart}, J.~S., {et~al.} 2013, \aap, 549,
  A45, \dodoi{10.1051/0004-6361/201220155}

\bibitem[{{Davis}(1951)}]{Davis1951}
{Davis}, L. 1951, Physical Review, 81, 890, \dodoi{10.1103/PhysRev.81.890.2}

\bibitem[{{Dotson} {et~al.}(2000){Dotson}, {Davidson}, {Dowell}, {Schleuning},
  \& {Hildebrand}}]{Dotson2000}
{Dotson}, J.~L., {Davidson}, J., {Dowell}, C.~D., {Schleuning}, D.~A., \&
  {Hildebrand}, R.~H. 2000, \apjs, 128, 335, \dodoi{10.1086/313384}

\bibitem[{{Dotson} {et~al.}(2010){Dotson}, {Vaillancourt}, {Kirby}, {Dowell},
  {Hildebrand}, \& {Davidson}}]{Dotson2010}
{Dotson}, J.~L., {Vaillancourt}, J.~E., {Kirby}, L., {et~al.} 2010, \apjs, 186,
  406, \dodoi{10.1088/0067-0049/186/2/406}

\bibitem[{{Ehlerov{\'a}} {et~al.}(2022){Ehlerov{\'a}}, {Palou{\v{s}}},
  {Morris}, {W{\"u}nsch}, {Barna}, \& {Vermot}}]{Ehlerova2022}
{Ehlerov{\'a}}, S., {Palou{\v{s}}}, J., {Morris}, M.~R., {et~al.} 2022, \aap,
  668, A124, \dodoi{10.1051/0004-6361/202244682}

\bibitem[{{Eimer} {et~al.}(2023){Eimer}, {Li}, {Brewer}, {Shi}, {Ali}, {Appel},
  {Bennett}, {Bruno}, {Bustos}, {Chuss}, {Cleary}, {Dahal}, {Datta}, {Denes
  Couto}, {Denis}, {D{\"u}nner}, {Essinger-Hileman}, {Flux{\'a}}, {Hubmayer},
  {Harrington}, {Iuliano}, {Karakla}, {Marriage}, {N{\'u}{\~n}ez}, {Parker},
  {Petroff}, {Reeves}, {Rostem}, {Valle}, {Watts}, {Weiland}, {Wollack}, {Xu},
  \& {Zeng}}]{Eimer2023}
{Eimer}, J.~R., {Li}, Y., {Brewer}, M.~K., {et~al.} 2023, arXiv e-prints,
  arXiv:2309.00675, \dodoi{10.48550/arXiv.2309.00675}

\bibitem[{{Fissel} {et~al.}(2016){Fissel}, {Ade}, {Angil{\`e}}, {Ashton},
  {Benton}, {Devlin}, {Dober}, {Fukui}, {Galitzki}, {Gandilo}, {Klein},
  {Korotkov}, {Li}, {Martin}, {Matthews}, {Moncelsi}, {Nakamura},
  {Netterfield}, {Novak}, {Pascale}, {Poidevin}, {Santos}, {Savini}, {Scott},
  {Shariff}, {Diego Soler}, {Thomas}, {Tucker}, {Tucker}, \&
  {Ward-Thompson}}]{Fissel2016}
{Fissel}, L.~M., {Ade}, P. A.~R., {Angil{\`e}}, F.~E., {et~al.} 2016, \apj,
  824, 134, \dodoi{10.3847/0004-637X/824/2/134}

\bibitem[{{Gordon} {et~al.}(2018){Gordon}, {Lopez-Rodriguez}, {Andersson},
  {Clarke}, {Coude}, {Moullet}, {Richards}, {Shuping}, {Vacca}, \&
  {Yorke}}]{Gordon2018}
{Gordon}, M.~S., {Lopez-Rodriguez}, E., {Andersson}, B.~G., {et~al.} 2018,
  arXiv e-prints, arXiv:1811.03100, \dodoi{10.48550/arXiv.1811.03100}

\bibitem[{{Greenfield} {et~al.}(2014){Greenfield}, {Tollerud}, {Robitaille}, \&
  {Developers}}]{Greenfield2014}
{Greenfield}, P., {Tollerud}, E.~J., {Robitaille}, T., \& {Developers}, A.
  2014, in American Astronomical Society Meeting Abstracts, Vol. 223, American
  Astronomical Society Meeting Abstracts \#223, 255.24

\bibitem[{{Guan} {et~al.}(2021){Guan}, {Clark}, {Hensley}, {Gallardo}, {Naess},
  {Duell}, {Aiola}, {Atkins}, {Calabrese}, {Choi}, {Cothard}, {Devlin},
  {Duivenvoorden}, {Dunkley}, {D{\"u}nner}, {Ferraro}, {Hasselfield}, {Hughes},
  {Koopman}, {Kosowsky}, {Madhavacheril}, {McMahon}, {Nati}, {Niemack}, {Page},
  {Salatino}, {Schaan}, {Sehgal}, {Sif{\'o}n}, {Staggs}, {Vavagiakis},
  {Wollack}, \& {Xu}}]{Guan2021}
{Guan}, Y., {Clark}, S.~E., {Hensley}, B.~S., {et~al.} 2021, \apj, 920, 6,
  \dodoi{10.3847/1538-4357/ac133f}

\bibitem[{{Guerra} {et~al.}(2023){Guerra}, {Lopez-Rodriguez}, {Chuss},
  {Butterfield}, \& {Schmelz}}]{Guerra2023}
{Guerra}, J.~A., {Lopez-Rodriguez}, E., {Chuss}, D.~T., {Butterfield}, N.~O.,
  \& {Schmelz}, J.~T. 2023, \aj, 166, 37, \dodoi{10.3847/1538-3881/acdacd}

\bibitem[{{Hankins} {et~al.}(2020){Hankins}, {Lau}, {Radomski}, {Cotera},
  {Morris}, {Mills}, {Walker}, {Barnes}, {Simpson}, {Herter}, {Longmore},
  {Bally}, {Kasliwal}, {Sabha}, \& {Garc{\'\i}a-Mar{\'\i}n}}]{Hankins2020}
{Hankins}, M.~J., {Lau}, R.~M., {Radomski}, J.~T., {et~al.} 2020, \apj, 894,
  55, \dodoi{10.3847/1538-4357/ab7c5d}

\bibitem[{{Harper} {et~al.}(2018){Harper}, {Runyan}, {Dowell}, {Wirth},
  {Amato}, {Ames}, {Amiri}, {Banks}, {Bartels}, {Benford}, {Berthoud},
  {Buchanan}, {Casey}, {Chapman}, {Chuss}, {Cook}, {Derro}, {Dotson}, {Evans},
  {Fixsen}, {Gatley}, {Guerra}, {Halpern}, {Hamilton}, {Hamlin}, {Hansen},
  {Heimsath}, {Hermida}, {Hilton}, {Hirsch}, {Hollister}, {Hostetter}, {Irwin},
  {Jhabvala}, {Jhabvala}, {Kastner}, {Kov{\'a}cs}, {Lin}, {Loewenstein},
  {Looney}, {Lopez-Rodriguez}, {Maher}, {Michail}, {Miller}, {Moseley},
  {Novak}, {Pernic}, {Rennick}, {Rhody}, {Sandberg}, {Sandford}, {Santos},
  {Shafer}, {Sharp}, {Shirron}, {Siah}, {Silverberg}, {Sparr}, {Spotz},
  {Staguhn}, {Toorian}, {Towey}, {Tuttle}, {Vaillancourt}, {Voellmer},
  {Volpert}, {Wang}, \& {Wollack}}]{Harper2018}
{Harper}, D.~A., {Runyan}, M.~C., {Dowell}, C.~D., {et~al.} 2018, Journal of
  Astronomical Instrumentation, 7, 1840008, \dodoi{10.1142/S2251171718400081}

\bibitem[{{Hatchfield} {et~al.}(2024){Hatchfield}, {Battersby}, {Barnes},
  {Butterfield}, {Ginsburg}, {Henshaw}, {Longmore}, {Lu}, {Svoboda}, {Walker},
  {Callanan}, {Mills}, {Ho}, {Kauffmann}, {Kruijssen}, {Ott}, {Pillai}, \&
  {Zhang}}]{Hatchfield2024}
{Hatchfield}, H.~P., {Battersby}, C., {Barnes}, A.~T., {et~al.} 2024, \apj,
  962, 14, \dodoi{10.3847/1538-4357/ad10af}

\bibitem[{{Henshaw} {et~al.}(2019){Henshaw}, {Ginsburg}, {Haworth}, {Longmore},
  {Kruijssen}, {Mills}, {Sokolov}, {Walker}, {Barnes}, {Contreras}, {Bally},
  {Battersby}, {Beuther}, {Butterfield}, {Dale}, {Henning}, {Jackson},
  {Kauffmann}, {Pillai}, {Ragan}, {Riener}, \& {Zhang}}]{Henshaw2019}
{Henshaw}, J.~D., {Ginsburg}, A., {Haworth}, T.~J., {et~al.} 2019, \mnras, 485,
  2457, \dodoi{10.1093/mnras/stz471}

\bibitem[{{Henshaw} {et~al.}(2022){Henshaw}, {Krumholz}, {Butterfield},
  {Mackey}, {Ginsburg}, {Haworth}, {Nogueras-Lara}, {Barnes}, {Longmore},
  {Bally}, {Kruijssen}, {Mills}, {Beuther}, {Walker}, {Battersby}, {Bulatek},
  {Henning}, {Ott}, \& {Soler}}]{Henshaw2022}
{Henshaw}, J.~D., {Krumholz}, M.~R., {Butterfield}, N.~O., {et~al.} 2022,
  \mnras, 509, 4758, \dodoi{10.1093/mnras/stab3039}

\bibitem[{{Heywood} {et~al.}(2022){Heywood}, {Rammala}, {Camilo}, {Cotton},
  {Yusef-Zadeh}, {Abbott}, {Adam}, {Adams}, {Aldera}, {Asad}, {Bauermeister},
  {Bennett}, {Bester}, {Bode}, {Botha}, {Botha}, {Brederode}, {Buchner},
  {Burger}, {Cheetham}, {de Villiers}, {Dikgale-Mahlakoana}, {du Toit},
  {Esterhuyse}, {Fanaroff}, {February}, {Fourie}, {Frank}, {Gamatham}, {Geyer},
  {Goedhart}, {Gouws}, {Gumede}, {Hlakola}, {Hokwana}, {Hoosen}, {Horrell},
  {Hugo}, {Isaacson}, {J{\'o}zsa}, {Jonas}, {Joubert}, {Julie}, {Kapp},
  {Kenyon}, {Kotz{\'e}}, {Kriek}, {Kriel}, {Krishnan}, {Lehmensiek},
  {Liebenberg}, {Lord}, {Lunsky}, {Madisa}, {Magnus}, {Mahgoub}, {Makhaba},
  {Makhathini}, {Malan}, {Manley}, {Marais}, {Martens}, {Mauch}, {Merry},
  {Millenaar}, {Mnyandu}, {Mokone}, {Monama}, {Mphego}, {New}, {Ngcebetsha},
  {Ngoasheng}, {Ockards}, {Oozeer}, {Otto}, {Passmoor}, {Patel}, {Peens-Hough},
  {Perkins}, {Ramaila}, {Ramanujam}, {Ramudzuli}, {Ratcliffe}, {Robyntjies},
  {Salie}, {Sambu}, {Schollar}, {Schwardt}, {Schwartz}, {Serylak}, {Siebrits},
  {Sirothia}, {Slabber}, {Smirnov}, {Sofeya}, {Taljaard}, {Tasse}, {Tiplady},
  {Toruvanda}, {Twum}, {van Balla}, {van der Byl}, {van der Merwe}, {Van
  Tonder}, {Van Wyk}, {Venter}, {Venter}, {Wallace}, {Welz}, {Williams}, \&
  {Xaia}}]{Heywood2022}
{Heywood}, I., {Rammala}, I., {Camilo}, F., {et~al.} 2022, \apj, 925, 165,
  \dodoi{10.3847/1538-4357/ac449a}

\bibitem[{{Hsieh} {et~al.}(2018){Hsieh}, {Koch}, {Kim}, {Ho}, {Tang}, \&
  {Wang}}]{Hsieh2018}
{Hsieh}, P.-Y., {Koch}, P.~M., {Kim}, W.-T., {et~al.} 2018, \apj, 862, 150,
  \dodoi{10.3847/1538-4357/aacb27}

\bibitem[{{Hunter}(2007)}]{Hunter2007}
{Hunter}, J.~D. 2007, Computing in Science and Engineering, 9, 90,
  \dodoi{10.1109/MCSE.2007.55}

\bibitem[{{Jow} {et~al.}(2018){Jow}, {Hill}, {Scott}, {Soler}, {Martin},
  {Devlin}, {Fissel}, \& {Poidevin}}]{Jow2018}
{Jow}, D.~L., {Hill}, R., {Scott}, D., {et~al.} 2018, \mnras, 474, 1018,
  \dodoi{10.1093/mnras/stx2736}

\bibitem[{{Kendrew} {et~al.}(2013){Kendrew}, {Ginsburg}, {Johnston}, {Beuther},
  {Bally}, {Cyganowski}, \& {Battersby}}]{Kendrew2013}
{Kendrew}, S., {Ginsburg}, A., {Johnston}, K., {et~al.} 2013, \apjl, 775, L50,
  \dodoi{10.1088/2041-8205/775/2/L50}

\bibitem[{{Kov{\'a}cs}(2008)}]{Kovacs2008}
{Kov{\'a}cs}, A. 2008, in Society of Photo-Optical Instrumentation Engineers
  (SPIE) Conference Series, Vol. 7020, Millimeter and Submillimeter Detectors
  and Instrumentation for Astronomy IV, ed. W.~D. {Duncan}, W.~S. {Holland},
  S.~{Withington}, \& J.~{Zmuidzinas}, 70201S, \dodoi{10.1117/12.790276}

\bibitem[{{Kruijssen} {et~al.}(2015){Kruijssen}, {Dale}, \&
  {Longmore}}]{Kruijssen2015}
{Kruijssen}, J.~M.~D., {Dale}, J.~E., \& {Longmore}, S.~N. 2015, \mnras, 447,
  1059, \dodoi{10.1093/mnras/stu2526}

\bibitem[{{Krumholz} \& {Kruijssen}(2015)}]{Krumholz2015}
{Krumholz}, M.~R., \& {Kruijssen}, J.~M.~D. 2015, \mnras, 453, 739,
  \dodoi{10.1093/mnras/stv1670}

\bibitem[{{Lada} {et~al.}(2012){Lada}, {Forbrich}, {Lombardi}, \&
  {Alves}}]{Lada2012}
{Lada}, C.~J., {Forbrich}, J., {Lombardi}, M., \& {Alves}, J.~F. 2012, \apj,
  745, 190, \dodoi{10.1088/0004-637X/745/2/190}

\bibitem[{{Lang} {et~al.}(2010){Lang}, {Goss}, {Cyganowski}, \&
  {Clubb}}]{Lang2010}
{Lang}, C.~C., {Goss}, W.~M., {Cyganowski}, C., \& {Clubb}, K.~I. 2010, \apjs,
  191, 275, \dodoi{10.1088/0067-0049/191/2/275}

\bibitem[{{Lang} {et~al.}(1999){Lang}, {Morris}, \& {Echevarria}}]{Lang1999b}
{Lang}, C.~C., {Morris}, M., \& {Echevarria}, L. 1999, \apj, 526, 727,
  \dodoi{10.1086/308012}

\bibitem[{{Lau} {et~al.}(2013){Lau}, {Herter}, {Morris}, {Becklin}, \&
  {Adams}}]{Lau2013}
{Lau}, R.~M., {Herter}, T.~L., {Morris}, M.~R., {Becklin}, E.~E., \& {Adams},
  J.~D. 2013, \apj, 775, 37, \dodoi{10.1088/0004-637X/775/1/37}

\bibitem[{Lazarian \& Hoang(2007)}]{Lazarian07}
Lazarian, A., \& Hoang, T. 2007, Monthly Notices of the Royal Astronomical
  Society, 378, 910, \dodoi{10.1111/j.1365-2966.2007.11817.x}

\bibitem[{{Le Gouellec} {et~al.}(2023){Le Gouellec}, {Andersson}, {Soam},
  {Schirmer}, {Michail}, {Lopez-Rodriguez}, {Flores}, {Chuss}, {Vaillancourt},
  {Hoang}, \& {Lazarian}}]{LeGouellec2023}
{Le Gouellec}, V. J.~M., {Andersson}, B.~G., {Soam}, A., {et~al.} 2023, \apj,
  951, 97, \dodoi{10.3847/1538-4357/accff7}

\bibitem[{{Li} {et~al.}(2023){Li}, {Eimer}, {Osumi}, {Appel}, {Brewer}, {Ali},
  {Bennett}, {Bruno}, {Bustos}, {Chuss}, {Cleary}, {Couto}, {Dahal}, {Datta},
  {Denis}, {D{\"u}nner}, {Espinoza}, {Essinger-Hileman}, {Flux{\'a} Rojas},
  {Harrington}, {Iuliano}, {Karakla}, {Marriage}, {Miller}, {Novack},
  {N{\'u}{\~n}ez}, {Petroff}, {Reeves}, {Rostem}, {Shi}, {Valle}, {Watts},
  {Weiland}, {Wollack}, {Xu}, {Zeng}, \& {Class Collaboration}}]{Li2023}
{Li}, Y., {Eimer}, J.~R., {Osumi}, K., {et~al.} 2023, \apj, 956, 77,
  \dodoi{10.3847/1538-4357/acf293}

\bibitem[{{Longmore} {et~al.}(2013){Longmore}, {Kruijssen}, {Bally}, {Ott},
  {Testi}, {Rathborne}, {Bastian}, {Bressert}, {Molinari}, {Battersby}, \&
  {Walsh}}]{Longmore2013}
{Longmore}, S.~N., {Kruijssen}, J.~M.~D., {Bally}, J., {et~al.} 2013, \mnras,
  433, L15, \dodoi{10.1093/mnrasl/slt048}

\bibitem[{{Lopez-Rodriguez} {et~al.}(2022){Lopez-Rodriguez}, {Clarke},
  {Shenoy}, {Vacca}, {Coude}, {Arneson}, {Ashton}, {Eftekharzadeh}, {Beck},
  {Beckman}, {Borlaff}, {Clark}, {Dale}, {Martin-Alvarez}, {Ntormousi},
  {Reach}, {Roman-Duval}, {Tassis}, {Harper}, \&
  {Marcum}}]{Lopez-Rodriguez2022}
{Lopez-Rodriguez}, E., {Clarke}, M., {Shenoy}, S., {et~al.} 2022, \apj, 936,
  65, \dodoi{10.3847/1538-4357/ac83ac}

\bibitem[{{Lu} {et~al.}(2015){Lu}, {Zhang}, {Kauffmann}, {Pillai}, {Longmore},
  {Kruijssen}, {Battersby}, \& {Gu}}]{Lu2015}
{Lu}, X., {Zhang}, Q., {Kauffmann}, J., {et~al.} 2015, \apjl, 814, L18,
  \dodoi{10.1088/2041-8205/814/2/L18}

\bibitem[{{Lu} {et~al.}(2019){Lu}, {Zhang}, {Kauffmann}, {Pillai}, {Ginsburg},
  {Mills}, {Kruijssen}, {Longmore}, {Battersby}, {Liu}, \& {Gu}}]{Lu2019}
---. 2019, \apj, 872, 171, \dodoi{10.3847/1538-4357/ab017d}

\bibitem[{{Lu} {et~al.}(2024){Lu}, {Liu}, {Pillai}, {Zhang}, {Liu}, {Gu},
  {Hasegawa}, {Li}, {Tang}, {Hatchfield}, {Issac}, {Liu}, {Luo}, {Mai}, \&
  {Shen}}]{Lu2023}
{Lu}, X., {Liu}, J., {Pillai}, T., {et~al.} 2024, \apj, 962, 39,
  \dodoi{10.3847/1538-4357/ad1395}

\bibitem[{{Mackey} \& {Lim}(2011)}]{Mackey2011}
{Mackey}, J., \& {Lim}, A.~J. 2011, \mnras, 412, 2079,
  \dodoi{10.1111/j.1365-2966.2010.18043.x}

\bibitem[{{Mangilli} {et~al.}(2019){Mangilli}, {Aumont}, {Bernard}, {Buzzelli},
  {de Gasperis}, {Durrive}, {Ferriere}, {Fo{\"e}nard}, {Hughes}, {Lacourt},
  {Misawa}, {Montier}, {Mot}, {Ristorcelli}, {Roussel}, {Ade}, {Alina}, {de
  Bernardis}, {de Gouveia Dal Pino}, {Dubois}, {Engel}, {Guillet}, {Hargrave},
  {Laureijs}, {Longval}, {Maffei}, {Magalhaes}, {Marty}, {Masi}, {Montel},
  {Pajot}, {P{\'e}rot}, {Rodriguez}, {Salatino}, {Saccoccio}, {Savini},
  {Stever}, {Tauber}, {Tibbs}, \& {Tucker}}]{Mangilli2019}
{Mangilli}, A., {Aumont}, J., {Bernard}, J.~P., {et~al.} 2019, \aap, 630, A74,
  \dodoi{10.1051/0004-6361/201935072}

\bibitem[{{Mills} {et~al.}(2018){Mills}, {Ginsburg}, {Immer}, {Barnes},
  {Wiesenfeld}, {Faure}, {Morris}, \& {Requena-Torres}}]{Mills2018}
{Mills}, E.~A.~C., {Ginsburg}, A., {Immer}, K., {et~al.} 2018, \apj, 868, 7,
  \dodoi{10.3847/1538-4357/aae581}

\bibitem[{{Molinari} {et~al.}(2011){Molinari}, {Bally}, {Noriega-Crespo},
  {Compi{\`e}gne}, {Bernard}, {Paradis}, {Martin}, {Testi}, {Barlow}, {Moore},
  {Plume}, {Swinyard}, {Zavagno}, {Calzoletti}, {Di Giorgio}, {Elia},
  {Faustini}, {Natoli}, {Pestalozzi}, {Pezzuto}, {Piacentini}, {Polenta},
  {Polychroni}, {Schisano}, {Traficante}, {Veneziani}, {Battersby}, {Burton},
  {Carey}, {Fukui}, {Li}, {Lord}, {Morgan}, {Motte}, {Schuller},
  {Stringfellow}, {Tan}, {Thompson}, {Ward-Thompson}, {White}, \&
  {Umana}}]{Molinari2011}
{Molinari}, S., {Bally}, J., {Noriega-Crespo}, A., {et~al.} 2011, \apjl, 735,
  L33, \dodoi{10.1088/2041-8205/735/2/L33}

\bibitem[{{Morris}(1989)}]{Morris1989iaus}
{Morris}, M. 1989, in The Center of the Galaxy, ed. M.~{Morris}, Vol. 136, 171

\bibitem[{{Morris}(1993)}]{Morris1993}
{Morris}, M. 1993, \apj, 408, 496, \dodoi{10.1086/172607}

\bibitem[{{Morris}(1996)}]{Morris1996a}
{Morris}, M. 1996, in Unsolved Problems of the Milky Way, ed. L.~{Blitz} \&
  P.~J. {Teuben}, Vol. 169, 247

\bibitem[{{Morris}(2006)}]{Morris2006sum}
{Morris}, M. 2006, in Journal of Physics Conference Series, Vol.~54, Journal of
  Physics Conference Series, 1--9, \dodoi{10.1088/1742-6596/54/1/001}

\bibitem[{{Morris} \& {Serabyn}(1996)}]{Morris1996b}
{Morris}, M., \& {Serabyn}, E. 1996, \araa, 34, 645,
  \dodoi{10.1146/annurev.astro.34.1.645}

\bibitem[{{Morris}(2023)}]{Morris2023}
{Morris}, M.~R. 2023, in Physics and Chemistry of Star Formation: The Dynamical
  ISM Across Time and Spatial Scales, 49, \dodoi{10.48550/arXiv.2301.13469}

\bibitem[{{Nakano} \& {Nakamura}(1978)}]{Nakano1978}
{Nakano}, T., \& {Nakamura}, T. 1978, \pasj, 30, 671

\bibitem[{{Nishiyama} {et~al.}(2009){Nishiyama}, {Tamura}, {Hatano}, {Kanai},
  {Kurita}, {Sato}, {Matsunaga}, {Nagata}, {Nagayama}, {Kandori}, {Nakajima},
  {Kusakabe}, {Sato}, {Hough}, {Sugitani}, \& {Okuda}}]{Nishiyama2009}
{Nishiyama}, S., {Tamura}, M., {Hatano}, H., {et~al.} 2009, \apj, 690, 1648,
  \dodoi{10.1088/0004-637X/690/2/1648}

\bibitem[{{Nishiyama} {et~al.}(2010){Nishiyama}, {Hatano}, {Tamura},
  {Matsunaga}, {Yoshikawa}, {Suenaga}, {Hough}, {Sugitani}, {Nagayama}, {Kato},
  \& {Nagata}}]{Nishiyama2010}
{Nishiyama}, S., {Hatano}, H., {Tamura}, M., {et~al.} 2010, \apjl, 722, L23,
  \dodoi{10.1088/2041-8205/722/1/L23}

\bibitem[{{Novak} {et~al.}(2000){Novak}, {Dotson}, {Dowell}, {Hildebrand},
  {Renbarger}, \& {Schleuning}}]{Novak2000}
{Novak}, G., {Dotson}, J.~L., {Dowell}, C.~D., {et~al.} 2000, \apj, 529, 241,
  \dodoi{10.1086/308231}

\bibitem[{{Novak} {et~al.}(2003){Novak}, {Chuss}, {Renbarger}, {Griffin},
  {Newcomb}, {Peterson}, {Loewenstein}, {Pernic}, \& {Dotson}}]{Novak2003b}
{Novak}, G., {Chuss}, D.~T., {Renbarger}, T., {et~al.} 2003, \apjl, 583, L83,
  \dodoi{10.1086/368156}

\bibitem[{{Par{\'e}} {et~al.}(2019){Par{\'e}}, {Lang}, {Morris}, {Moore}, \&
  {Mao}}]{Pare2019}
{Par{\'e}}, D.~M., {Lang}, C.~C., {Morris}, M.~R., {Moore}, H., \& {Mao}, S.~A.
  2019, \apj, 884, 170, \dodoi{10.3847/1538-4357/ab45ed}

\bibitem[{{Pillai} {et~al.}(2015){Pillai}, {Kauffmann}, {Tan}, {Goldsmith},
  {Carey}, \& {Menten}}]{Pillai2015}
{Pillai}, T., {Kauffmann}, J., {Tan}, J.~C., {et~al.} 2015, \apj, 799, 74,
  \dodoi{10.1088/0004-637X/799/1/74}

\bibitem[{{Plante} {et~al.}(1995){Plante}, {Lo}, \& {Crutcher}}]{Plante1995}
{Plante}, R.~L., {Lo}, K.~Y., \& {Crutcher}, R.~M. 1995, \apjl, 445, L113,
  \dodoi{10.1086/187902}

\bibitem[{{Qu} {et~al.}(2023){Qu}, {Sherwin}, {Madhavacheril}, {Han},
  {Crowley}, {Abril-Cabezas}, {Ade}, {Aiola}, {Alford}, {Amiri}, {Amodeo},
  {An}, {Atkins}, {Austermann}, {Battaglia}, {Battistelli}, {Beall}, {Bean},
  {Beringue}, {Bhandarkar}, {Biermann}, {Bolliet}, {Bond}, {Cai}, {Calabrese},
  {Calafut}, {Capalbo}, {Carrero}, {Carron}, {Challinor}, {Chesmore}, {Cho},
  {Choi}, {Clark}, {C{\'o}rdova Rosado}, {Cothard}, {Coughlin}, {Coulton},
  {Dalal}, {Darwish}, {Devlin}, {Dicker}, {Doze}, {Duell}, {Duff},
  {Duivenvoorden}, {Dunkley}, {D{\"u}nner}, {Fanfani}, {Fankhanel}, {Farren},
  {Ferraro}, {Freundt}, {Fuzia}, {Gallardo}, {Garrido}, {Gluscevic}, {Golec},
  {Guan}, {Halpern}, {Harrison}, {Hasselfield}, {Healy}, {Henderson},
  {Hensley}, {Herv{\'\i}as-Caimapo}, {Hill}, {Hilton}, {Hilton}, {Hincks},
  {Hlo{\v{z}}ek}, {Ho}, {Huber}, {Hubmayr}, {Huffenberger}, {Hughes}, {Irwin},
  {Isopi}, {Jense}, {Keller}, {Kim}, {Knowles}, {Koopman}, {Kosowsky},
  {Kramer}, {Kusiak}, {La Posta}, {Lague}, {Lakey}, {Lee}, {Li}, {Li}, {Limon},
  {Lokken}, {Louis}, {Lungu}, {MacCrann}, {MacInnis}, {Maldonado}, {Maldonado},
  {Mallaby-Kay}, {Marques}, {McMahon}, {Mehta}, {Menanteau}, {Moodley},
  {Morris}, {Mroczkowski}, {Naess}, {Namikawa}, {Nati}, {Newburgh}, {Nicola},
  {Niemack}, {Nolta}, {Orlowski-Scherer}, {Page}, {Pandey}, {Partridge},
  {Prince}, {Puddu}, {Radiconi}, {Robertson}, {Rojas}, {Sakuma}, {Salatino},
  {Schaan}, {Schmitt}, {Sehgal}, {Shaikh}, {Sierra}, {Sievers}, {Sif{\'o}n},
  {Simon}, {Sonka}, {Spergel}, {Staggs}, {Storer}, {Switzer}, {Tampier},
  {Thornton}, {Trac}, {Treu}, {Tucker}, {Ulluom}, {Vale}, {Van Engelen}, {Van
  Lanen}, {van Marrewijk}, {Vargas}, {Vavagiakis}, {Wagoner}, {Wang}, {Wenzl},
  {Wollack}, {Xu}, {Zago}, \& {Zhang}}]{Qu2023}
{Qu}, F.~J., {Sherwin}, B.~D., {Madhavacheril}, M.~S., {et~al.} 2023, arXiv
  e-prints, arXiv:2304.05202, \dodoi{10.48550/arXiv.2304.05202}

\bibitem[{{Reissl} {et~al.}(2020){Reissl}, {Guillet}, {Brauer}, {Levrier},
  {Boulanger}, \& {Klessen}}]{Reissl2020}
{Reissl}, S., {Guillet}, V., {Brauer}, R., {et~al.} 2020, \aap, 640, A118,
  \dodoi{10.1051/0004-6361/201937177}

\bibitem[{{Santos} {et~al.}(2019){Santos}, {Chuss}, {Dowell}, {Houde},
  {Looney}, {Lopez Rodriguez}, {Novak}, {Ward-Thompson}, {Berthoud}, {Dale},
  {Guerra}, {Hamilton}, {Hanany}, {Harper}, {Henning}, {Jones}, {Lazarian},
  {Michail}, {Morris}, {Staguhn}, {Stephens}, {Tassis}, {Trinh}, {Van Camp},
  {Volpert}, \& {Wollack}}]{Santos2019}
{Santos}, F.~P., {Chuss}, D.~T., {Dowell}, C.~D., {et~al.} 2019, \apj, 882,
  113, \dodoi{10.3847/1538-4357/ab3407}

\bibitem[{{Serabyn} {et~al.}(1992){Serabyn}, {Lacy}, \&
  {Achtermann}}]{Serabyn1992}
{Serabyn}, E., {Lacy}, J.~H., \& {Achtermann}, J.~M. 1992, \apj, 395, 166,
  \dodoi{10.1086/171640}

\bibitem[{{Soler} {et~al.}(2013){Soler}, {Hennebelle}, {Martin},
  {Miville-Desch{\^e}nes}, {Netterfield}, \& {Fissel}}]{Soler2013}
{Soler}, J.~D., {Hennebelle}, P., {Martin}, P.~G., {et~al.} 2013, \apj, 774,
  128, \dodoi{10.1088/0004-637X/774/2/128}

\bibitem[{{Tegmark}(1997)}]{Tegmark1997}
{Tegmark}, M. 1997, \apjl, 480, L87, \dodoi{10.1086/310631}

\bibitem[{{Wright} {et~al.}(1996){Wright}, {Hinshaw}, \&
  {Bennett}}]{Wright1996}
{Wright}, E.~L., {Hinshaw}, G., \& {Bennett}, C.~L. 1996, \apjl, 458, L53,
  \dodoi{10.1086/309927}

\bibitem[{{Yusef-Zadeh} {et~al.}(2022){Yusef-Zadeh}, {Arendt}, {Wardle},
  {Heywood}, {Cotton}, \& {Camilo}}]{Yusef-Zadeh2022}
{Yusef-Zadeh}, F., {Arendt}, R.~G., {Wardle}, M., {et~al.} 2022, \apjl, 925,
  L18, \dodoi{10.3847/2041-8213/ac4802}

\bibitem[{{Yusef-Zadeh} {et~al.}(2004){Yusef-Zadeh}, {Hewitt}, \&
  {Cotton}}]{Yusef-Zadeh2004}
{Yusef-Zadeh}, F., {Hewitt}, J.~W., \& {Cotton}, W. 2004, \apjs, 155, 421,
  \dodoi{10.1086/425257}

\bibitem[{{Yusef-Zadeh} \& {Morris}(1987{\natexlab{a}})}]{Yusef-Zadeh1987a}
{Yusef-Zadeh}, F., \& {Morris}, M. 1987{\natexlab{a}}, \apj, 320, 545,
  \dodoi{10.1086/165572}

\bibitem[{{Yusef-Zadeh} \& {Morris}(1987{\natexlab{b}})}]{Yusef-Zadeh1987}
---. 1987{\natexlab{b}}, \aj, 94, 1178, \dodoi{10.1086/114555}

\bibitem[{{Yusef-Zadeh} \& {Morris}(1987{\natexlab{c}})}]{YM1987}
---. 1987{\natexlab{c}}, \apj, 322, 721, \dodoi{10.1086/165767}

\bibitem[{{Yusef-Zadeh} {et~al.}(1984){Yusef-Zadeh}, {Morris}, \&
  {Chance}}]{YMC1984}
{Yusef-Zadeh}, F., {Morris}, M., \& {Chance}, D. 1984, \nat, 310, 557,
  \dodoi{10.1038/310557a0}

\bibitem[{{Yusef-Zadeh} {et~al.}(2009){Yusef-Zadeh}, {Hewitt}, {Arendt},
  {Whitney}, {Rieke}, {Wardle}, {Hinz}, {Stolovy}, {Lang}, {Burton}, \&
  {Ramirez}}]{Yusef-Zadeh2009}
{Yusef-Zadeh}, F., {Hewitt}, J.~W., {Arendt}, R.~G., {et~al.} 2009, \apj, 702,
  178, \dodoi{10.1088/0004-637X/702/1/178}

\bibitem[{{Zhao} {et~al.}(2016){Zhao}, {Morris}, \& {Goss}}]{Zhao2016}
{Zhao}, J.-H., {Morris}, M.~R., \& {Goss}, W.~M. 2016, \apj, 817, 171,
  \dodoi{10.3847/0004-637X/817/2/171}

\end{thebibliography}
\bibliographystyle{aasjournal}

\end{document}